\documentclass[useAMS,usenatbib,fleqn]{mn2e}
\usepackage{graphicx,color,pdfpages,amsmath,float,subfig}
\usepackage{makecell}
\usepackage[table, dvipsnames]{xcolor}
\newcommand{\eagle}{\textsc{eagle}{}}

\newcommand{\cloudy}{\textsc{cloudy}{ }}
\newcommand{\subfind}{\textsc{subfind}{ }}

\def\ltsima{$\; \buildrel < \over \sim \;$}
\def\simlt{\lower.5ex\hbox{\ltsima}}
\def\gtsima{$\; \buildrel > \over \sim \;$}
\def\simgt{\lower.5ex\hbox{\gtsima}}
\def\la{$\langle$}

\renewcommand{\vec}[1]{ {\bmath #1} } 

\title[Numerical convergence of simulated galaxies]{Numerical convergence of hydrodynamical simulations of galaxy formation: the abundance and internal structure of galaxies and their cold dark matter haloes}

\author[Ludlow et al.] {\parbox{18cm}{
    Aaron D. Ludlow$^{1,\star}$,
    Joop Schaye$^{2}$,
    Matthieu Schaller$^{2}$,
    Richard Bower$^{3}$
  }\vspace{0.3cm}\\
  $^{1}${International Centre for Radio Astronomy Research, University of Western Australia, 35 Stirling Highway, Crawley,}\\
  {Western Australia, 6009, Australia}\\
  $^{2}${Leiden Observatory, Leiden University, PO Box 9513, 2300 RA Leiden, the Netherlands}\\
  $^{3}${Institute for Computational Cosmology, Department of Physics, Durham University, Durham DH1 3LE, U.K.}\\
}

\begin{document}

\maketitle 

\begin{abstract}
 We address the issue of numerical convergence in cosmological smoothed particle hydrodynamics simulations using 
 a suite of runs drawn from the \eagle{} project. Our simulations adopt subgrid models that produce realistic galaxy
 populations at a fiducial mass and force resolution, but systematically vary the latter in order to study their
 impact on galaxy properties. We provide several analytic criteria that help guide the selection of gravitational
 softening for hydrodynamical simulations, and present results from runs that both adhere to and deviate from them.
 Unlike dark matter-only simulations, hydrodynamical simulations exhibit a strong sensitivity to gravitational
 softening, and care must be taken when selecting numerical parameters. Our results--which focus mainly on star
 formation histories, galaxy stellar mass functions and sizes--illuminate three main considerations. First, softening
 imposes a minimum resolved escape speed, $v_\epsilon$, due to the binding energy between gas particles. Runs that
 adopt such small softening lengths that $v_\epsilon \simgt 10\,{\rm km s^{-1}}$ (the sound speed in ionised
 $\sim 10^4\,{\rm K}$ gas) suffer from reduced effects of photo-heating. Second, feedback from stars or active
 galactic nuclei may suffer from numerical over-cooling if the gravitational softening length is chosen below a
 critical value, $\epsilon_{\rm eFB}$. Third, we note that small softening lengths exacerbate the segregation of
 stars and dark matter particles in halo centres, often leading to the counter-intuitive result that galaxy sizes
 {\em increase} as softening is reduced. The structure of dark matter haloes in hydrodynamical runs respond to
 softening in a way that reflects the sensitivity of their galaxy populations to numerical parameters.
\end{abstract}

\renewcommand{\thefootnote}{\fnsymbol{footnote}}
\footnotetext[1]{E-mail: aaron.ludlow@icrar.org} 

\section{Introduction}
\label{SecIntro}

Collisionless N-body simulations offer a stable platform for modelling the non-linear
structure of dark matter-dominated systems. Once a cosmological model has been specified
and initial conditions chosen, only a small number of {\em numerical} parameters govern
the system's evolution. This has enabled numerical studies to elucidate the impact that these
parameters have on the outcome of simulations
\citep[e.g.][]{Power2003,Springel2008b,Stadel2009,Navarro2010,VeraCiro2011,Hopkins2018,vandenBosch2018a},
leading to well-established ``convergence criteria'' that help unravel numerically robust results
from those more susceptible to the details of the underlying numerical framework.

As a result of these efforts there is now concurrence on a number issues. For example,
the detailed inner structure of dark matter (DM) haloes can be trusted down to radii 
of order 5 per cent of their virial radius provided they contain several thousand particles
\citep[e.g.][hereafter Paper I]{Ludlow2019a}, while mass functions converge to better
than $\approx 5$ or 10 per cent (depending on how mass is defined) for haloes composed of
$\simgt 100$ particles \citep[see, e.g.][]{Efstathiou1988,Jenkins2001,Tinker2008,Ludlow2019a}.
Such convergence is important: it suggests that cosmological simulations simultaneously resolve
halo formation on a variety of scales and across a range of redshifts. This is essential for
modelling galaxy formation, particularly in {\rm cold} dark matter (CDM) models 
where haloes and galaxies assemble hierarchically through mergers of their (poorly-resolved)
ascendants.

Yet the Universe is not all dark, and any convincing model of structure formation must
also robustly predict the evolution of its visible constituents. Hydrodynamical
simulations co-evolve collisionless dark matter with the baryonic component necessary
for the formation of luminous galaxies. Current efforts in this area mainly follow one of three
numerical approaches: Lagrangian smoothed particle hydrodynamics \citep[SPH;][]{Gingold1977,Monaghan1992,Katz1996,Price2007},
Eulerian adaptive mesh refinement methods \citep[AMR; e.g.][]{Kravtsov1999,Teyssier2002,Knebe2001,ENZO}
or, more recently, Lagrangian ``moving mesh''  \citep[e.g.][]{Springel2010,Vandenbroucke2016} or
Godunov-type methods \citep[e.g.][]{Gaburov2011,Hopkins2015}. Each of these have their own virtues and
weaknesses, and often yield conflicting solutions even when applied to simple test problems
\citep{OShea2005,Agertz2007,Tasker2008,Hubber2013}.

Inconsistency between numerical methods, however, is likely overwhelmed by uncertainties
in the physical processes responsible for galaxy formation
\citep[e.g.][]{Scannapieco2012,Schaller2015b,Sembolini2016}. Star formation, feedback
from supernovae (SNe) and active galactic nuclei (AGN), metal enrichment and viscosity,
for example, all occur on physical scales  not resolved by simulations. Instead, they must
be modelled using ``subgrid'' techniques -- parameterisations of the unresolved processes
employed within each resolution element or smoothing kernel
\citep[see, e.g.,][for recent discussions]{SomervilleDave2015,Schaye2015}.
To make matters worse, the physical processes that these models attempt to emulate are often poorly constrained by
observation and models must be calibrated so that simulations sensibly reproduce some
desired set of galaxy observations.

This necessarily imparts large ambiguities on galaxy formation models since different
groups employ not only different numerical methods but also different implementations of
subgrid processes and the parameters that govern them. As a result, even simulations
starting from the {\em same} initial conditions, but carried out using different
hydrodynamical codes, do not converge to similar solutions.
This was pointed out by \citet{Scannapieco2012}, who used 9 different 
hydrodynamical codes to simulate the formation of a Milky Way-mass galaxy ($\sim 10^{12}\,{\rm M_\odot}$)
from two sets of initial conditions that differed only in numerical resolution.
Despite sharing initial conditions,
galaxies simulated with different codes possessed a wide variety of visual morphologies, from star-forming
disks to red-and-dead ellipticals, had final stellar masses that varied by almost an order
of magnitude, and instantaneous star formation rates (SFRs) that differed by a factor $\sim 10^3$;
some ``observables'' even spanned the extremes of available observational data.
Although the majority of the galaxy-to-galaxy variation was driven by different implementations
of feedback, even the same code did not produce consistent galaxy properties when run at
different mass resolution. Systematic tests such as these highlight the uncertainties inherent
to galaxy formation simulations.

Nevertheless, improvements are being made, both to numerical algorithms and
to subgrid models, and studies continue to illuminate their complex inter-dependence 
and impact on simulation results.
For example, \citet{Hopkins2018} recently carried out a systematic study of the impact of
subgrid physics and numerical parameters (primarily mass and force resolution) on the
outcome of ``zoomed'' simulations of {\em individual} galaxies, concluding that convergence is more easily achieved
in some galaxy properties than in others. For example, stellar mass and baryonic
mass profiles are more robust to changes in resolution than metal abundance and visual
morphology, whereas gaseous winds and properties of the circum-galactic medium (CGM) are more
sensitive. They conclude that high-resolution simulations of galaxy formation yield numerically
robust results provided they: resolve 1) Toomre masses and 2) individual SNe,
and 3) ensure (energy, mass, momentum) conservation during feedback coupling.

One may assume, naively, that improving our understanding of galaxy formation through
simulation merely requires improvements to hydrodynamic algorithms and a better understanding
and treatment of the complex (subgrid) physical processes involved. Yet it was recently 
pointed out that minute and apparently superficial changes to a simulation's set-up 
cultivate rather large, stochastic differences in the properties of individual galaxies
at some later stage. \citet{Keller2018}, for example, showed that floating-point round-off
errors and random number generators can give rise to large differences in the star formation
histories (SFHs), properties of the CGM and stellar masses of individual galaxies, particularly when mergers are involved.
Building on this, \citet{Genel2018} quantified the ``butterfly effect'' by comparing the outcome of
pairs of cosmological simulations that differed from one another by small, random perturbations to
initial particle positions, but were identical in all other respects. Although the {\em statistical}
properties of the galaxies remained unchanged, properties of {\em individual} galaxies diverged from
one another stochastically. They conclude that chaotic galaxy-to-galaxy variations can account
for up to 50 per cent of the variance in simulated galaxy scaling relations, such as the star
formation-mass and Tully-Fisher relations. Both studies emphasized that, while seemingly
deterministic, galaxy formation simulations are by no means immune chaos and stochasticity
\citep[see also, e.g.,][]{Thiebaut2008,Sellwood2009,ElZant2018}, and that care must be taken when
interpreting the impact that different physical models have on galaxy properties, particularly
for studies relying on ``zoomed'' simulations of a small number of objects. 

These studies should inspire future efforts to better understand the role that numerical
techniques, subgrid models and their associated parameters have on galaxy formation simulations. 
The task will not be simple, but will be worthwhile.
Motivated by this, we report here initial results from a new suite of hydrodynamical simulations
designed to illuminate the impact of {\em numerical} parameters (primarily mass and force resolution)
on simulation results. We focus our analysis on the statistical properties of
galaxies and their dark matter haloes, considering separately their internal structure and total
abundance. 

We stress that the results of our study mainly target SPH simulations
whose subgrid physics models resemble those adopted for the \eagle{} programme. It will become
clear in later sections that numerical and subgrid parameters are strongly coupled, implying that
simulation results affected by numerical resolution may be compensated by modifying one or
more parameters controlling the subgrid physics. Our intent for this paper is not to provide a road map
to navigate these changes, but rather to emphasize the often detrimental effects that numerical
parameters, if improperly chosen, can have on well-calibrated subgrid models. For that reason, the
majority of our runs will employ the ``Reference'' or ``Recalibrated'' subgrid models adopted
for the \eagle{} project \citep[see][for details]{Schaye2015,Crain2015}, and test how sensitive the predictions
of these models are to changing the numerical resolution. Unlike \citet{Hopkins2018}, our simulations
target (small) cosmological volumes (12.5 cubic Mpc) as opposed to individual objects, allowing us to
assess convergence across a broad
range of scales, and in haloes resolved with a few hundred to many thousands of particles. We
consider this an essential exercise given the hierarchical nature of galaxy formation in the
standard CDM model, where poorly-resolved low-mass galaxies may influence the later formation and
evolution of more massive systems.

We organize our paper as follows. In section~\ref{SecS} we introduce our simulation
suite, providing details of the relevant numerical parameters and subgrid models, and pertinent aspects of
their post-processing. In section~\ref{Expectations} we provide the analytic background
that will help guide the interpretation of our results, which we present in section~\ref{Results}.
We end with a discussion of our findings and outlook for future work in section~\ref{conclusions}.

\section{Simulations and analysis}
\label{SecS}

\begin{center}
  \begin{table*}
    \caption{Numerical parameters used in our simulations. The first column provides a label for
      each run; LXXNXXX encodes the comoving box side-length $L$ and number of particles (gas or DM) $N_{\rm p}$ on
      a side. This is followed by an identifier for the run type: Recal and Ref are, respectively,
      the ``recalibrated'' and ``Reference'' \eagle{} subgrid models; NR refers to runs carried out
      with non-radiative hydrodynamics. The dark matter, $m_{\rm DM}$, and gas particle masses,
      $m_{\rm g}$, are also provided. Softening lengths $\epsilon_{\rm CM}$, initially kept fixed in
      co-moving coordinates, reach a maximum physical value $\epsilon_{\rm phys}$ at redshift
      $z_{\rm phys}$, after which they remain constant in proper coordinates. {\tt{ErrTolIntAcc}}
      and the Courant Factor control the timestep size for orbit integration for gravity and 
      hydrodynamics. Rows corresponding to our fiducial models have been highlighted using grey shading.}
    \begin{tabular}{c c c c c c c c c c c c c}\hline \hline
      & Name & ${\rm Model}$ &$N_{\rm p}$ &  $m_{\rm DM}$          &  $m_{\rm g}$           &$\epsilon_{\rm phys}$  & $\epsilon_{\rm CM}$ & $z_{\rm phys}$ & \tt{ErrTolIntAcc} & Courant \\
      &      &               &            &[$10^5\, {\rm M}_\odot$]&[$10^5\, {\rm M}_\odot$]& $[{\rm pc}]$          & $[{\rm pc}]$        &                &                   & Factor  \\\hline
      & L12N376 & Recal & $376$ & 12.1 & 2.26 &  2800.0  & 10640.0 & 2.8 & 0.025  & 0.15  \\
      & L12N376 & Recal & $376$ & 12.1 & 2.26 &  1400.0  & 5320.0  & 2.8 & 0.025  & 0.15  \\
      & L12N376 & Recal & $376$ & 12.1 & 2.26 &   700.0  & 2660.0  & 2.8 & 0.025  & 0.15  \\
      & L12N376 & Recal & $376$ & 12.1 & 2.26 &   700.0  & 1660.0  & 1.3 & 0.025  & 0.15  \\
      \rowcolor{lightgray}
      & L12N376 & Recal & $376$ & 12.1 & 2.26 &   350.0  & 1330.0  & 2.8 & 0.025  & 0.15  \\
      & L12N376 & Recal & $376$ & 12.1 & 2.26 &   175.0  & 665.0   & 2.8 & 0.025  & 0.15  \\
      & L12N376 & Recal & $376$ & 12.1 & 2.26 &   87.5   & 332.5   & 2.8 & 0.025  & 0.15  \\
      & L12N376 & Recal & $376$ & 12.1 & 2.26 &   43.75  & 166.25  & 2.8 & 0.025  & 0.15  \\\vspace{0.25cm}
      & L12N376 & Recal & $376$ & 12.1 & 2.26 &   21.88  & 83.13   & 2.8 & 0.025  & 0.15  \\

      & L12N376 & Recal & $376$ & 12.1 & 2.26 &   700.0  & 10500.0 & 14 & 0.025 & 0.15  \\ 
      & L12N376 & Recal & $376$ & 12.1 & 2.26 &   43.75  & 656.25  & 14 & 0.025 & 0.15  \\ \vspace{0.25cm}
      & L12N376 & Recal & $376$ & 12.1 & 2.26 &   21.88  & 328.1   & 14 & 0.025 & 0.15  \\

      & L12N188 & Ref   & $188$ & 97.0 & 18.1 &  2800.0  & 10640.0 & 2.8 & 0.025  & 0.15  \\
      & L12N188 & Ref   & $188$ & 97.0 & 18.1 &  1400.0  & 5320.0  & 2.8 & 0.025  & 0.15  \\
      & L12N188 & Ref   & $188$ & 97.0 & 18.1 &  1400.0  & 3420.0  & 1.3 & 0.025  & 0.15  \\
      \rowcolor{lightgray}
      & L12N188 & Ref   & $188$ & 97.0 & 18.1 &   700.0  & 2660.0  & 2.8 & 0.025  & 0.15  \\
      & L12N188 & Ref   & $188$ & 97.0 & 18.1 &   350.0  & 1330.0  & 2.8 & 0.025  & 0.15  \\
      & L12N188 & Ref   & $188$ & 97.0 & 18.1 &   175.0  & 665.0   & 2.8 & 0.025  & 0.15  \\
      & L12N188 & Ref   & $188$ & 97.0 & 18.1 &   87.5   & 332.5   & 2.8 & 0.025  & 0.15  \\
      & L12N188 & Ref   & $188$ & 97.0 & 18.1 &   43.75  & 166.25  & 2.8 & 0.025  & 0.15  \\\vspace{0.25cm}
      & L12N188 & Ref   & $188$ & 97.0 & 18.1 &   21.88  & 83.13   & 2.8 & 0.025  & 0.15  \\

      & L12N188 & NR & $188$ & 97.0 & 18.1 &$4.48\times 10^4$ &$4.48\times 10^4$ & 0 & 0.025  & 0.15 \\
      & L12N188 & NR & $188$ & 97.0 & 18.1 &$2.24\times 10^4$ &$2.24\times 10^4$ & 0 & 0.025  & 0.15 \\
      & L12N188 & NR & $188$ & 97.0 & 18.1 &$1.12\times 10^4$ &$1.12\times 10^4$ & 0 & 0.025  & 0.15 \\
      & L12N188 & NR & $188$ & 97.0 & 18.1 &   5600.0         &   5600.0         & 0 & 0.025  & 0.15 \\
      & L12N188 & NR & $188$ & 97.0 & 18.1 &   2800.0         &   2800.0         & 0 & 0.025  & 0.15 \\
      & L12N188 & NR & $188$ & 97.0 & 18.1 &   1400.0         &   1400.0         & 0 & 0.025  & 0.15 \\
      & L12N188 & NR & $188$ & 97.0 & 18.1 &   700.0          &   700.0          & 0 & 0.025  & 0.15 \\
      & L12N188 & NR & $188$ & 97.0 & 18.1 &   700.0          &   700.0          & 0 & 0.0025 & 0.05 \\
      & L12N188 & NR & $188$ & 97.0 & 18.1 &   350.0          &   350.0          & 0 & 0.025  & 0.15 \\
      & L12N188 & NR & $188$ & 97.0 & 18.1 &   175.0          &   175.0          & 0 & 0.025  & 0.15 \\
      & L12N188 & NR & $188$ & 97.0 & 18.1 &   87.5           &   87.5           & 0 & 0.025  & 0.15 \\\hline
\end{tabular}
    \label{TabSimParam}
  \end{table*}
\end{center}

\subsection{Simulation set-up}
\label{SSecSimSetup}

All simulations were carried out with a modified version of the N-body
SPH code \textsc{gadget3}, which includes substantial improvements to the hydrodynamics
scheme, time-stepping criteria and subgrid physics models \citep{Springel2005b,Schaye2015}.
Our suite includes a number of runs that follow {\em only} collisionless DM and adiabatic
hydrodynamics, as well as others that invoke the full subgrid physics of the \eagle{} code.
Our entire suite of simulations was carried out in the same $L=12.5$ (cubic comoving) Mpc
volume described in Paper I; we reiterate the most important aspects of the simulation here,
for completeness.

Cosmological parameters are taken from the Planck I data release \citep{Planck2014};
their values are as follows: 
1) the dimensionless Hubble constant is $h\equiv H_0/(100\, {\rm km\, s^{-1}\, Mpc})=0.6777$;
2) the $z=0$ linear rms density fluctuation in 8 $h^{-1}{\rm Mpc}$ spheres, $\sigma_8=0.8288$;
3) the power-law index of primordial adiabatic perturbations, $n_s=0.9661$;
4) the primordial abundance of helium, $Y=0.248$;
5) finally, the cosmic density parameters, $\Omega_{\rm M}=1-\Omega_\Lambda=0.307$ and
$\Omega_{\rm bar}=\Omega_{\rm M}-\Omega_{\rm DM}=0.04825$, give the total energy density
of species $i$ in units of the redshift-dependent closure density,
$\rho_{\rm crit}(z)\equiv 3 H^2(z)/8\pi G$ (for convenience, we define the present-day
critical density as $\rho_0\equiv\rho_{\rm crit}[z=0]$).

Two sets of initial conditions were generated by perturbing a glass of DM particles using second-order
Lagrangian perturbation theory consistent with a starting redshift of $z=127$; the Lagrangian
particle loads were sampled with $N^3_{\rm p}=188^3$ and $376^3$ particles of DM using the same
amplitudes and phases for mutually-resolved modes (note that $N_{\rm p}$ refers to the number
  of baryonic or DM particles along a side of the simulation box). Gas particles
were created by duplicating those of DM, whose masses were reduced by a factor $1-\Omega_{\rm bar}/\Omega_{\rm M}$.
The resulting DM particle masses are $m_{\rm DM}=1.21\times 10^6\,{\rm M_\odot}$
and $9.70\times 10^6\,{\rm M_\odot}$ for the high- and low-resolution runs, respectively;
the primordial gas particle masses\footnote{Because of
  star formation and stellar mass loss, gas and star particle masses do not, in general, remain constant
  throughout the simulation. To keep matters simple, we will, when necessary, compare the masses of
  baryonic structures to the equivalent mass of {\em primordial} gas particles, which we denote $m_{\rm g}$.}
are $m_{\rm g}=2.26\times 10^5\,{\rm M_\odot}$ and $1.81\times 10^6\,{\rm M_\odot}$, respectively, and are
equivalent to those used for the ``high-'' and ``intermediate-resolution'' runs of the original
\eagle{} project. In the nomenclature of \citet{Schaye2015}, these runs are labelled L0012N0188 and L0012N00376,
but we will often distinguish them simply by quoting the number of particles per species, $N^3_{\rm p}$. 

Our simulations have uniform mass resolution and adopt equal gravitational softening
lengths for all particles types (DM, gas and stars), thus fixing the ratio of softening to the mean inter-particle spacing,
$\epsilon/(L/N_{\rm p})$. As in Paper I, Plummer-equivalent softening lengths adopted for the original
\eagle{} programme will be referred to as $\epsilon_{\rm fid}$, the ``fiducial softening''. For {\textsc{eagle}},
$\epsilon/(L/N_{\rm p})\approx 0.011$ in proper coordinates at $z\leq 2.8$, corresponding to $\epsilon_{\rm fid}=700\,{\rm pc}$
and $350\,{\rm pc}$ for our $N^3_{\rm p}=188^3$ and $376^3$ runs, respectively. For $z>z_{\rm phys}=2.8$,
$\epsilon$ was fixed in comoving coordinates to a value of $\epsilon_{\rm fid}/(L/N_{\rm p})=0.04$; or
$2660\,{\rm pc}$ and $1330\,{\rm pc}$, respectively. When necessary, we will quote softening lengths at
specific redshifts, which will be specified explicitly, but will often distinguish runs by simply quoting
the {\em present-day} physical softening length, hereafter denoted $\epsilon_0\equiv \epsilon(z=0)$.

We report results for a series of runs that vary $\epsilon_0$ by sequent factors of 2 above and below \eagle's
fiducial values; a limited number of these--used mainly for illustration--also vary $z_{\rm phys}$. In all cases,
the SPH kernel support size is restricted to values $l_{\rm hsml}^{\rm min}\geq 0.1\times \epsilon_{\rm sp}$, where
\begin{equation}
  \epsilon_{\rm sp}\equiv 2.8\times\epsilon
\end{equation}
is the {\em spline} softening length.

The majority of our runs adopt
the same timestep criteria for gravity ({\tt{ErrTolIntAcc}}=0.025) and hydrodynamics (a Courant factor of 0.05),
but we have verified that our results are insensitive to this choice (Figure~\ref{fig1} and Appendix~\ref{sA2}).
Table~\ref{TabSimParam} provides a detailed summary of all numerical parameters relevant to our runs. 

All of the results presented in the following sections are specific to runs that adopt equal
numbers of DM and (primordial) gas particles, and the same softening lengths for all particle types.
Simulations that employ different or adaptive softening lengths for baryon or DM particles,
or different {\em numbers} of each particle species, may yield different results
\citep[see][for an example of the latter]{Ludlow2019b}. We defer a thorough investigation of
numerical convergence in these regimes to future work.

\subsection{Summary of subgrid physics}
\label{SSecSubgrid}

Cosmological simulations of large volumes lack the mass and spatial
resolution to directly capture the main physical processes responsible for galaxy formation and
evolution; these processes must be implemented using subgrid models. For completeness, we
summarize below the subgrid physics adopted for the \eagle{} project that is most relevant to our work.

Radiative heating and cooling rates are modelled on a per-particle
basis assuming gas is optically thin and in ionization equilibrium \citep{Wiersma2009a}. The rates are
interpolated from tables generated using \cloudy \citep{Ferland1998} and depend on the temperature and
density of a gas particle, its elemental abundance, and on redshift.
We implement hydrogen reionization by invoking a spatially-constant, time-dependent photo-ionizing
background \citep{Haardt_and_Madau_01} at $z=11.5$ and injecting 2 eV per Hydrogen atom
instantaneously (note that reionization is neglected in all non-radiative simulations, as well
as in several full-physics runs).

Because our simulations do not resolve the cold phase of the interstellar medium (ISM; ${\rm T}\ll 10^4\,{\rm K}$), we impose a
temperature floor corresponding to a ($\gamma_{\rm eos}=4/3$) polytrope with ${\rm T_{eos}}=8000\,{\rm K}$
at $n_{\rm H}=0.1\,{\rm cm^{-3}}$. Star formation occurs stochastically in dense gas at a rate given by
the pressure law of \citet{Schaye2008}, which is based on the Kennicutt-Schmidt (KS) star formation law
\citep{Kennicutt1989},
i.e. $\dot{\Sigma_\star}\propto \Sigma_{\rm gas}^n$, where $\Sigma_{\rm gas}$ and $\dot{\Sigma}_\star$ are the
gas and star formation rate surface densities, respectively. The slope of the KS-law is set to $n=1.4$ for densities
$n_{\rm H}\leq n_{\rm SB}=10^3\,{\rm cm^{-3}}$ and (to approximate the star-burst phase) to $n=2.0$ at higher densities. 
We adopt the \citet{Schaye2004} metallicity-dependent threshold for star formation, $n_{\rm H}^\star$, in order to emulate
the more efficient
transition from a warm to a cold, molecular phase in the metal-enriched ISM. Star
formation is also limited to gas particles whose overdensity exceeds 57.7 times the cosmic mean.

Each star particle is assumed to host a simple stellar population whose initial mass function is consistent with
the proposal of \citet{Chabrier2003b}; stellar evolution and mass loss are as described by \citet{Wiersma2009b}.
Stars reaching the ends of their lives inject energy into their surroundings stochastically in line with the thermal feedback model of
\citet{DallaVecchia2012}. The model circumvents numerical radiative losses of feedback energy by imposing
temperature increments, $\Delta T_{\rm SF}$, upon fluid resolution elements affected by SNe rather than
by injecting energy directly. In the latter approach, the SNe energy is spread across a far
larger mass (the mass of {\em at least} one gas particle) than in reality (of order a few ${\rm M_\odot}$), 
resulting in a smaller temperature increase and a shorter
cooling time. This weakens--or in some cases, suppresses entirely--the pressure gradients that give rise to
feedback-driven outflows. All of our runs adopt $\Delta T_{\rm SF}=10^{7.5}\,{\rm K}$.

Our simulations also include a subgrid prescription for the formation and growth of black holes (BHs)
and their associated feedback. The former is based on the approach of \citet{Springel2005c}
but includes modifications advocated by \citet{BoothSchaye2009} and by \citet{Rosas-Guevara2015}. BHs are
seeded in FOF haloes (identified on-the-fly using a linking length $b=0.2$) the first time they exceed a
mass threshold of $10^{10}h^{-1}{\rm M_\odot}$. This is done by converting the highest-density gas particle into a
(collisionless) BH, which is assigned an initial seed mass of $10^{5}h^{-1}{\rm M_\odot}$. BHs grow by
(stochastically) accreting nearby gas particles and those with mass $\leq 10^2\,m_{\rm gas}$ are regularly
steered towards the centres of their host haloes. Feedback energy associated with gas accretion and BH growth
couples to the ISM using a method based on \citet{BoothSchaye2009}. In Appendix ~\ref{sA1} we compare results
from two Recal models that include or ignore BH formation and feedback from AGN.

As discussed extensively in \citet{Schaye2015} \citep[see also][]{Crain2015}, subgrid parameters
that govern these processes must be calibrated
so that simulations match certain fundamental aspects of the observed galaxy population. The \eagle{}
programme owes part of its success to the calibration of AGN and stellar feedback model parameters 
which ensures that simulated galaxies match the observed mass-size relation, galaxy stellar
mass function (GSMF) and the relation between the masses of galaxies and their
BHs at $z=0$. Subgrid parameters were optimized independently at different mass resolutions,
though variations between them were small. In \citet{Schaye2015}, these are referred to as the ``Reference''
(or Ref, for $N^3_{\rm p}=188^3$) and ``Recalibrated'' subgrid models (Recal, for $N^3_{\rm p}=376^3$).
We note that, at the mass scales accessible to our simulations and for the fiducial numerical parameters
adopted for \textsc{eagle}, the Ref and Recal models yield similar results for the GSMF and galaxy mass-size relation.  

Motivated by this, we adopt the ``Recalibrated'' subgrid parameters for our $N^3_{\rm p}=376^3$ runs, and
``Reference'' parameters for $N^3_{\rm p}=188^3$, but have verified that none of
the results or conclusions presented in the following sections are effected by this
choice\footnote{All of our $N^3_{\rm p}=188^3$ simulations were carried out using both sets of 
  parameters--Reference {\em and} Recalibrated--but we have chosen to present results only for Ref.
  The similarities between Ref and Recal models at fixed mass resolution is briefly discuss in
  Appendix~\ref{sA1}.}
For a given particle mass and fixed set of subgrid parameters, we repeat the same simulation changing {\em only}
the gravitational softening length.
This permits {\em strong} convergence\footnote{Strong convergence in galaxy formation
  simulations is obtained when results are robust to arbitrary changes in numerical parameters,
  such as mass or force resolution. Weak convergence requires recalibrating subgrid parameters
  to offset differences brought about by modified numerics, such as increased mass resolution
  \citep[see][for discussion of strong versus weak convergence]{Schaye2015}.}
tests of the properties of galaxies and their haloes, allowing us to directly
assess the robustness of well-calibrated galaxy formation models to changes in mass and force
resolution. Simulations should be insensitive to these parameters, within reason, if they are
to enjoy predictive power. 

\subsection{Halo identification}
\label{SSecSubfind}

Haloes were identified in all simulation outputs using \subfind \citep{Springel2001b,Dolag2009},
which initially links dark matter particles into friends-of-friends (FOF) groups before
separating them into self-bound ``subhaloes'' by removing unbound material. If present, star and gas particles
are associated with the FOF halo of their nearest DM particle. Note that FOF haloes are defined
entirely by the DM density field, whereas the gravitational unbinding of substructure within them uses all
particle types. 

FOF groups contain a dominant subhalo (referred to as the ``main'' halo) that contains
the majority of its mass, and a collection of lower-mass substructures, many of which host ``satellite''
galaxies. The galaxy that inhabits the main halo will be referred to as the ``central'' galaxy.

\subfind records a number of attributes of each FOF halo and its entire hierarchy of subhaloes
and satellite galaxies. These include--but are not limited to--their position $\vec{x}_p$ (defined as
the coordinate of the DM particle with the minimum potential energy), the radius and magnitude
of the peak circular speed, $r_{\rm max}$ and $V_{\rm max}$ (which include contributions from all
particle types), as well as the self-bound mass of each particle type. 

For FOF haloes, \subfind also records a variety of common mass definitions, notably $M_{\rm 200}$,
the total mass contained within a sphere centred on $\vec{x}_p$ and enclosing an average density of 200$\times\rho_{\rm crit}(z)$. When
necessary we will use $M_{200}$ as our default halo mass, but acknowledge that other mass definitions
may differ by up to 20 per cent (Paper I). The virial mass implicitly defines the virial radius,
$r_{200}=(3\,M_{200}/800\,\pi\,\rho_{\rm crit})^{1/3}$, and corresponding virial velocity,
$V_{200}=\sqrt{G\, M_{200}/r_{200}}$. 

\subsection{Analysis}
\label{SSecAnalysis}

Our analysis focuses on the statistical properties of main haloes and the galaxies that
occupy them, including their circular velocity profiles, baryon fractions, SFRs,
stellar masses and sizes.
Baryon fractions are defined as the ratio of the baryonic (stars+gas) to total mass that \subfind
deems bound to each halo. SFRs measure the total
instantaneous rate of star formation for the same gas particles. Both are taken directly
from \subfind outputs. Following \citet{Schaye2015}, stellar masses, $M_\star$, are calculated by summing the mass of
star particles {\em bound} to each galaxy that also lie within a spherical aperture extending
$30\,{\rm (physical)}\, {\rm kpc}$ from its centre. The {\em projected} radius enclosing half of $M_\star$ defines
the effective ``half-mass'' radius, which we denote $R_{50}$.

For a few of our runs we have used the \subfind halo catalogues to construct merger trees
following the procedure outlined in \citet{Jiang2014}. The algorithm locates descendants of
each subhalo by tracking their self-bound particles forward through simulations outputs,
starting when they first appear and continuing either until $z=0$ or until they have fully
merged with a more massive system. 

\section{Analytic expectations}
\label{Expectations}

We showed in Paper I that simulations of collisionless dark matter are largely insensitive to
gravitational force softening, provided it is smaller than the minimum resolved radius
dictated by 2-body relaxation. This enables a robust assessment of convergence in the properties
of dark matter haloes, particularly in aspects pertaining to their abundance and internal structure
-- mass profiles, structural scaling relations and mass functions, to name a few. Hydrodynamic
simulations that mix baryonic fluids with collisionless stellar and DM particles of unequal mass
do not necessarily adhere to the intuition built upon DM-only simulations
\citep[see][for one important example]{Ludlow2019b}. Motivated by this, we here provide
some simple analytic estimates of how $\epsilon$-dependent effects may manifest in cosmological,
hydrodynamical SPH simulations. 

Throughout this section we assume a cubic simulation box of comoving side-length $L$ that is sampled with $N^3_{\rm p}$ particles
of both gas and dark matter. Gas and DM particles have masses $m_{\rm g}=\rho_{0}\, \Omega_{\rm bar}\, (L/N_{\rm p})^3$
and $m_{\rm DM}=(\Omega_{\rm M}-\Omega_{\rm bar})/\Omega_{\rm bar}\times m_{\rm g}$, respectively;
we also define $m_{\rm p}=m_{\rm DM}+m_{\rm g}$. We further assume a flat universe and define
$E^2(z)\equiv (H(z)/H_0)^2=\Omega_{\rm M}(1+z)^3 + \Omega_\Lambda$. The virial radius of a halo, denoted
$r_{\Delta}$, encloses a mean density equal to $\Delta\times\rho_{\rm crit}(z)$; the corresponding
virial mass and circular velocity are $M_{\Delta}$ and $V_\Delta\equiv\sqrt{G\, M_\Delta/r_\Delta}$.
Using the above definitions, $M_\Delta$ and $r_\Delta$ can be conveniently expressed as
\begin{equation}
  M_\Delta=N_\Delta\,\rho_0\,\Omega_{\rm M}\biggl(\frac{L}{N_{\rm p}}\biggr)^3,
  \label{eq:M200}
\end{equation}
where $N_\Delta=M_\Delta/m_{\rm p}$ is the typical number of (gas or DM) particles within $r_\Delta$, and
\begin{equation}
  r_\Delta(z)=\biggl(\frac{3\, \Omega_{\rm M}}{4\,\pi\,E^2(z)\,\Delta}\biggr)^{1/3}\, N_\Delta^{1/3}\biggl(\frac{L}{N_{\rm p}}\biggr).
  \label{eq:r200}
\end{equation}

All numerical values quoted below assume cosmological parameters given in section~\ref{SSecSimSetup}.

\begin{figure*}
  \includegraphics[width=0.99\textwidth]{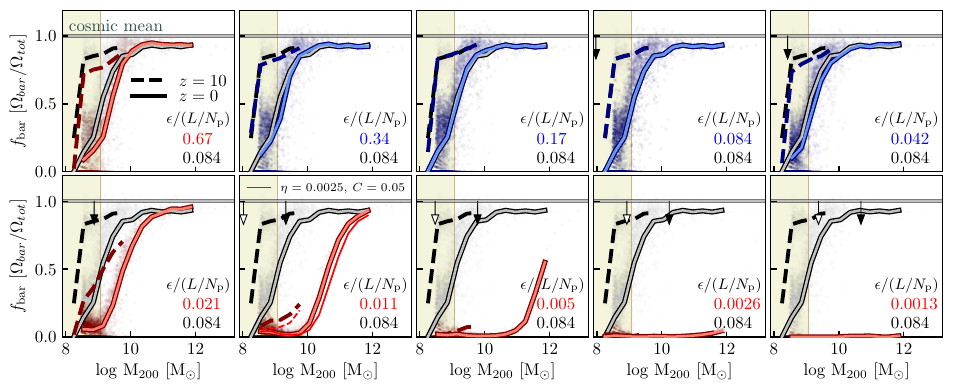}
  \caption{Halo baryon fractions as a function of virial mass for adiabatic (i.e. non-radiative) runs.
    We show results for $N^3_p=188^3$ with softening lengths that vary from
    $\epsilon/(L/N_{\rm p})\approx 0.67$ to $\approx 0.0013$ in units of the mean inter-particle
    spacing (for all runs $z_{\rm phys}=0$, which enforces fixed co-moving softening at all times).
    Dashed curves show results for haloes at $z=10$ and solid curves at $z=0$. Blue curves
    distinguish values of $\epsilon$ that fall within the range $0.02\simlt \epsilon/(L/N_{\rm p})\simlt 0.5$
    derived in section~\ref{SS_AdiabaticLim}: these runs are expected to be unaffected by
    force biasing ($\epsilon_{200}\simlt r_{200}(z)$; eq~\ref{eq:maxeps}) and stochastic heating due to particle collisions
    ($\epsilon/(L/N_{\rm p})\simgt 0.02$; eq.~\ref{eq:mineps_adiabatic}). Red curves are for runs that lie
    above or below these limits. For comparison, grey lines, repeated in each panel,
    show results for $\epsilon/(L/N_{\rm p})=0.084$, for which baryon fractions typically reach a maximum
    at all masses. The horizontal line indicates the cosmic mean baryon fraction, and beige shaded regions highlight
    $M_{200}\simlt 100\times m_{\rm p}$. Thin lines in the panel for which $\epsilon/(L/N_{\rm p})=0.011$
    show results from a run where the number of timesteps was increased by a factor $\approx 3$. Finally,
    downward pointing arrows mark the virial masses at which at which the characteristic velocity
    perturbation due to 2-body encounters is $\delta v_g=V_{200}$, with open and solid symbols corresponding
    to $z=10$ and $0$, respectively. }
  \label{fig1}
\end{figure*}

\subsection{Constraints from adiabatic hydrodynamics}
\label{SS_AdiabaticLim}

As discussed in Paper I, the softening length can be chosen so that it does not hamper the formation of the
lowest-mass structures resolved by a simulation. A plausible upper limit requires that $\epsilon$ remain small
relative to the virial radius of the lowest-mass haloes, implying a maximum {\em physical} softening length of
\begin{equation}
  \begin{split}
    \epsilon^{\rm max}_{\Delta}(z) &\ll r_\Delta(z) \\
    & \approx C \times E^{-2/3}(z) \biggl(\frac{N_{\Delta}}{100}\biggr)^{1/3} \biggl(\frac{L}{N_p}\biggr),
    \label{eq:maxeps}
  \end{split}
\end{equation}
where the constant $C=(75\,\Omega_{\rm M}/4\,\pi\,\Delta)^{1/3}$ depends on $\Delta$ and $\Omega_{\rm M}$:
for our chosen cosmology, $C\approx 0.332$ for $\Delta=200$ and $\approx 0.512$ for $\Delta=18\,\pi^2\Omega_{\rm M}$.
Note that in the matter-dominated epoch, $z\gg 1$, $\epsilon^{\rm max}_{\Delta}\propto (1+z)^{-1}$ and
this corresponds to a fixed {\em comoving} softening length. 
For a minimum resolved halo mass corresponding to $N_\Delta=100$, eq.~\ref{eq:maxeps} implies a conservative
{\em upper limit} to the comoving softening length of much less than one-half of the mean inter-particle separation
($l\equiv L/N_p=66.5\,{\rm kpc}$ and $33.3\,{\rm kpc}$ for our $N^3_{\rm p}=188^3$ and $376^3$ runs, respectively).

Close encounters with dark matter particles can deflect
gas particles to velocities of order $\delta v_g\approx 2\, G\, m_{\rm DM}/ (b\, v)$,
where $v=\sqrt{G\, m_{\rm DM}/b}$ is their relative velocity at separation $b$ \citep[e.g.][]{BinneyTremaine87}.
In the absence of radiative cooling, the kinetic energy associated with such perturbations
will dissipate to heat either
adiabatically, or through a combination of shocks and artificial viscosity. Assuming an
impact parameter $b=\epsilon/\sqrt{2}$ that maximizes the force perpendicular to the particle's
direction of motion (see Paper I for details), the square velocity change due to such encounters is
\begin{equation}
  \delta v_g^2\approx \frac{4\, \sqrt{2}G}{\epsilon}\frac{\Omega_{\rm DM}}{\Omega_{\rm bar}} m_{\rm g}.
  \label{eq:delv2}
\end{equation}
Such collisions should be unimportant provided the associated kinetic energy remains small
compared to the specific binding energy of the lowest-mass haloes resolved by the simulation, which
is of order
\begin{equation}
  V_\Delta^2=\frac{G\, M_\Delta}{r_\Delta}=G\,\frac{\Omega_{\rm M}}{\Omega_{\rm bar}}\frac{N_\Delta}{r_\Delta}\, m_{\rm g}\,.
  \label{eq:vvirmin}
\end{equation}
Requiring $\delta v_g^2\ll V_\Delta^2$ imposes a {\em lower} limit on the physical gravitational softening of
\begin{equation}
  \begin{split}
    \epsilon^{\rm min}_{\rm 2body}(z) &\gg \biggl(\frac{3\sqrt{8}}{625\, \pi\,E^2(z)\,\Delta}\biggr)^{1/3}
    \frac{\Omega_{\rm DM}}{\Omega_{\rm M}^{2/3}}\biggl(\frac{N_\Delta}{100}\biggr)^{-2/3}
    \biggl(\frac{L}{N_p}\biggr)\\
    & \approx C^\prime\times E^{-2/3}(z)\biggl(\frac{N_\Delta}{100}\biggr)^{-2/3}\biggl(\frac{L}{N_p}\biggr),
  \end{split}
  \label{eq:mineps_adiabatic}
\end{equation}
where in our case $C^\prime\approx 0.016$ for $\Delta=200$ and $\approx 0.024$ for $\Delta=18\,\pi^2\Omega_{\rm M}$. 
Note that, like eq.~\ref{eq:maxeps}, this corresponds to a fixed comoving scale at high redshift. 

These results suggest that strict sanctions should be imposed on the force softening used for
adiabatic simulations in order to simultaneously eliminate strongly biased forces and to suppress spurious
collisional heating of gas particles in poorly resolved systems: the maximum dynamic range in $\epsilon$ is only of order
$\epsilon^{\rm max}_{\rm 200}/\epsilon^{\rm min}_{\rm 2body}\approx 20$, and should be considerably
smaller to ensure $N_\Delta\approx 100$ haloes are unaffected. For our $N^3_{\rm p}=188^3$ and
$N^3_{\rm p}=376^3$ runs, eq.~\ref{eq:mineps_adiabatic} suggests that, in the absence of radiative cooling,
gravitational heating of gas particles should be apparent in low-mass haloes (i.e. $N_{200}\approx 100$)
for softening lengths below $\approx 1100\,{\rm pc}$ and $\approx 550\,{\rm pc}$,
respectively. We will test these limits explicitly in section~\ref{Verification}.

\subsection{The minimum resolved escape velocity}
\label{SS_MinVel}

We expect the above results to be applicable to adiabatic hydrodynamics, where gas heating occurs
only through gravitational interactions and radiative cooling is neglected. In more realistic galaxy
formation scenarios the thermal energy associated with inter-particle collisions can be rapidly radiated
away  due to the short cooling times of dense fluid elements. This enables gas particles to reach higher
densities and increases their specific binding energies to of order $v_\epsilon^2=v_{\rm esc}^2=2\,G\, m_{\rm g}/\epsilon$.
Given a particle mass, $m_{\rm g}$, and a desired {\em minimum resolved velocity}, $v_\epsilon$, we can
express this as a constraint on the physical softening length. Choosing convenient units:
\begin{equation}
  \begin{split}
    \epsilon_v & > 8.6\,{\rm pc}\,\biggl(\frac{m_{\rm g}}{10^5\, {\rm M}_\odot}\biggr)
    \biggl(\frac{v_\epsilon}{10\,{\rm km\,s^{-1}}}\biggr)^{-2}\\
    & \approx 5.2\times 10^5\,{\rm pc}\, \biggl(\frac{L/N_{\rm p}}{{\rm Mpc}}\biggr)^3
    \biggl(\frac{v_\epsilon}{10\,{\rm km\,s^{-1}}}\biggr)^{-2}.
  \end{split}
  \label{eq:minveps}  
\end{equation}
Note that we have deliberately expressed $v_\epsilon$ in units of $10\,{\rm km\, s^{-1}}$ to highlight
two important feedback effects occurring at comparable energies: photo-electric heating of gas associated
with cosmic reionization and HII regions, which corresponds to typical temperatures of $\sim 10^4\,{\rm K}$
and sound speeds of $c_s\approx 10\,{\rm km\,s^{-1}}$. For a given mass resolution, softening lengths
{\em smaller} than $\epsilon_v$ may conceal these important physical processes and have adverse
effects on galaxy formation models, particularly when applied to low-mass, high-redshift haloes. For example,
reionization may be suppressed if $\epsilon\simlt \epsilon_v$. For our runs, this corresponds to physical softening
lengths of $\epsilon_v\approx 156\,{\rm pc}$ for $N^3_{\rm p}=188^3$, and to $\epsilon_v\approx 19\,{\rm pc}$ for
$N^3_{\rm p}=376^3$, both smaller (by factors $\approx 1.4$ and $5.5$, respectively) than $\epsilon_{\rm fid}$ at
$z_{\rm reion}=11.5$.

\begin{figure*}
  \includegraphics[width=0.99\textwidth]{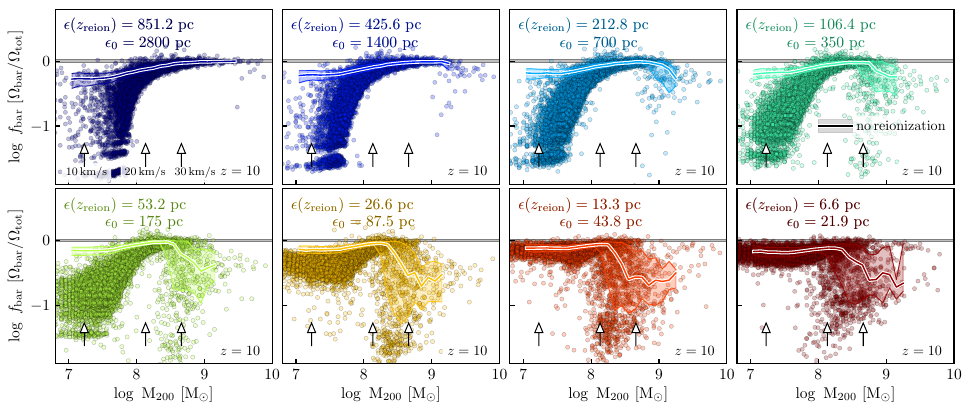}
  \caption{Halo baryon fractions as a function of virial mass for $N^3_{\rm p}=376^3$ runs at $z=10$. Different
    panels show results for different softening lengths; the physical values are quoted at both $z=0$ ($\epsilon_0$) and at
    $z_{\rm reion}=11.5$
    in each panel. Points show results for individual galaxies in our full-physics runs; solid lines show,
    for comparison, the median trends from the same runs but without photo-heating from reionization (the
    shaded regions surrounding these curves indicate the 20$^{\rm th}$ and 80$^{\rm th}$ percentiles). The
    horizontal grey lines show the cosmic mean baryon fraction; upward pointing arrows correspond to the virial
    masses of haloes with $V_{200}=10$, 20 and 30 ${\rm km\,s^{-1}}$. Note the high baryon fractions for haloes with
    $V_{200}\simlt 20\, {\rm km\,s^{-1}}$ in runs for which $\epsilon\simlt \epsilon_v$ ($\approx 19\,{\rm pc}$ in
    these runs; see Section \ref{SS_MinVel}
    for details).}
  \label{fig3}
\end{figure*}

\subsection{Ensuring efficient feedback}
\label{SS_EfficientFeedback}

Our simulations enforce a softening-dependent minimum SPH smoothing length of
$l_{\rm hsml}^{\rm min}=f\times \epsilon_{\rm sp}$, where $f=0.1$ is a constant for all of our runs\footnote{In Appendix~\ref{A3}
  we study how varying $l_{\rm hsml}^{\rm min}$ affects the results of our simulations. We find that
  $l_{\rm hsml}^{\rm min}=0.1\times \epsilon_{\rm sp}$, i.e. $f=0.1$,  gives robust results for the full range
  of softening parameters considered in our paper.} (included explicitly in what follows). The maximum gas
densities resolved in our runs are therefore of order 
\begin{equation}
  n_{\rm H}^{\rm max}=N_{\rm ngb} \frac{m_{\rm g}\, X}{m_{\rm H}}\biggl(\frac{1}{l_{\rm hsml}^{\rm min}}\biggr)^3=
    N_{\rm ngb} \frac{m_{\rm g}}{m_{\rm H}} \frac{X}{f^3} \frac{1}{\epsilon_{\rm sp}^3},
  \label{eq:nmax1}
\end{equation}
where $X=1-Y=0.752$ is the primordial hydrogen abundance, $\epsilon_{\rm sp}$ the
{\em spline} softening length, and $N_{\rm ngb}$ is the number of SPH neighbours used to weight
hydrodynamical variables. For cosmological simulations we can use the relations above to
cast this into a more convenient form:
\begin{equation}
  \begin{split}
    n_{\rm H}^{\rm max}\approx 180\,\, {\rm cm^{-3}}\,\biggr(\frac{N_{\rm ngb}}{58}\biggl)\biggr(\frac{X}{0.75}\biggl)\biggr(\frac{f}{0.1}\biggl)^{-3}\\
    \times \biggl(\frac{m_{\rm g}}{10^5\,{\rm M}_\odot}\biggr)\biggl(\frac{\epsilon}{350\,{\rm pc}}\biggr)^{-3},
    \label{eq:nmax2}
  \end{split}
\end{equation}
where $\epsilon$ now refers to the Plummer-equivalent softening length.
In practice, densities much higher than $n_{\rm H}^{\rm max}$ can occur but correspond to
length-scales smaller than $l_{\rm hsml}^{\rm min}$; in that regime hydrodynamic forces are not properly modelled.

As discussed in \citet{DallaVecchia2012}, thermal feedback will be inefficient
if a heated resolution element radiates its energy before it can expand and do work
on surrounding gas (this occurs when the cooling time is shorter than the sound crossing
time across the resolution element). Similarly, kinetic feedback (with coupled hydrodynamic forces) will not properly capture
the energy-conserving phase of a feedback event if the post-shock temperature renders the
cooling time too short. Real supernova remnants reach maximum temperatures that are sufficient
to ensure an energy-driven phase, but in cosmological simulations subject to numerical radiative
losses this is not
necessarily the case. Indeed, some simulations resort to suppressing cooling entirely
\citep[e.g.][]{Thacker2001,Stinson2006,Brook2012b} in SNe-heated resolution elements, or to
injecting {\em momentum} into those particles, which are then temporarily decoupled from hydrodynamic
forces \citep[e.g.][]{Springel2003,Pillepich2018,Dave2019}

\citet{DallaVecchia2012} derived an estimate of the critical density, $n_{\rm H,tc}$, below
which the gas cooling time exceeds the sound crossing time. For a given gas particle mass
and post-heating temperature, $T$, $n_{\rm H,tc}$ can be expressed
\begin{equation}
  n_{\rm H,tc}=26\,\,{\rm cm^{-3}}\biggl(\frac{T}{10^{7.5}\,{\rm K}}\biggr)^{3/2}
  \biggl(\frac{m_{\rm g}}{10^5\,M_\odot}\biggr)^{-1/2},
  \label{eq:nHtc}
\end{equation}
where $T\sim 10^{7.5}{\rm K}$ is the resulting temperature when all of the energy available
from core-collapse SNe is injected into a gas mass similar to that of the simple stellar
population from which the SNe are drawn. 
Equation~\ref{eq:nmax2} suggests that, as $\epsilon$ is decreased at fixed $m_{\rm g}$,
significant numbers of gas particles may reach densities that exceed $n_{\rm H,tc}$,
weakening the effects of stellar feedback from nearby stellar particles. This undesirable complication can be avoided
by demanding $n_{\rm H,tc}\simgt n_{\rm H}^{\rm max}$, which can be written as a
constraint on (the Plummer-equivalent) $\epsilon$:

\begin{align}
\epsilon_{\rm eFB} & \approx 350\,\,{\rm pc}\,\biggr(\frac{N_{\rm ngb}}{58}\biggl)^{1/3}\,\biggr(\frac{X}{0.75}\biggl)^{1/3}\biggr(\frac{f}{0.1}\biggl)^{-1} \nonumber \\
  & \hspace{2cm} \times \biggl(\frac{T}{10^{7.5}\,{\rm K}}\biggr)^{-1/2} \biggl(\frac{m_{\rm g}}{10^5\,M_\odot}\biggr)^{1/2} \\
  & = 8.6\times 10^4\,\,{\rm pc}\, \biggr(\frac{N_{\rm ngb}}{58}\biggl)^{1/3}\, \biggr(\frac{X}{0.75}\biggl)^{1/3} \biggr(\frac{f}{0.1}\biggl)^{-1}  \nonumber \\
  & \hspace{2cm} \times  \biggl(\frac{T}{10^{7.5}\,{\rm K}}\biggr)^{-1/2} \biggl(\frac{L/N_{\rm p}}{{\rm Mpc}}\biggr)^{3/2}.
    \label{eq:minnHeps}
\end{align}
We have used the subscript ``eFB'' to denote its relation to the efficiency
of stellar feedback.

The lower limits on the proper softening length implied by
eq.~\ref{eq:minnHeps} (for $f=0.1$) are $\epsilon_{\rm eFB}\approx 1.5\,{\rm kpc}$ 
and $0.5\,{\rm kpc}$ for $N^3_{\rm p}=188^3$ and $376^3$, respectively, which are
larger than our fiducial (maximum physical) softening lengths at $z=0$ by factors of
$\approx 2$ and $1.4$. While eq.~\ref{eq:minnHeps} may also apply to simulations adopting
kinetic feedback, we acknowledge that it may not be an important constraint
for runs invoking momentum injection.

\section{Results}
\label{Results}

\subsection{Verification of analytic constraints}
\label{Verification}

The analytic results of the previous section indicate that, for a given gas particle
mass or mean inter-particle spacing, insidious numerical effects may plague simulations
if softening lengths are chosen below certain thresholds. In the next three subsections we
validate these analytic expectations. For clarity, the results below are presented
mainly for our high resolution runs, but we have verified their validity at both
available resolutions.

\subsubsection{Collisional heating of adiabatic gas}
\label{collheat}

Collisional heating, when significant, may suppress the clustering of gas within
DM haloes. As $\epsilon$ decreases collisions become more
pronounced and will affect haloes of increasing mass.
Figure~\ref{fig1} shows the baryon fractions of haloes as a function of their virial mass for a 
suite of non-radiative ($N^3_{\rm p}=188^3$) simulations. Each panel corresponds to a different value of
$\epsilon$ (decreasing from top-left to bottom-right), which varies from
$\epsilon/(L/N_{\rm p})\approx 0.67$ (exceeding the {\em upper limit} of eq.~\ref{eq:maxeps} for
$N_{200}=100$) to $\epsilon/(L/N_{\rm p})\approx 1.3\times 10^{-3}$ (below the {\em lower limit}
implied by eq.~\ref{eq:mineps_adiabatic}). We use blue lines to indicate runs for
which $\epsilon$ falls within the limits derived in section~\ref{SS_AdiabaticLim}
(i.e. for $\epsilon^{\rm min}_{\rm 2body}\simlt \epsilon \simlt \epsilon_\Delta^{\rm max}$);
red curves show runs outside this range. Results are shown for haloes identified at $z=10$
(dashed lines) and $z=0$ (solid lines). For comparison, we have plotted the results for
$\epsilon/(L/N_{\rm p})\approx 0.084$ in each panel using grey lines (this run resulted in the
maximum baryon fractions for any value of $\epsilon$ we tested).

When $\epsilon/(L/N_{\rm p})\simgt 0.5$ baryon fractions are suppressed slightly in haloes
resolved with $\simlt 100$ particles (the beige shaded region in each panel indicates the
100-particle limit). This is expected: large values of $\epsilon$ lead to biased forces on
these scales, preventing gas from reaching high densities in halo centres. However, the effect is
weak since we have not investigated cases for which $\epsilon \gg \epsilon_\Delta^{\rm max}$.

As $\epsilon$ decreases baryon fractions increase, and are approximately converged provided
$0.34\simlt \epsilon/(L/N_{\rm p})\simlt 0.042$. For smaller $\epsilon$ the effects of collisional heating
are readily apparent: baryon fractions decrease, first only in low-mass haloes, but when small enough,
the effect extends to {\em all} resolved mass scales. The value of
$\epsilon$ below which collisional heating is first apparent agrees well with our
previous analytic estimate of $\epsilon/(L/N_{\rm p})\approx 0.024$ (eq.~\ref{eq:mineps_adiabatic}).
Note, as well, that our results are not unduly influenced by the timestep size. Thin lines in the 
panel for which $\epsilon/(L/N_{\rm p})= 0.011$ show results from a run which used ${\tt ErrTolIntAcc}=0.0025$ 
and a Courant factor of 0.05 (these parameters control the gravity and hydrodynamic timesteps, and correspond
to one tenth and one third of our default values, respectively). 

Finally, note that the mass scale below which baryon fractions are numerically suppressed exceeds
that at which particle collisions are expected to efficiently unbind gas from haloes. The
downward pointing arrows in each panel of Figure~\ref{fig1} mark the virial masses at which $\delta v_g=V_{200}$
(open and filled arrow correspond to $z=10$ and 0, respectively). This is presumably because, as $\epsilon$
decreases, the hierarchical assembly of haloes becomes increasingly biased to gas-poor mergers, resulting
in descendants that have $f_{\rm bar}\ll \Omega_{\rm bar}/\Omega_{\rm tot}$ at scales $V_{200}\gg \delta v_g$. 

\subsubsection{Baryon fractions and the minimum resolved escape velocity}
\label{cosmicSFH}

Radiative cooling can overcome the collisional heating of gas that may occur in 
non-radiative simulations.
This is particularly true if velocity perturbations from 2-body collisions heat fluid
elements to temperatures $T\simgt 10^4\,{\rm K}$ (which occurs when $\epsilon\simlt\epsilon_v$
with $v_\epsilon=10\,{\rm km s^{-1}}$;
eq.~\ref{eq:minveps}) where cooling rates are considerably higher than those normally encountered
in halo centres. Softening, if too small, can therefore have other unwanted effects 
by elevating the minimum resolved escape velocity of gas particles, and may render impotent
the photo-heating associated with reionization or HII regions.

We test this by comparing the baryon fractions of haloes in runs carried out using a variety of
softening lengths, both with and without photo-heating from reionization. 
The results are shown in Figure~\ref{fig3}, where we plot $f_{\rm bar}$ for haloes identified at
$z=10$ (the first available snapshot after reionization) in our
suite of $N^3_{\rm p}=376^3$ runs. Different panels show results for different $z=0$
softening lengths, $\epsilon_0$ (note that we quote {\em physical} values of $\epsilon$ at
$z_{\rm reion}=11.5$ in each panel in addition to $\epsilon_0$).
Before reionization all haloes, regardless of $\epsilon$, have similar baryon
fractions, close to the universal value $f_{\rm bar}=\Omega_{\rm bar}/\Omega_{\rm tot}$ (to avoid clutter,
we do not show this explicitly in Figure~\ref{fig3}). After reionization, however, systematic differences
are evident. The upper panels, for which
$\epsilon(z_{\rm reion})\simgt 50\,{\rm pc}$, show a sharp reduction in the baryon fractions of haloes whose circular
velocities at $r_{200}$ are less than about $20\,{\rm km\,s^{-1}}$ (middle vertical arrow), reaching an average
$f_{\rm bar}/(\Omega_{\rm bar}/\Omega_{\rm tot})\approx 0.05$ for $V_{200}\approx 10\,{\rm km\,s^{-1}}$
(left-most vertical arrow). The evaporation of gas from low-mass haloes is a well-known consequence of photo-heating due to reionization.
As $\epsilon$ is further reduced, $f_{\rm bar}$ increases systematically and approaches the cosmic mean
for all $\epsilon\simlt 26.6\,{\rm pc}$, at which point the baryon fractions {\em at all
  resolved mass scales} lie within 60--70 per cent of the cosmic mean, even among haloes with circular
velocities as low as $10-20\,{\rm km\,s^{-1}}$. This threshold ($\epsilon_0\approx 26.6\,{\rm pc}$)
agrees well with our simple analytic estimate of
$\epsilon_v\approx 19 \,{\rm pc}$. Softening lengths $\simlt \epsilon_{v}$ clearly suppress the effects
of reionization, as anticipated in Section~\ref{SS_MinVel}.
The solid lines in Figure~\ref{fig3} emphasize this point. These curves show the median baryon fractions
after repeating the same runs {\em without} photo-heating due to reionization (shaded regions indicate the
$20^{\rm th}$ and $80^{\rm th}$ percentiles of the scatter).

The excess baryons bound to low-mass haloes in these runs encourages star formation in systems
that would otherwise remain ``dark''. This spurious SF, and the associated feedback and chemical
enrichment, may in turn affect the evolution of the halo's descendants. 

\begin{figure*}
  \includegraphics[width=0.99\textwidth]{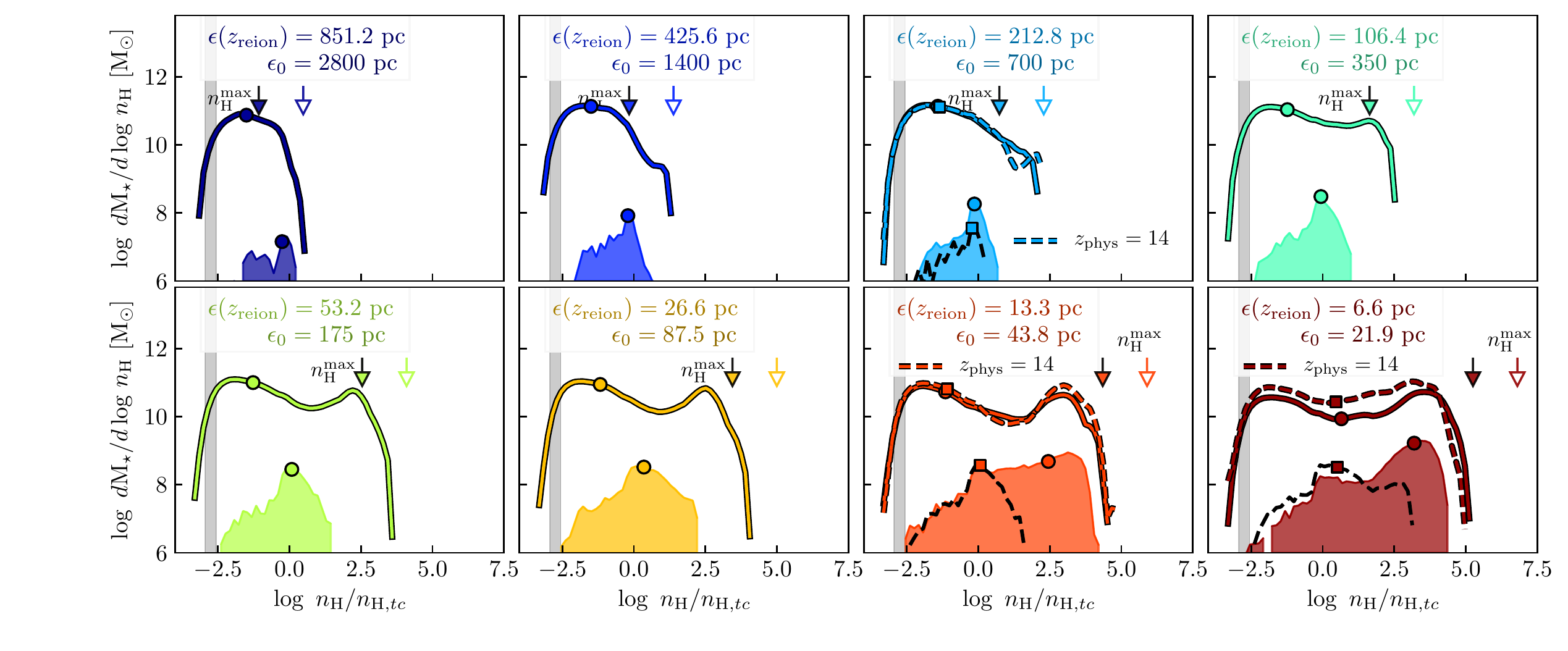}
  \caption{Differential mass distribution of fluid element densities at the
    moment they were converted into star particles. Results are shown for our $N^3_{\rm p}=376^3$
    runs, with softening decreasing from top-left to bottom-right (softening lengths at $z=0$ and
    $z_{\rm reion}=11.5$ are quoted in each panel). Densities are normalized to $n_{\rm H,tc}$,
    the maximum density for numerically
    efficient feedback (eq.~\ref{eq:nHtc}). In all panels, shaded regions correspond to stars born prior to
    $z=11.5$; lines to the entire stellar population. Outsized circles mark the median
    birth density of each (sub-)sample. All runs adopted a fixed physical softening below $z_{\rm phys}=2.8$ 
    and a fixed comoving for higher $z$, apart from those shown as dashed lines; these used $z_{\rm phys}=14$ 
    but have the same $z=0$ softening length (median birth densities are indicated by squares in these cases). 
    Downward pointing arrows mark the maximum density $n_{\rm H}^{\rm max}$ estimated from eq.~\ref{eq:nmax2}
    at $z=0$ (filled) and $z=11.5$ (open). The grey shaded region shows the SF density threshold for gas with
    a metallicity of 0.5--2 times the solar value. Apart from $\epsilon$ and 
    $z_{\rm phys}$, all other numerical and subgrid parameters are identical for all runs.}
  \label{fig5}
\end{figure*}

\subsubsection{Constraints from feedback efficiency: physical conditions at stellar birth}
\label{StellarBirth}

The \eagle{} code records the density of the fluid element from which each star particle is born.
Comparing this density to the maximum value $n_{\rm H,tc}$ (eq.~\ref{eq:nHtc}) 
for efficient feedback (eq.~\ref{eq:minnHeps}) provides a useful diagnostic of the prevalence of numerical over-cooling.

Figure~\ref{fig5} shows the mass distribution of fluid densities at the moment of stellar birth for
the same $N^3_{\rm p}=376^3$ runs plotted in Figures~\ref{fig3} (but limited to those including photoheating
from reionization). In all cases, densities
have been normalized to $n_{\rm H,tc}$. The (coloured) shaded regions and solid lines indicate, respectively, 
stars formed prior to $z_{\rm reion}=11.5$ and the entire population. For the $\epsilon_0=700\,{\rm pc}$,
$43.8\,{\rm pc}$ and $21.8\,{\rm pc}$ panels we use dashed lines to show the impact of increasing $z_{\rm phys}$
from $2.8$ to $14$. Note that birth densities taper off quickly above some maximum density:
the downward pointing arrows show $n_{\rm H}^{\rm max}$ estimated from eq.~\ref{eq:nmax2}
for each value of $\epsilon$ (filled and open arrows correspond to $z=0$ and 11.5, respectively).

\begin{figure*}
  \includegraphics[width=0.99\textwidth]{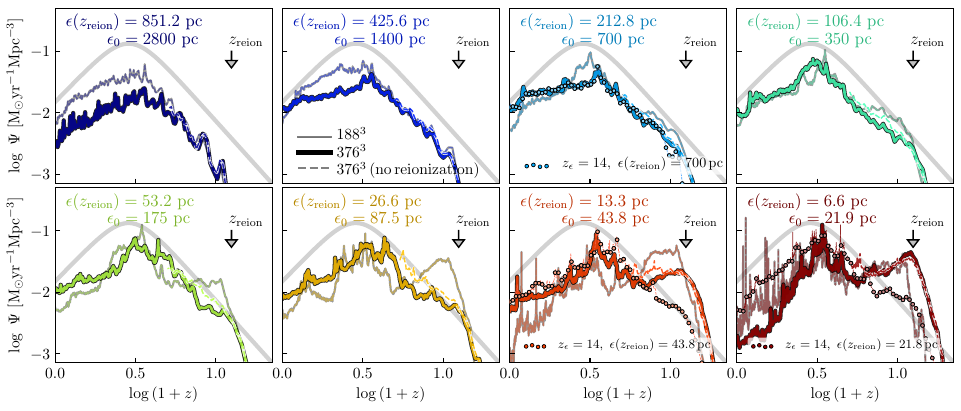}
  \caption{Cosmic star formation histories (SFH) for runs with different $z=0$ softening lengths.
    Thin (faint) lines distinguish our $N^3_{\rm p}=188^3$ Reference runs from the higher resolution ones
    (heavy lines). Downward pointing arrows mark the reionization redshift, $z_{\rm reion}=11.5$.
    All runs use $z_{\rm phys}=2.8$ apart from three: connected circles show three specific ($N^3_{\rm p}=376^3$) cases
    for which $z_{\rm phys}=14$, which leads to larger softening lengths for all $z>2.8$. The {\em physical}
    softening lengths at $z=0$ and $z=11.5$ are quoted in each panel. Dashed lines (which terminate at $z=4$)
    show results from runs in which photo-heating due to reionization was turned off. Note that as $\epsilon$ is
    reduced, the cosmic SFH becomes increasingly biased toward high-redshift SF, and eventually develops a strong
    ``peak'' roughly coincident with $z_{\rm reion}$. For comparison, we show the empirical relation of
      \citet{MadauDickinsons2014} as a thick grey line in each panel.}
  \label{fig2}
\end{figure*}

Numerical radiative losses become an increasing threat as $\epsilon$ is reduced below $\epsilon_{\rm eFB}$
(recall that $\epsilon_{\rm eFB}\approx 500\,{\rm pc}$ for $N^3_{\rm p}=376^3$). Consider, for example,
stars formed at $z\simgt 11.5$, when softening lengths were roughly a factor of 3 smaller than at $z=0$.
Above this redshift, the fraction of stars forming at 
densities $\simgt n_{\rm H,tc}$ increases steadily from $\approx 27$ per cent for $\epsilon=700\,{\rm pc}$
to $\approx 97$ per cent for $\epsilon=21.9\,{\rm pc}$. For $\epsilon\simlt \epsilon_{\rm eFB}$, the energy
injected by this initial
burst of star formation is insufficient to halt immediate numerical losses, and the distribution of
birth densities develops a strong peak at $n_{\rm H}\sim n_{\rm H}^{\rm max}\gg n_{\rm H,tc}$. This is particularly problematic
in the runs with the two smallest softening lengths, where between $\approx 32$  and 54 per cent of all
stars form at densities that stifle the effects of feedback. For
$\epsilon=700\,{\rm pc}\simgt \epsilon_{\rm eFB}$ (larger than the lower limit for efficient feedback, eq.~\ref{eq:minnHeps}),
on the other hand, only $\approx 9$ per cent of all stars form at $n\simgt n_{\rm H,tc}$.

For the most heavily affected runs, increasing $z_{\rm phys}$ from 2.8 (our fiducial value) to 14
palliates star formation in high density gas at early times, but is not a cure: a strong second peak 
inevitably develops at $n_{\rm H}\sim n^{\rm max}$ at later times, resulting in stellar populations whose feedback,
for numerical reasons, is prone to radiative losses. It is easy to understand why. For a given $\epsilon_0$, increasing
$z_{\rm phys}$ relative to our fiducial value implies larger {\em physical} softening lengths at $z>2.8$, but not at
lower $z$.
For particular values of $\epsilon_0$ and $z_{\rm phys}$, $\epsilon(z_{\rm reion})$ may exceed the
lower limit ($\epsilon_v$ with $10\,{\rm km s^{-1}}$; eq.~\ref{eq:minveps}) required for efficient photo-heating from reionization,
but still fall short of the more conservative lower limit required for efficient stellar feedback, $\epsilon_{\rm eFB}$ (eq.~\ref{eq:minnHeps}).
If this is the case, reionization will effectively remove baryons from low-mass haloes and quell SF at 
high-redshift, but stars will nevertheless form from high-density ($n_{\rm H}\gg n_{\rm H,tc}$) gas
at later times in their more massive descendants, and their associated feedback will be largely radiated away.

This is indeed the case for the runs with $\epsilon_0=21.8\,{\rm pc}$ and $43.8\,{\rm pc}$. When
$z_{\rm phys}=14$, the physical softening length at $z_{\rm reion}$ in these runs is equal to
$\epsilon_0\simgt \epsilon_v =19\,{\rm pc}$, and star formation is suppressed at $z\geq z_{\rm reion}$
(these effects will also be apparent in the star formation histories, shown later in Figure~\ref{fig2}), but,
because $\epsilon_0< \epsilon_{\rm eFB}$, still occurs at densities $\simgt  n_{\rm H,tc}$.
Although the number of stars formed at $z>11.5$ drops substantially when $z_{\rm phys}=14$, SF from
high density ($n_{\rm H}>n_{\rm H,tc}$) gas remains widespread. Preventing this requires a {\em physical} softening
length of order $\epsilon_{\rm eFB}\simgt 500\,{\rm pc}$ (eq.~\ref{eq:minnHeps} with $N^3_{\rm p}=376$) at essentially all redshifts. This is
only the case for one run, plotted in the upper-left panel of Figure~\ref{fig5}.

Finally, we note that softening is expected to influence star formation in other ways. If $\epsilon$ is {\em large},
$n_{\rm H}^{\rm max}$ may fall short of the density threshold for star formation, $n_\star$ (shown as a grey shaded band in
Figure~\ref{fig5} for metallicities spanning $0.5\leq Z/Z_\odot\leq 2$), although most of our runs are not in this regime.
One possible exception is our $N^3_{\rm p}=376^3$ run with $\epsilon=2800\,{\rm pc}$ where, at {\em low} redshift,
$n_{\rm H}^{\rm max}$ is comparable to $n_\star$ in metal-poor gas ($Z\approx 0.05-0.01\,Z_\odot$). Simulations
that adopt higher SF density thresholds \citep[e.g.][]{Brook2012a,Wang2015} may suffer these effects for comparatively
smaller softening lengths.

\subsection{Impact on galaxy formation models}
\label{Impact}

As mentioned in 
Section~\ref{SSecSimSetup}, the subgrid feedback modules in \eagle{} were calibrated at {\em fixed resolution}
so that simulations reproduced the observed present-day size-mass relation of galaxies, and their GSMF.
This may be problematic, as good agreement is no longer guaranteed if numerical parameters are modified. 
For example, the softening length can affect the range of gas densities over which
star formation occurs: if too small it may push gas particles into the regime
of inefficient feedback, and if too large may prohibit a meaningful application of the subgrid model
(if, for example, the physically-motivated SF density threshold is not resolved). Clearly the results of
simulations cannot converge for arbitrary numerical parameters when the subgrid model is held fixed.
It is nevertheless useful to establish the range of parameters--particularly those
governing mass and force resolution--over which the results can be considered reliably
modelled.

What impact does changing $\epsilon$ at fixed $m_{\rm g}$ have on our calibration
diagnostics? We turn our attention to this question in the following sections.

\subsubsection{The cosmic star formation history}
\label{CSFH}

The cosmic SFHs for our full simulation suite are plotted in Figure~\ref{fig2}.
As above, different panels show results for runs with different softening lengths; thick and thin
solid lines distinguish our $N^3_{\rm p}=376^3$ and $188^3$ runs, respectively. All runs
used $z_{\rm phys}=2.8$, with three exceptions: connected circles in panels corresponding to
$\epsilon_0=700\,{\rm pc}$, $43.8\,{\rm pc}$ and $21.8\,{\rm pc}$ instead used $z_{\rm phys}=14$,
but the same $\epsilon_0$ (these runs are limited to $N^3_{\rm p}=376^3$). 

The SFHs are reasonably well-converged for the two different mass resolutions, but only for a narrow range of $\epsilon$,
$350\, {\rm pc}\,\simlt \epsilon_0\simlt \,700\,{\rm pc}$ or so (equivalently,
$\epsilon(z_{\rm reion})\approx 106-213\,{\rm pc}$). This is perhaps not surprising: the \eagle{}
subgrid models were calibrated using comparable values ($\epsilon_0=700\,{\rm pc}$ for $N^3_{\rm p}=188^3$
and $\epsilon_0=350\,{\rm pc}$ for $N^3_{\rm p}=376^3$). At fixed {\em mass resolution}, SFHs are reasonably
robust to changes in $\epsilon$ provided it remains within about $1/2$ to 4 times the value adopted
for calibration. For larger or smaller values resolution effects are evident.

For both mass resolutions, increasing $\epsilon$ by more than a factor of $\approx 4$ relative to
the fiducial values results in an overall suppression of SF at essentially all $z$. This is seen
most clearly in the $N^3_{\rm p}=376^3$ run carried out with  $\epsilon=2800\,{\rm pc}$ 
(8 times the fiducial softening length for this mass resolution), for which the SFR is suppressed at virtually {\em all}
redshifts. The maximum resolved density (eqs.~\ref{eq:nmax1} and~\ref{eq:nmax2}) in this run is only of order
$n_{\rm H}^{\rm max}\approx 0.31\,{\rm cm^{-3}}$, which is comparable to the SF density threshold in
metal-poor gas ($n_\star\approx 0.21-0.13\,{\rm cm^{-3}}$ for $Z=0.05-0.1\,Z_\odot$).
Although many gas particles will reach densities that surpass $n_{\rm H}^{\rm max}$, increasing $\epsilon$ clearly
places limits on the dynamic range of gas densities that are eligible to form stars (Figure~\ref{fig5}),
and the global SFR drops as a consequence.   
Such large softening lengths are therefore inappropriate for the physical model employed. Note, for example, that the
SF density threshold, $n_\star$, is physically motivated \citep[see][]{Schaye2004}, and {\em not} determined
by calibration. Softening lengths sufficiently large to prevent gas densities from reaching $n_\star$ should
therefore be avoided.

Conversely, for the same $\epsilon_0$ our low-mass resolution run reaches maximum densities
$n_{\rm H}^{\rm max}\approx 2.5\,{\rm cm^{-3}}$,
which is roughly an order of magnitude larger than $n_\star$. As a result, the SFH
is less affected in this case. It is important to note, however, that simply resolving $n_\star$ does
not in itself guarantee good convergence in SFHs. According to the KS law, the SFR per-particle
depends on $n$: the {\em global} SFR is therefore also modulated by the maximum resolved density, a result
that was already apparent in Figure~\ref{fig5}. We acknowledge, however, that, provided $\epsilon$ is
sufficiently small so that $n_{\rm H}^{\rm max}\gg n_\star$ but sufficiently large so that
$v_\epsilon\ll 10\,{\rm km s^{-1}}$ (i.e. $\epsilon\gg \epsilon_v$ for $v_\epsilon=10\,{\rm km s^{-1}}$),
careful calibration may compensate for the softening dependence of the SFHs seen in Figure~\ref{fig2}.

Decreasing $\epsilon$ results in a systematic {\em increase} in the SFR at early
times, eventually leading to a ``peak'' in the cosmic SFH at $z\approx z_{\rm reion}=11.5$ (marked by a downward
pointing arrow). This initial burst of star formation--clearly {\em numerical} in origin--is first
noticeable in the low mass resolution runs, and occurs when $\epsilon$ falls below $\approx 53.3 \,{\rm pc}$.
At high mass resolution, early SFHs grow slowly until $\epsilon\approx 13.3\,{\rm pc}$, but develop a similar
peak for smaller values (also apparent in the distribution stellar birth densities for these runs; Figure~\ref{fig5}).
It is tempting to relate the early peak in SF to the suppression of reionization in runs for which
$\epsilon\simlt \epsilon_v$ (Figure~\ref{fig3}), which {\em enhances} SF in low-mass haloes at high redshift.
Indeed, the two effects occur for similar softening lengths. This possibility, however, is easily ruled out. The thin dashed lines
plotted in each panel of Figure~\ref{fig2} show the SFHs for an additional set of $N^3_{\rm p}=376^3$ runs
(stopped at $z=4$) in which reionization was not implemented. The global SFHs are only slightly affected by the presence of a photo-ionizing
background, despite having a considerable impact on the baryon fractions of low-mass systems (Figure~\ref{fig3}).
The excess baryons in these low-mass haloes--and the associated excess star formation--contributes negligibly
to the ``peak'' in the global SFR, which is dominated by the most-massive haloes in the
volume\footnote{Additional tests (not presented here) suggest that the magnitude and
  redshift of the peak SFR at early times is sensitive to not only softening, but also to subgrid parameters
  that control star formation and stellar feedback. This is not surprising, but emphasizes the difficulty
  of disentangling the impact of numerical and subgrid parameters on the results of cosmological hydrodynamical simulations. }.
The fact that the SFR peaks of $z\approx z_{\rm reion}$ is thus a coincidence.

Nevertheless, the crest in star formation at early times can be quelled by increasing $z_{\rm phys}$, which
decreases the maximum resolved density at early times and suppresses rampant star formation in high-density gas
(recall Figure~\ref{fig5}). The connected circles shown in panels for which $\epsilon_0=700\,{\rm pc}$,
$43.8\,{\rm pc}$ and $21.8\,{\rm pc}$ show the SFHs in three $N^3_{\rm p}=376^3$ runs that use the same $z=0$
softening, but $z_{\rm phys}=14$. 

Note that the softening dependence of the cosmic SFHs presented in Figure~\ref{fig2} is {\em not}
driven by its explicit connection to the minimum SPH smoothing length, $l_{\rm hsml}^{\rm min}$, a point
that we demonstrate explicitly in Appendix~\ref{sA3}.

\begin{figure*}
  \includegraphics[width=0.99\textwidth]{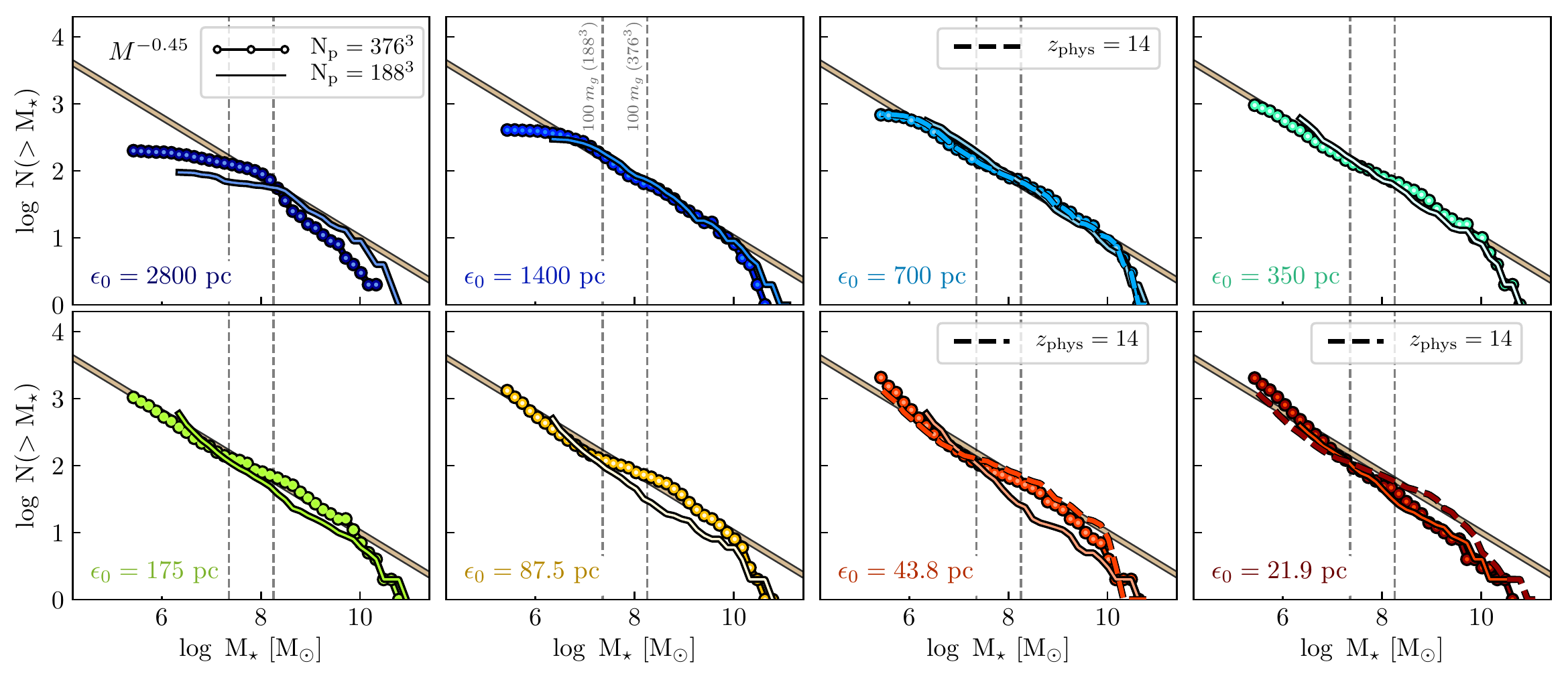}
  \caption{Comparison of the cumulative $z=0$ galaxy stellar mass functions for all models. Different panels
    correspond to runs carried out with different maximum physical softening
    lengths, colour coded as in previous figures. Solid lines show runs with
    $N^3_{\rm p}=188^3$ particles and connected circles with $N^3_{\rm p}=376^3$, both use $z_{\rm phys}=2.8$;
    dashed lines show results for a subset of the high mass resolution runs that adopted $z_{\rm phys}=14$
    (but the same $z=0$ softening lengths). Vertical lines in each
    panel mark mass scales corresponding to 100 primordial gas particles at each resolution.
    In all cases, ${\rm M_\star}$ is defined as the total bound stellar mass enclosed by a 30
    (physical) kpc aperture coincident with the galaxy position. For comparison, we plot a
    power-law $N\propto M^{-0.45}$ in each panel.}
  \label{fig7}
\end{figure*}

\subsubsection{The galaxy stellar mass function}
\label{GSMF}

The global SFHs presented in Figures~\ref{fig2} suggest that 
changing to numerical parameters without recalibrating may impact the properties of 
galaxies that form in simulations. We investigate one such possibility in 
Figure~\ref{fig7}, which shows the $z=0$ {\em cumulative} galaxy stellar mass functions (GSMF)
for our simulation suite (note that this the total number of galaxies that exceeds a given
stellar mass, and differs from the more conventional differential GSMF).
Different panels correspond to different maximum physical softening lengths,
as indicated. In each panel, solid lines and connected points are used for runs
carried out with $N^3_{\rm p}=188^3$ and $376^3$ particles, respectively; dashed lines, where
present, correspond to $N^3_{\rm p}=376^3$ runs that used $z_{\rm phys}=14$, the rest
used $z_{\rm phys}=2.8$. 

Reflecting similarities in the SFHs, the cumulative GSMFs converge reasonably well over a narrow range of softening,
$350\,{\rm pc}\simlt \epsilon\simlt 1400\,{\rm pc}$, and can be approximated by a single power-law,
$N\propto M^{-0.45}$ for $10^7 \simlt M_\star/{\rm M_\odot}\simlt 10^{10}$
(shown as a thick grey line in each panel). For higher and lower values of softening
departures from this shape are evident. For large values, the abundance of low-mass galaxies is noticeably
suppressed. This is clearly seen in the upper-left panel ($\epsilon_0=2800\,{\rm pc}$) where GSMFs
flatten below $M_\star \approx 10^8\,{\rm M_\odot}$. A similar effect occurs in runs
with $\epsilon_0=1400\,{\rm pc}$ and $700\,{\rm pc}$, but in these cases it is shifted to lower masses. We
will show in Section~\ref{DMconc} that this is a consequence of {\em over}-softening the inner regions
of DM haloes, which lowers the typical densities that can be resolved in their central regions.
Galaxy formation is suppressed in haloes for
which the softening length is comparable to the physical scale over which galaxies typically form (which is
of order $0.1-0.15\,r_{200}$). 

At low masses the abundance of galaxies in our $N^3_{\rm p}=376^3$  run {\em increases systematically} with
decreasing $\epsilon$ for all $\epsilon_0\simgt 43.3\,{\rm pc}$. The same is true of our $N^3_{\rm p}=188^3$ runs
provided $\epsilon_0\simgt 350\,{\rm pc}$. The systematic dependence on $\epsilon$ suggests that poor convergence in the
abundance of low-mass galaxies is {\em not} a result of the stochastic nature of our SF
prescription, but of numerical effects. For $N^3_{\rm p}=376^3$,
for example, the number of galaxies resolved with fewer than 10 stellar particles (i.e. $M_\star\simlt 10\times m_g$)
increases by a factor of $\approx 57$ between runs with the smallest and largest softening lengths. Their increased
abundance is a direct consequence of the systematic increase in early SF that accompanies smaller $\epsilon$;
and it gives rise to a {\em steepening} of the GSMFs at low ${\rm M_\star}$. 

The high-mass end of the cumulative GSMF also depends on $\epsilon$, but non-monotonically:
for ${\rm M\simgt 10^8\,M_\odot}$, it is steeper than $M^{-0.45}$
at both the highest and lowest values of $\epsilon$ we study (note that, for intermediate values of $\epsilon$, the
GSMF drops for ${\rm M\simgt 10^{10}\,M_\odot}$ due to the finite box-size of our simulations).
Comparing Figures~\ref{fig2} and \ref{fig7}
provides some clues to the origin of the effect. At high mass, GSMFs are steeper in essentially {\em all} runs that
exhibit a reduced SFR at late times (relative to the fiducial run). For our
high mass-resolution runs this occurs only for the largest and smallest softening lengths tested, but is
evident for all $\epsilon\leq 87.5\,{\rm pc}$ at low mass-resolution. 

It is unlikely that the numerical processes suppressing SFRs at late times are the same at large and small
$\epsilon$, but the exact cause is not clear. We speculate that, when large, softening suppresses SF not
only in low-mass haloes, but also in the central regions of massive ones, giving rise to an overall reduction
in SF across all halo masses (we will return to this point in Section~\ref{ImpactDM}). When $\epsilon$ is small,
gas particles typically reach high densities before forming stars, which increases both their SFR and the burstiness
of SF. Strong star bursts may overcome inefficient feedback and gradually expel gas from massive haloes, reducing
the global SFR. Other possibilities include: differences in gas consumption timescales; an earlier onset of BH
formation and associated differences in AGN feedback (but see Appendix~\ref{sA1} for examples of runs {\em without}
AGN feedback); or spurious energy transfer between DM and baryonic particles that gradually reduce gas densities
\citep{Ludlow2019b}. These possibilities require further investigation. 

\begin{figure*}
  \includegraphics[width=0.99\textwidth]{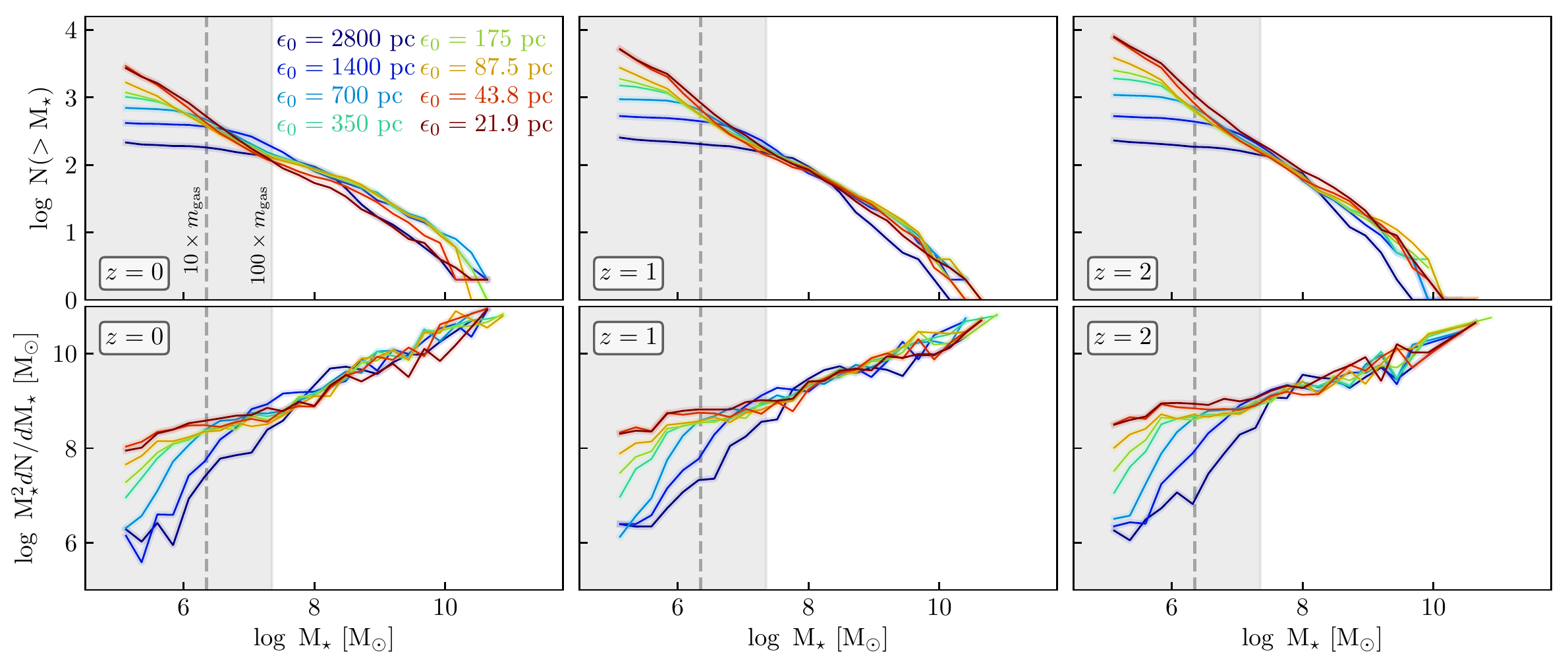}
  \caption{{\em Upper panels:} Cumulative galaxy stellar mass functions at $z=0$ (left), $z=1$ (middle)
    and $z=2$ (right) for our $N^3_{\rm p}=376^3$ runs carried out with different softening lengths, $\epsilon$.
    The total number of galaxies resolved by these simulations increases systematically with decreasing $\epsilon$,
    and differs by as much as an order of magnitude between the smallest ($\epsilon=21.8\,{\rm pc}$) and largest
    ($\epsilon=2800\,{\rm pc}$) values. {\em Lower panels:} Total stellar mass of
    galaxies in fixed logarithmic intervals of ${\rm M_\star}$. At lowest masses plotted--corresponding to
    galaxies composed of a {\rm single} stellar particle--the total stellar mass differs by as much as a factor of
    $100$ between the lowest and smallest softening lengths. In all panels, the grey shaded region marks a stellar
    mass corresponding to 100 primordial gas particles, which is clearly a minimum requirement for convergence
    in stellar mass functions.}
  \label{StellarMFz}
\end{figure*}

In the upper panels of Figure~\ref{StellarMFz} we plot the cumulative GSMFs in our $N^3_{\rm p}=376^3$ runs at
different redshifts, ranging from $z=0$ (upper-left) to $z=2$ (upper-right). The shaded regions in each
panel indicates $M_\star\leq 100\times m_{\rm gas}$, and the vertical dashed line corresponds to $10\times m_{\rm gas}$. 
The GSMFs agree well at each $z$ provided $\epsilon$ does not veer too far
from the fiducial value. Runs carried out with $\epsilon_{\rm fid}/8\simlt \epsilon \simlt 2\times \epsilon_{\rm fid}$,
for example, yield similar GSMFs at all redshifts considered. Differences greater than a few per cent are only noticeable
at low masses, where galaxies are resolved with $\simlt 10$ stellar particles. This is consistent with the
similarity in the SFHs of these runs (see Figure~\ref{fig2}), though convergence is better for the GSMFs than for the
SFHs. For softening lengths either larger or smaller than these
values, however, differences in the GSMFs are evident at all three redshifts. When $\epsilon$ is larger,
the number of low-mass galaxies is suppressed relative to our fiducial run, an effect that (for fixed $\epsilon$)
occurs at roughly the {\em same} stellar mass at each $z$ considered. The systematic increase in SF that accompanies
smaller $\epsilon$ is dominated by low-mass haloes at high-redshift. This results in GSMFs that, at each $z$, become
noticeably {\em steeper} at low mass than in our fiducial run when $\epsilon<\epsilon_{\rm fid}/4$.

The lower panels of Figure~\ref{StellarMFz} plot, for the same runs and redshifts, the {\em total} stellar
mass contained in {\em all} galaxies in a given logarithmic interval of stellar mass. Above $M_\star\approx 100\times m_{\rm g}$,
these curves converge reasonably well for all $\epsilon$ and $z$, and their shape suggests that at any $z$
the vast majority of stars are found in the most massive galaxies in the simulation (but recall that the volume
is only $(12.5\,{\rm Mpc})^3$). For example, we find that roughly 50 per cent of the {\em total}
stellar mass formed by $z=0$ is contained in the 3 to 5 most massive galaxies, depending on $\epsilon$.
Convergence at masses $\simlt 100\times m_{\rm gas}$ is poor; at $z=0$, for example, the total mass in such
galaxies varies by as much as a factor of 7 between runs with the largest and smallest softening. 

\subsubsection{Galaxy sizes}
\label{Sizes}

\citet{Crain2015} showed that simulations with subgrid physics calibrated to match observations of the galaxy stellar mass
function may fail to reproduce other observations of the galaxy population. In particular, they noted that models with feedback
prescriptions allowing significant star formation to occur in high-density ($n\gg n_{\rm H,tc}$) gas, despite having plausible
stellar masses, tend to result in unrealistically compact massive galaxies
whose sizes do not match those observed. The birth conditions of stars highlighted in section~\ref{StellarBirth}
suggest that reducing the gravitational softening, like modifying the feedback implementation,
may have a similar effect.

We quantify galaxy sizes using $R_{50}$, the projected ``half-stellar mass'' radius enclosing 50 per
cent of the galaxy's total stellar mass. $R_{50}$ is estimated directly from the (randomly projected)
surface mass-density profiles of galaxies, rather than by fitting a parameterized model, such as
a S$\acute{e}$rsic profile. 

\begin{figure*}
  \includegraphics[width=0.99\textwidth]{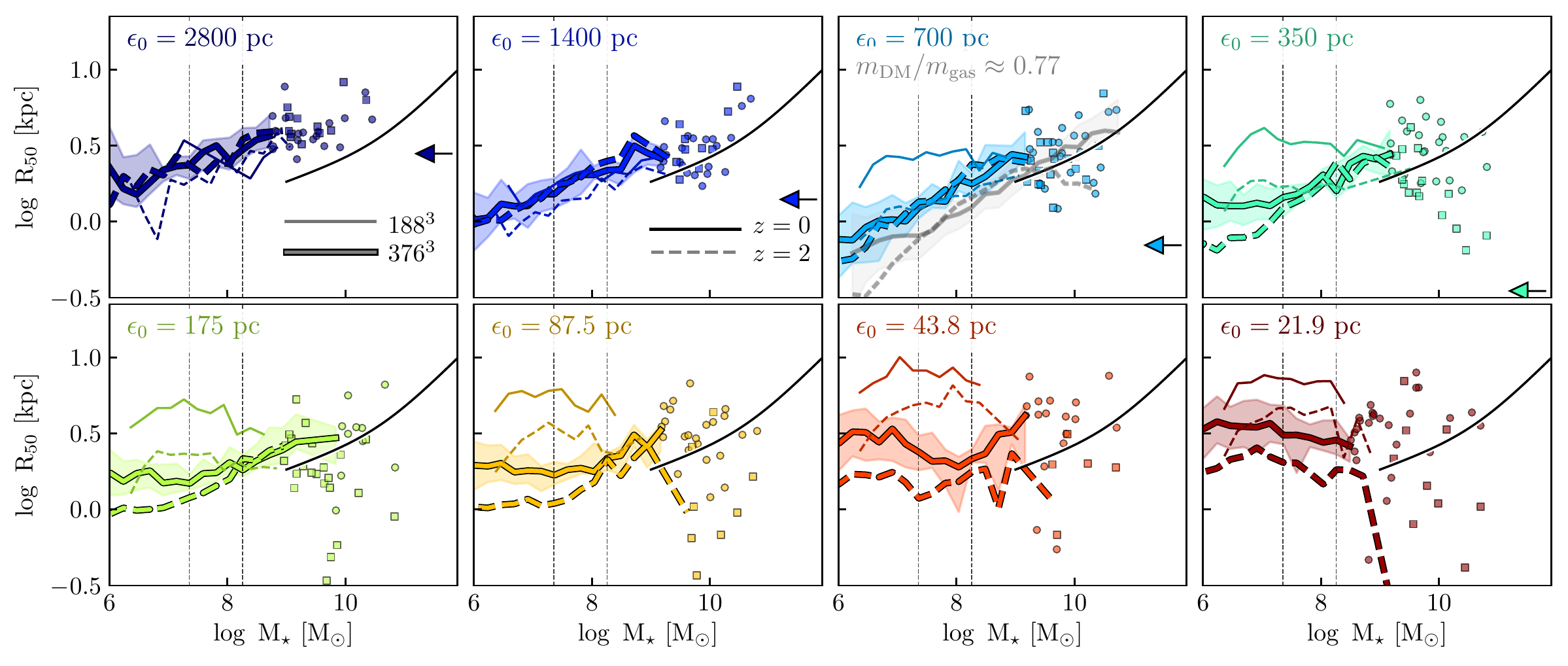}
  \caption{Projected (physical) half stellar-mass radii as a function of galaxy stellar mass
    at $z=0$ (solid lines) and $z=2$ (dashed lines) for our $N^3_{\rm p}=376^3$ 
    (thick lines) and $N^3_{\rm p}=188^3$ runs (thin lines). The shaded coloured regions
    indicate the $20^{\rm th}$ and $80^{\rm th}$ percentiles of the size
    distribution, but, for clarity, are only shown for the $z=0$ outputs of
    the high-resolution runs. As in previous figures,
    different panels show results for different softening lengths (indicated
    using arrows on the right-hand side of each panel). The solid black line
    shows a fit to the half-light radii for early-type galaxies observed in SDSS
    \citep{Shen2003}. Vertical dashed lines indicate the mass
    scale of 100 primordial gas particles in our low- and high-resolution
    runs. The faint grey lines in the upper middle-right panel show sizes obtained
    from a simulation that used the same baryonic particle mass and subgrid
    physics as our $N^3_{\rm p}=188^3$ runs, but had higher resolution in DM
    particles, so that $\mu\equiv m_{\rm DM}/m_{\rm gas}\approx 0.77$ (instead of
    $\approx 5.3$). This
    run minimizes the spurious transfer of energy from DM to gas particles
    that artificially inflates galaxy sizes \citep[see][for details]{Ludlow2019b}.}
  \label{fig10}
\end{figure*}

Figure~\ref{fig10} compares the $R_{50}-M_\star$ relations obtained from our runs. Different panels are used for
different softening lengths, and results are shown at $z=0$ (solid lines)
and $z=2$ (dashed lines). Thick lines shown the median trends for $N^3_{\rm p}=376^3$, thin lines for
$N^3_{\rm p}=188^3$. Only mass bins including $\geq 10$ objects are plotted as medians; for other bins individual
galaxies in our $N^3_{\rm p}=376^3$ runs are shown using filled circles (for $z=0$) or squares (for $z=2$).
For comparison, a fit to the \citet{Shen2003} size-mass relation for early-type galaxies in SDSS is shown using
a thick black line.

\citet{Ludlow2019b} showed that galaxy sizes in hydrodynamical simulations that use unequal-mass baryonic
and DM particles, as is the case here, are affected by energy equipartition. Galaxies of a given stellar mass grow in size as a
result of spurious energy transfer from DM to stars, but the effect can be suppressed by adopting 
stellar and DM particles of approximately equal mass. A similar effect has also been reported
in the central regions of haloes formed in simulations that adopt two {\em collisionless} particle species of unequal
mass \citep{BinneyKnebe2002,Ludlow2019b}. In the upper middle-right panel of
Figure~\ref{fig10} (corresponding to $\epsilon_0=700\,{\rm pc}$) we plot for comparison the
$z=0$ and $z=2$ size-mass relations obtained from a simulation carried out with $\mu\equiv m_{\rm DM}/m_{\rm gas}\approx 0.77$;
these are shown using grey lines. This simulation was carried out in a periodic box with $L=25$ (comoving) Mpc
and used $N_{\rm gas}=376^3$ particles of gas, but 7 times as many DM particles. The baryonic mass resolution
is identical to that of our $N^3_{\rm p}=188^3$ runs, whereas the DM mass resolution is 7 times higher. The
run used a softening length of $\epsilon_0=700\, {\rm pc}$, $z_{\rm phys}=2.8$ and subgrid parameters consistent
with the Reference model (see \citealt{Ludlow2019b} for details).

Figure~\ref{fig10} is worth a few comments. Notice first that galaxy sizes converge between different
mass resolutions, but only for the two {\em largest} values of $\epsilon$, $\simgt 1400\,{\rm pc}$ or
so. The effects of mass segregation driven by energy-equipartition are also minimized in these runs: the size-mass relations in both
are approximately independent of redshift provided $\epsilon\simgt 1400\,{\rm pc}$, which is also the case for
$\epsilon=700\,{\rm pc}$ in our high mass-resolution run. (The redshift independence of the size-mass relation is a salient
feature of the $\mu \approx 1$ run.) Despite this, galaxy sizes are {\em not} converged with
respect to softening. Note, for example, that half-mass radii become systematically smaller (by $\approx 40-50$
per cent) at essentially {\em all} stellar masses when $\epsilon$ is reduced from $2800\,{\rm pc}$ to $1400\,{\rm pc}$; 
on average, sizes are reduced even further when $\epsilon=700\,{\rm pc}$ (by an additional $\approx 20$ per cent), but
only for  $N^3_{\rm p}=376^3$ run. For $M_\star\simgt 100\, m_{\rm gas}$, the median $z=0$ half-mass radii in our
$N^3_{\rm p}=376^3$ runs are quite stable provided $175\simlt \epsilon/{\rm [pc]}\simlt 1400$, although there is evidence of
increased scatter amongst the highest-mass galaxies. 

It is interesting to note, however, that reducing $\epsilon$ {\em does not} necessarily
result in more compact galaxies, at least not at the mass
scales resolved by our simulations. Despite the fact that small softening promotes inefficient feedback
(Figure~\ref{fig5}) and centrally concentrated star formation in high-density gas,
{\em the median sizes of galaxies increase systematically} as $\epsilon$ decreases, an effect that
is particularly evident at low stellar mass (but may reverse at the highest values of ${\rm M_\star}$).
In fact, the most diffuse galaxies in any run are found in the one 
carried out with the {\em smallest} softening parameter. Note also that when $\epsilon_0$ drops below $1400\,{\rm pc}$
(or $<700\,{\rm pc}$ for our $N^3_{\rm p}=376^3$ run), the size-mass relations in our $N^3_{\rm p}=188^3$ runs develop a redshift
dependence that is not observed for larger $\epsilon_0$. This redshift-dependence is evidence of mass segregation
driven by energy equipartition, which is suppressed when $\epsilon$ is large, and exacerbated when $\epsilon$ is small. 
Indeed, the effect is absent from our $\mu\approx 1$ run, despite having identical subgrid physics,
baryonic mass resolution and gravitational softening.

The segregation of stars and DM particles in our simulations is a consequence of 2-body scattering, and
can therefore be suppressed by either increasing the number of collisionless (stellar and DM) particles, or adopting
$\mu\approx 1$. By comparing otherwise
identical runs with $\mu\approx 1$ and $\mu\approx 5.3$, \citet{Ludlow2019b} concluded that ${\rm N_\star}\approx 2000$
stellar particles per galaxy are required in order to mitigate the effects of mass segregation at galaxy half
mass radii. For larger ${\rm N_\star}$, the effect will remain, but will be pushed to smaller scales.

The net result is that, if too large, softening will ``puff up'' galaxies, leading to larger sizes,
but if too small it will escalate energy equipartition, driving spurious growth in size. The best
compromise between these two regimes appears to favour softening lengths roughly a factor of two larger
than \eagle's fiducial value, but we stress that further tests are required to fully disentangle
the impact of softening and mass segregation on galaxy sizes. For example, simple analytic estimates suggest
that the effects of 2-body scattering--and thus mass segregation--should exhibit a weak (logarithmic) dependence
on $\epsilon$; our results appear to favour a stronger dependence.

\begin{figure*}
  \includegraphics[width=0.99\textwidth]{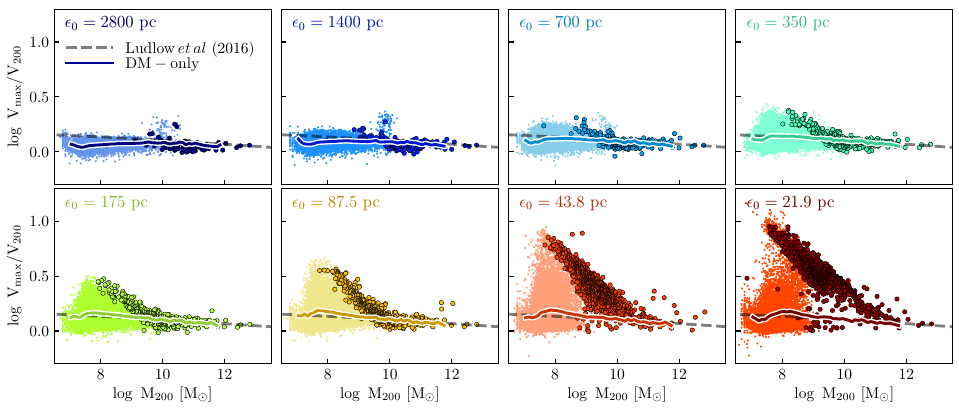}
  \caption{Maximum circular velocity, ${\rm V_{max}/V_{200}}$, versus virial mass, ${\rm M_{200}}$, for
    central galaxies and DM haloes in our $N^3_{\rm p}=376^3$ runs. Dots show results for main haloes that contain no stellar component; circles
    show haloes that contain at least one star particle. The solid lines of corresponding colour show median
    trends from a series of DM-only simulations carried out using the same softening lengths
    \citep[see][for details]{Ludlow2019a}; dashed lines show predictions from the analytic model of
    \citet{Ludlow2016}, which is based on DM-only simulations. As $\epsilon$ decreases, haloes become
    more concentrated; the effect, which is most pronounced at low virial mass, is {\em not} limited to
    systems that contain an embedded stellar component. }
  \label{VmaxM200}
\end{figure*}

\subsection{Impact on the internal structure of DM haloes}
\label{ImpactDM}
Cosmological hydrodynamical simulations are not only useful for elucidating the complex physics of galaxy formation,
but also for clarifying the impact that baryons have on the underlying distribution of dark matter.

Galaxy formation occurs deep in the central regions of haloes, where high gas densities enable efficient cooling and
initiate SF. The build-up of baryonic gas and its conversion to stars in the central regions of haloes can influence the
distribution of DM there \citep[e.g.][]{Duffy2010,Maccio2012,Bryan2013,Teyssier2013,Ogiya2014}, but exactly how is still debated.
For example, in the absence of feedback-driven outflows, the slow growth of a central galaxy may deepen the gravitational potential
adiabatically, resulting in a contraction of DM in the innermost regions of a halo \citep{Blumenthal1986}.
Violent, episodic SF, on the other hand, can drive gaseous outflows and give rise to strong fluctuations in the total gravitational
potential \citep{NEF1996}. If star formation occurs in sufficiently dense gas--and outflows are sufficiently violent--galaxy
formation {\em may} reduce DM densities in halo centres, giving rise to {\em cores} in the DM
distribution \citep{Governato2010,Pontzen2012}, although the outcome of such simulations is model-dependent 
\citep{Benitez-Llambay2018,Dutton2019}. Given these uncertainties, and the numerically-driven results above, is it possible to
make robust predictions for the DM distribution using hydrodynamical simulations?

In Paper I, we determined the
conditions under which DM-only simulations can make reliable and reproducible predictions for the innermost structure
of DM haloes in the absence of baryons. Briefly, circular velocity profiles converge to better than $\approx 10$
per cent at radii that exceed a {\em comoving} convergence radius that can be approximated by
$r_{\rm conv}\approx 0.055\, (L/N_p)$; this corresponds to $\approx 1.8$ and $\approx 3.7\,$ (comoving)
${\rm kpc}$ in our high- and low-mass resolution runs. All runs for which $\epsilon_0>\epsilon_{\rm fid}$
therefore have softening lengths that exceed $r_{\rm conv}$ for $z\geq 2.8$, which may compromise the innermost
structure of high-redshift haloes in these runs. The maximum {\em physical} softening lengths, however,
are $\simlt r_{\rm conv}$ in all runs except the one with $N^3_{\rm p}=376^3$ and $\epsilon_0=2800\,{\rm pc}$. We
will return to this discussion in Section~\ref{recc}.

\subsubsection{Dark matter halo concentrations}
\label{DMconc}

In Figure~\ref{fig5} we showed that the typical density of star-forming gas increases
systematically with decreasing $\epsilon$, reaching maximum values that scale approximately as
$n_{\rm H}^{\rm max}\propto \epsilon^{-3}$ (eq.~\ref{eq:nmax1}), often exceeding the critical threshold 
for efficient feedback (eq.~\ref{eq:minnHeps}). We may therefore expect the DM distribution in our simulations
to respond differently to galaxy formation in our runs, given the widely different gas densities
and feedback efficiencies attained for different $\epsilon$.

\begin{figure*}
  \includegraphics[width=0.99\textwidth]{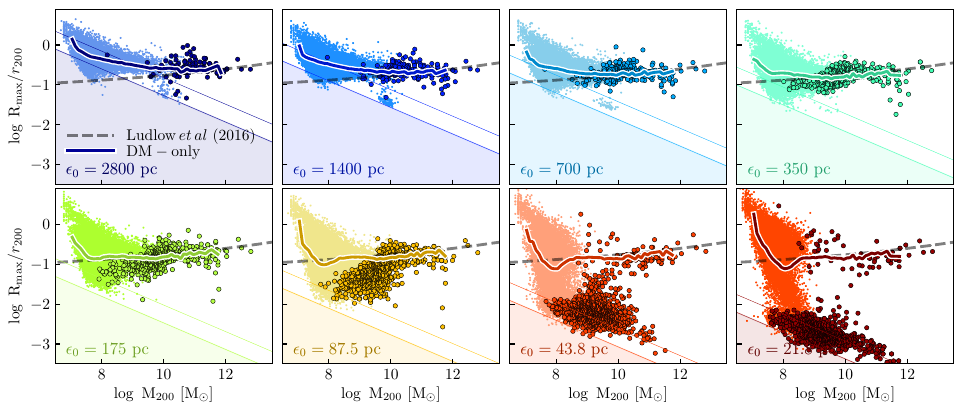}
  \caption{As in Figure~\ref{VmaxM200}, but for ${\rm R_{max}/r_{200}}$, a proxy for halo concentration. The
    shaded region highlight $r\leq \epsilon_0$; diagonal lines mark $r=\epsilon_{\rm sp}$, the spline softening
    length above which inter-particle forces become exactly Newtonian. Note that larger $\epsilon$ implies
    lower concentration, and that, provided $\epsilon_0\simgt 87.5\,{\rm pc}$, star formation is inhibited
    in haloes for which, on average, ${\rm R_{max}}\simlt \epsilon_{\rm sp}$. As $\epsilon$ is reduced, DM haloes
    become increasingly compact regardless of ${\rm M_{200}}$, particularly those hosting central galaxies. }
  \label{RmaxM200}
\end{figure*}

We explore the impact of galaxy formation on the internal structure of DM haloes in Figures~\ref{VmaxM200}
and~\ref{RmaxM200}, where we plot the ratios ${\rm V_{max}/V_{200}}$ and ${\rm R_{max}}/r_{200}$ (proxies for halo
concentration) versus ${\rm M_{200}}$ for all {\em main} haloes in our simulations. Dots distinguish those that do 
not host a stellar component from those that do, which are plotted using outsized circles. The thick solid line shows the median
trend for main haloes in the corresponding DM-only simulation (see Paper I for details);
the dashed line shows the prediction of the analytic model of \citet{Ludlow2016}, which is based on DM-only simulations. 

From Figure~\ref{VmaxM200} it is clear that
for large values of $\epsilon$ ($\simgt 700\,{\rm pc}$) the ${\rm V_{max}/V_{200}-M_{200}}$ relations in our
full hydrodynamical runs trace reasonably well the median relations in the DMO ones; but both yield haloes
that, at low mass, are systematically less concentrated than anticipated by the simple DM-only analytic model
(Figure~\ref{RmaxM200} suggests that this is a result of over-softening the central regions of haloes).
The agreement between our hydrodynamical and DM-only runs is perhaps
not surprising given that, at this resolution, only the more massive haloes host a central galaxy capable of modifying the
DM distribution. For smaller softening lengths, however, clear differences emerge. SF becomes increasingly
prevalent in low-mass haloes, leading to a systematic increase in their concentration. For
$\epsilon=175\,{\rm pc}$, for example, ${\rm V_{max}}$ exceeds ${\rm V_{200}}$ by as much as a factor
of 3 in the lowest-mass haloes hosting galaxies; for $\epsilon=43.8\,(21.8)\, {\rm pc}$ haloes become
even {\em more} concentrated, reaching ${\rm V_{max}/V_{200}}\approx 8\, (10)$ at the lowest masses. 
Intermediate- and high-mass haloes are also affected, but less so. Only the most massive haloes found in the
simulations have peak circular speeds that remain relatively insensitive to softening. 

Figure~\ref{RmaxM200} reveals other interesting trends. Similar to Figure~\ref{VmaxM200}, the
${\rm R_{max}}/r_{\rm 200}-{\rm M_{200}}$ relation paints a picture in which low-mass haloes become increasingly compact and
amenable to star formation as $\epsilon$ decreases. Notice as well that, regardless of $\epsilon$,
the least massive haloes capable of hosting a central galaxy typically have ${\rm R_{max}}\simgt \epsilon_{\rm sp}$
(recall that $\epsilon_{\rm sp}$ is the spline softening length beyond which
inter-particle forces become exactly Newtonian), although not all haloes that satisfy this criterion host galaxies.
A closer inspection reveals that SF is heavily suppressed in haloes for which
${\rm R_{max}}\simlt 2\times \epsilon_{\rm sp}=5.6\times \epsilon$ (except in extreme cases: when $\epsilon_0\simlt 43.8\,{\rm pc}$
stars readily form in haloes with ${\rm R_{max}}\simgt \epsilon$).
Nevertheless, as $\epsilon$ decreases, so does the limiting halo mass below which SF can occur. This, in turn,
affects the shape of the GSMF (see Figure~\ref{fig7}).

For example, an NFW profile with concentration $\approx 8 - 12$ (typical values for the mass
scales resolved by our simulation) has ${\rm R_{max}}\approx 0.18-0.27\times r_{200}$, which is comparable to,
or larger than, the spatial scale within which galaxy formation occurs\footnote{In fact, galaxies are
  often identified in simulations by first locating their DM haloes, and then assigning stellar and gas particles
  to the central galaxy provided they fall within a radius of $\approx (0.1-0.15)\times r_{200}$, which typically contains
  most of its baryonic mass \citep[see][for an in-depth discussion]{Stevens2014}.}. Because densities are suppressed
on scales smaller than the softening length, galaxy formation is pacified in haloes for which
${\rm R_{max}}\simlt \epsilon_{\rm sp}$. This explains the suppression of the GSMF at low $M_\star$ seen in
Figure~\ref{fig7}. For $\epsilon_0=2800\,{\rm pc}$ (or equivalently, $\epsilon_{\rm sp}=7840\,{\rm pc}$), we find
that ${\rm R_{max}}\simlt 2\times \epsilon_{\rm sp}$ for haloes with virial mass
$M_{200}\simlt 5.5\times 10^{10}\,{\rm M_\odot}$ and stellar masses $\simlt 10^8\,{\rm M}_\odot$, which agrees
well with the flattening of the cumulative GSMFs plotted in the upper-left panel of Figure~\ref{fig7}. 

It is perhaps not surprising that centrally-concentrated SF--when occurring in dense gas prone to
numerical over-cooling--gives rise to increasingly compact halo mass profiles. But the results
presented above, and in Figure~\ref{fig10}, raise additional questions. First, why do DM haloes become
increasingly compact when $\epsilon$ is reduced (Figure~\ref{RmaxM200}) and yet galaxies more diffuse
(Figure~\ref{fig10})? A closer inspection of Figure~\ref{RmaxM200} also reveals a substantial population
of ``dark'' haloes that contain no stellar component, and yet have exceptionally high concentrations that
exhibit a strong softening-dependence; this seems at odds with convergence tests based on DM-only
simulations (see Paper I for details). We turn our attention to these issues next. 

\subsubsection{Circular velocity profiles}
\label{DMVcirc}

\begin{figure*}
  \includegraphics[width=0.99\textwidth]{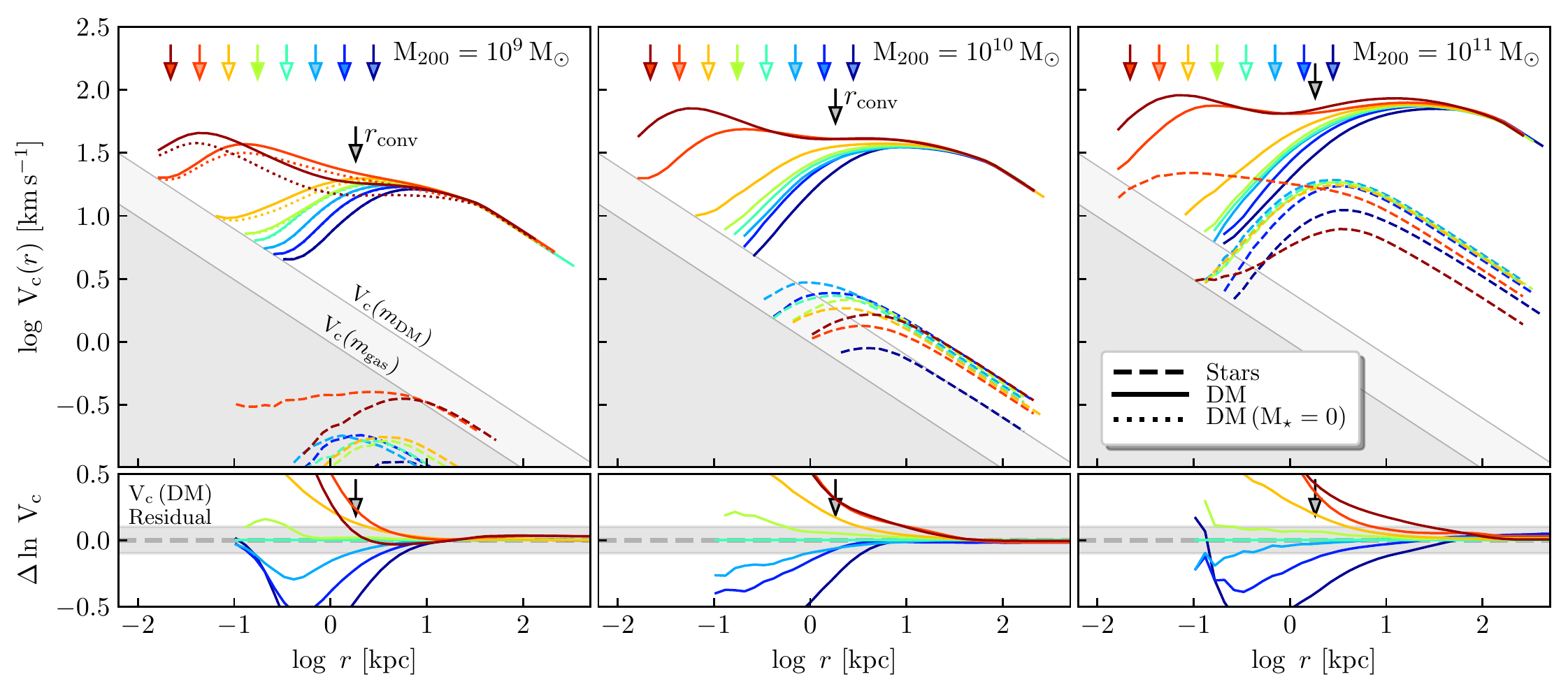}
  \caption{{\em Top panels:} Average ($z=0$) circular velocity profiles for DM haloes and galaxies in
    three separate bins of virial mass: ${\rm M_{200}}=10^9\,{\rm M_\odot}$ (left), $10^{10}\,{\rm M_\odot}$
    (middle) and $10^{11}\,{\rm M_\odot}$ (right). Results are shown for $N^3_{\rm p}=376^3$. Solid curves show the mean DM ${\rm V_c}(r)$ profiles for
    all haloes regardless of their stellar content; dashed curves for stars; dotted lines (where
    present) for DM haloes that contain no baryonic component. Lines are colour coded by softening
    length, the values of $\epsilon_0$ are indicated by downward arrows. Black arrows mark the convergence
    radius, $r_{\rm conv}=0.055\,(L/N_{\rm p})\approx 1.8\,{\rm kpc}$ \citep{Ludlow2019a}, expected from
    DM-only simulations. The light and dark shaded regions correspond to the Keplerian ${\rm V_c}(r)$ profiles
    of a single DM or (primordial) gas particle, respectively. {\em Bottom panels:} Residuals of the DM circular
    velocity profiles relative to the fiducial model.}
  \label{VcStacked}
\end{figure*}

Figure~\ref{VcStacked} plots the mean circular velocity profiles, ${\rm V_c}(r)=\sqrt{G M(r)/r}$,
of main DM haloes (solid lines; these curves
{\em do not} include the contribution of baryons) and their
central galaxies (dashed lines) in three bins of virial mass centred on
${\rm M_{200}}\approx 10^9\,{\rm M_\odot}$, $10^{10}\,{\rm M_\odot}$ and $10^{11}\,{\rm M_\odot}$. Results
are shown for $N^3_{\rm p}=376^3$ and different colours discriminate runs carried out with different softening
lengths (downward arrows mark the Plummer-equivalent lengths). 

The central mass profiles of haloes in all three mass bins exhibit a strong softening-dependence, becoming
increasingly concentrated as $\epsilon$ is reduced. As $r$ decreases, the $V_c(r)$ profiles diverge from one
another at radii that approach or exceed the nominal convergence radius of their DM halo (the black arrows
mark $r_{\rm conv}\approx 0.055\,(L/N_{\rm p})\approx 1.8\,{\rm kpc}$; see Paper I
for details), suggesting that convergence in the central structure of DM haloes--even well-resolved ones--is
more difficult to achieve in hydrodynamical simulations than in DM-only runs \citep[see also][]{Schaller2015a}.
However, not all runs are deviant. For a range of softening lengths, $175\simlt \epsilon_0/[{\rm pc}]\simlt 700$,
the $V_c(r)$ profiles do, in fact, converge to better than $\approx 10$ per cent at radii $\simgt r_{\rm conv}$. 
Those that do not correspond to runs with $\epsilon_0\simgt r_{\rm conv}\approx 1.8\,{\rm kpc}$ (dark blue curve),
or to those with $\epsilon_0\simlt 43.8\,{\rm pc}$ (or equivalently,
$\epsilon(z_{\rm reion})\simlt 13.3\,{\rm pc}\approx \epsilon_v$), in which inefficient photo-heating during
reionization stimulates star formation at high redshift in low-mass haloes. Agreement between these
profiles at $r\simgt r_{\rm conv}$ does not, however, imply convergence; that can only be established by
comparing runs of different {\em mass resolution}. We will return to this point in Section~\ref{recc}.

Interestingly, the increase in DM concentration that attends smaller softening lengths {\em is not}, 
in general, a consequence of a compact central stellar component. Stars comprise only a small fraction of the total mass 
in the central regions of haloes. Consider as an example the middle panel of Figure~\ref{VcStacked}, where
$M_{200}=10^{10}\,{\rm M_\odot}$ and ${\rm R_{max}}\approx 8\,{\rm kpc}$ (note that this is the value of ${\rm R_{max}}$
predicted by the analytic model of \citealt{Ludlow2016}). Considering only haloes that host a central galaxy, 
we find that the average stellar mass fraction within $\simlt {\rm R_{max}}=8\,{\rm kpc}$ is typically between
$f_{\rm bar}(<{\rm R_{max}})\approx 0.001$
to 0.011 for runs carried out with the smallest and largest softening lengths, respectively; for our fiducial softening,
$f_{\rm bar}(<{\rm R_{max}})\approx 0.002$. Note too, as expected from Figure~\ref{fig10}, the concentration of the {\em stellar}
mass profile depends non-monotonically on $\epsilon$, being least concentrated at the highest and lowest values of $\epsilon$
and most concentrated at intermediate values. 

Neither are the DM cusps a consequence of a dense, central concentration of non-SF gas. The dotted
curves in the left-most panel of Figure~\ref{VcStacked} show the circular velocity profiles of DM haloes that, at $z=0$,
contain no gas or stellar particles within their virial radius. The circular velocity profiles of haloes that {\em do not}
host central galaxies are remarkably similar to those that do. These puzzling results suggest that a stellar or
gaseous baryonic component capable of modifying the DM distribution may have once dominated the central regions of these
haloes, but has since been lost or diluted due to numerical relaxation between the DM and baryonic particles, potentially
accelerated by small softening \citep{Ludlow2019b}.

\begin{figure*}
  \includegraphics[width=0.75\textwidth]{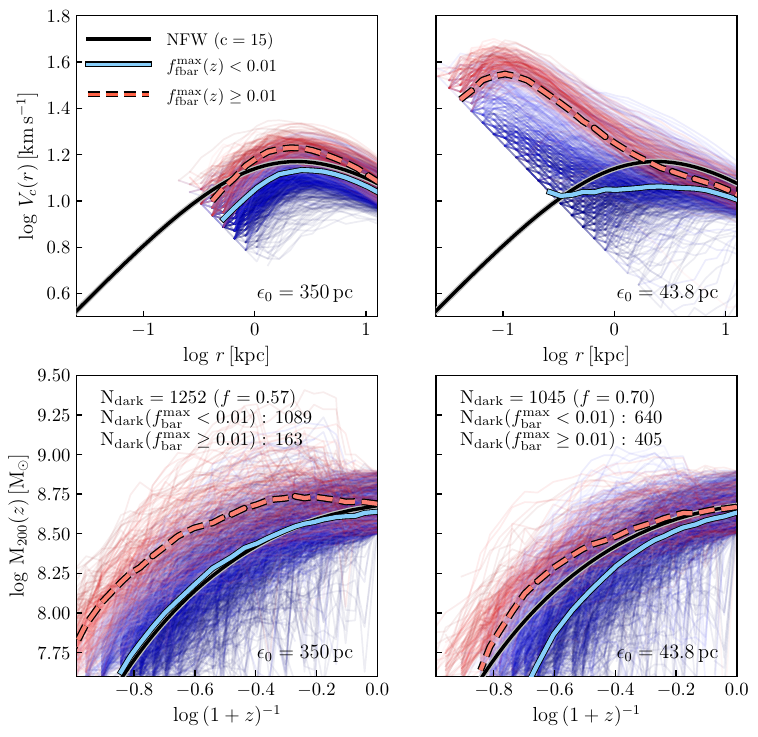}
  \caption{Circular velocity profiles (upper panels) and mass accretion histories (MAHs; lower panels) for
    dark matter main haloes that contain no baryons at $z=0$. Haloes are selected to lie in a narrow mass
    bin centred on ${\rm M_{200}}=5\times 10^8\,{\rm M_\odot}$ of width $\Delta\log {\rm M_{200}}=0.4$.
    Results are shown for $N^3_{\rm p}=376^3$ and for two softening lengths: $\epsilon_0=350\,{\rm pc}$ (our fiducial value; left panels)
    and $43.8\,{\rm pc}$ (right panels). We use the MAHs to divide the full halo sample into two sub-samples:
    blue curves correspond to those whose baryon fractions {\em never} exceeded 0.01 at any redshift;
    orange curves to those that did. ${\rm N_{dark}}$ is the total number of {\em dark} haloes that
    lie in this mass bin, which make up a fraction $f$ of the total. For both values of $\epsilon_0$,
    haloes that have once contained baryons form earlier than average, and have higher concentrations.}
  \label{VcMAH}
\end{figure*}

We explore this possibility further in Figure~\ref{VcMAH}, where we plot the ${\rm V_c}(r)$ profiles
(upper panels) and main progenitor mass accretion histories (MAHs; lower panels) for a sub-sample of ``dark''
DM haloes in a narrow range of ($z=0$) mass spanning $8.5\simlt \log {\rm M_{200}}/[{\rm M_\odot}]\simlt 8.9$. These
haloes contain no baryonic particles within their virial radii at $z=0$.
Results are shown for $N^3_{\rm p}=376^3$ and for two values of softening: $\epsilon=350\,{\rm pc}$
(our fiducial value; left-hand panels) and $\epsilon=43.8\,{\rm pc}$ (right-hand panels). In all cases,
profiles extend down to the radius enclosing 5 DM particles.
The total number of {\em dark} haloes, ${\rm N_{dark}}$, in each sample is provided in the lower panels;
these make up 57 per cent ($\epsilon=350\,{\rm pc}$) and 70 per cent ($\epsilon=43.8\,{\rm pc}$) of all
(dark plus luminous) haloes of the same mass. For comparison, an NFW profile with a concentration
$c=15$ (the value anticipated by the model of \citealt{Ludlow2016} for similar mass haloes) is shown using a thick
black line; the corresponding NFW MAH is shown in the lower panel \citep[see][for details]{Ludlow2014}.

The circular velocity profiles exhibit considerable scatter, particularly for $\epsilon=43.8\,{\rm pc}$. In that case,
two {\em distinct} sets of curves can be distinguished: one dominated by haloes with lower than average
concentrations, and another by much more concentrated systems, whose mass profiles are not well described by
an NFW profile. This dichotomy in profile shapes was already evident in the bimodal distribution of ${\rm R_{max}}$
seen in Figure~\ref{RmaxM200}.

How can baryon-free haloes exhibit mass profiles that deviate so strongly from one another? The colour coding
provides some clues. With values decreasing from red to blue, curves are coloured according
to the {\em maximum} baryon fraction, $f_{\rm bar}^{\rm max}$, that the halo's main progenitor ever reached during
its lifetime. Clearly a large fraction of these haloes have not always been dark. Star particles forming in these
haloes do so at higher densities as $\epsilon$ is decreased, giving rise a compact baryonic core capable of drawing in
dark matter, increasing its central concentration; the effect is more striking for small softening. 
For $\epsilon=350\,{\rm pc}$, for example, we find that $\approx 87$ per cent of these dark haloes have never hosted a substantial
baryonic component (defined here as $f_{\rm bar}^{\rm max}\geq 0.01$), but those that have form systematically earlier and
have slightly higher concentrations than those that have not. For
$\epsilon=43.8\,{\rm pc}$, a larger fraction of haloes have once been baryon rich. About 39 per cent have
$f_{\rm bar}^{\rm max}\geq 0.01$, half of which reached baryon fractions as high as 0.05. 
Median curves for these sub-samples are shown explicitly in Figure~\ref{VcMAH}: blue corresponds to 
those with $f_{\rm bar}^{\rm max}<0.01$ and red to those with $f_{\rm bar}^{\rm max}\geq 0.01$; the total number
of haloes in each sub-sample is provided in the lower panels. 
Although difficult to confirm with the available simulations, we speculate that the loss of baryons from low-mass
systems is a symptom of mass segregation driven by energy equipartition between baryonic and dark matter particles.
This occurs naturally in simulations such as our that model DM and baryons with unequal mass particles, and will be
more problematic in runs that adopt small softening lengths.

The ``mass segregation'' of DM and baryonic particles will also cause the former to accumulate in halo centres, possibly
explaining the steeply-rising central DM cusps in haloes that were once baryon rich, but this is difficult to establish
conclusively. For example, the increasing densities of star formation that accompanies smaller softening lengths may also
contribute to the high central DM densities in these runs due to the contraction of the DM halo driven by the temporary
presence of a compact baryonic component. The latter interpretation seems most likely given the softening-dependence of both
the DM mass profiles and SF rates. 

Although the haloes plotted in  Figure~\ref{VcMAH} are poorly resolved, they are nevertheless an important step on the
hierarchical ladder of structure formation and should not be ignored in convergence studies such as ours. As the
progenitors of later generations of haloes, poor convergence in the structure and SFHs of low-mass systems may seep
through their merger trees and have undesirable consequences for more massive systems at later times. Importantly, these
numerical artefacts are not limited to low-mass systems, but are also expected to be present in the centres of massive,
well-resolved galaxies. In this case,
mass segregation results in an expansion of the stellar component of a galaxy, and a contraction of its DM halo. Clearly
these issues require careful consideration if hydrodynamical simulations are to provide a meaningful assessment of the
distribution of DM on small scales. 

\section{Discussion}
\label{conclusions}

We carried out a suite of cosmological smoothed particle hydrodynamics simulations in order to 
clarify the impact of numerical parameters on the baryon content of dark matter haloes and
the properties of the galaxies that form within them. Our work complements previous studies that 
mainly focused on subgrid parameters (particularly those pertaining to stellar and AGN feedback efficiency)
and how they influence the statistics of the galaxy population
\citep[e.g.][]{Yepes1997,Springel2000,Stinson2006,Scannapieco2008a,Haas2013a,Haas2013b,Agertz2013}.
Instead, we employ a well-tested code and adopt subgrid parameters
that have been calibrated to ensure a realistic galaxy population at
some ``fiducial'' mass and force resolution and test the sensitivity of the model predictions
to changes in {\em numerical parameters}.

Our runs were carried out in (comoving) $L=12.5\,{\rm Mpc}$ boxes at two separate mass resolutions, one corresponding
to the ``intermediate-'' ($N^3_{\rm p}=188^3$) and another to the ``high-resolution'' $(N_p=376^3)$ simulations of
the \eagle{} project \citep[see][for details]{Schaye2015}. Each run used the same softening length for DM and baryonic
particles, which we varied from run to run by factors of two above and below the ``fiducial'' values adopted for \eagle~
($700\,{\rm pc}$ and $350\,{\rm pc}$ for our low- and high mass-resolution runs, respectively). Some runs used only
adiabatic hydrodynamics for the gas, whereas others adopted the full-physics implementation of the \eagle{} subgrid
model. Relevant numerical aspects of our runs are summarized in Table~\ref{TabSimParam}.

Using these simulations we tested plausible constraints on gravitational softening that were
estimated analytically in section~\ref{Expectations}. For adiabatic (non-radiative) simulations these
restrictions--imposed to ensure that the lowest-mass haloes resolved by the simulation are not unduly
influenced by softening--confine the dynamic range of the gravitational softening length, 
$\epsilon$, to roughly a factor of 20. The upper limit ensures $\epsilon$ remains {\em smaller}
than the virial radius of the lowest-mass haloes resolved by the simulation, and the lower limit
ensures that $\epsilon$ is {\em large} enough
to impede collisional heating of gas particles in those haloes. Satisfying both constraints limits
$\epsilon$ to a rather narrow window, roughly $0.024\simlt \epsilon/(L/N)\simlt 0.52$ (see eqs.~\ref{eq:maxeps} and
\ref{eq:mineps_adiabatic}). Imposing a minimum resolved escape velocity of $\approx 10\,{\rm km\,s^{-1}}$
(required for effective photo-heating during reionization or in HII regions; eq.~\ref{eq:minveps}) or a requirement for
numerically efficient feedback (eq.~\ref{eq:minnHeps}) yield even more conservative lower limits.
Simulations that do not comply with these restrictions are subject to numerical artefact.

Although our study reveals several important numerical effects in hydrodynamical 
simulations, it leaves a number of important issues unsettled. In the remainder of this section we 
provide a summary of our findings before highlighting several avenues for future progress.

\subsection{Summary}
\label{summary}

\begin{enumerate}
  \setlength\itemsep{1em}

\item Gas particles in non-radiative hydrodynamical simulations fall victim to significant 2-body scattering
  for $\epsilon$ less than the critical value, $\epsilon^{\rm min}_{\rm 2body}$ (eq.~\ref{eq:mineps_adiabatic}).
  Smaller values incite collisional heating, which reduces halo baryon fractions in low-mass
  systems, at least in the absence of radiative cooling (Figure~\ref{fig1}).
  The further $\epsilon$ veers below $\epsilon^{\rm min}_{\rm 2body}$, the more noticeable
  the heating effects. Our analytic estimate suggests that the critical value below which the baryon
  fractions of $N_{200}\approx 100$ haloes will be affected is of order $\epsilon \simlt 0.024\,(L/N_{\rm p})$,
  which agrees remarkably well with our numerical results. 

\item Collisional heating is therefore a potentially important but commonly neglected source of
  ``numerical feedback'', but may be suppressed if
  gas is allowed to cool radiatively. In this case, however, softening imposes a {\em minimum resolved
    escape velocity}, $v_\epsilon\propto (m_{\rm g}/\epsilon_v)^{1/2}$ (eq.~\ref{eq:minveps}), due to the self-binding
  energy of gas particles. Physical processes that occur on scales smaller than $v_\epsilon$ may be suppressed in runs
  that adopt $\epsilon\simlt \epsilon_v$. This result has important implications for accurately modelling HII regions
  and photo-heating due to reionization: for example, if $\epsilon$ is chosen such that $v_\epsilon\simgt 10\,{\rm km\,s^{-1}}$
  (approximately the sound speed in ionized gas at temperature $\sim 10^4\,{\rm K}$), the effects reionization are suppressed.
  We showed this in Figure~\ref{fig3}, where we compared the baryon fractions of haloes identified at $z=10$
  in a suite of simulations, some of which had $\epsilon\ll\epsilon_v$ while others had $\epsilon\gg\epsilon_v$. 
  Those with $\epsilon \simlt \epsilon_v$ exhibit clear symptoms: immediately following reionization, low-mass
  ($\simlt 10^8\,{\rm M_\odot}$, or $V_{200}\simlt 20\,{\rm km\,s^{-1}}$) haloes are replete with baryons despite the
  photo-heating effects; this provides fuel for star formation (SF) in a population of haloes that would otherwise
  have remained dark.

\item The softening length places an upper bound on the density of gas particles eligible to form stars,
  with maximum values scaling as $n_{\rm H}^{\rm max}\propto \epsilon^{-3}$ (eq.~\ref{eq:nmax2};
  Figure~\ref{fig5}). This may be 
  problematic if significant numbers of stars form from gas particles whose densities exceed the critical
  value, $n_{\rm H,tc}$, required for numerically efficient feedback (eq.~\ref{eq:nHtc}), which occurs for
  softening lengths smaller than $\epsilon_{\rm eFB}$ (eq.~\ref{eq:minnHeps}).
  Previous work has shown that galaxies whose stars form under such conditions are unrealistically compact
  \citep{Crain2015}, even when their stellar masses are sensible. Figure~\ref{fig5} shows that,
  as $\epsilon$ decreases, a growing number of stars form above this critical threshold,
  rising from about 9 per cent for $\epsilon/\epsilon_{\rm eFB}\approx 1.4$ ($\epsilon_{\rm eFB}\approx 500\, {\rm pc}$ in our
  $N^3_{\rm p}=376^3$ run) to $\simgt 54$ per cent for $\epsilon/\epsilon_{\rm eFB}\approx 0.044$
  (equivalent to a present-day softening length $\epsilon_0=21.9\,{\rm pc}$). Suppressing SF above $n_{\rm H,tc}$ requires requires
  $\epsilon$ to be chosen in accord with eq.~\ref{eq:minnHeps}. Imposing a sufficiently steep
    equation of state, or pressure floor, to
    impede such high gas densities is a possible alternative to avoiding inefficient feedback.

\item These results suggest that galaxies may inherit fatal characteristics 
  if $\epsilon$ falls below certain thresholds. At fixed particle mass, for example, the low-mass end of
  the galaxy stellar mass function (GSMF) becomes systematically {\em steeper} if $\epsilon$ is decreased
  by more than a factor of $\approx 4$ below our fiducial value. This is driven by two main effects.
  First, SF is systematically enhanced in low-mass haloes at early times,
  increasing the number of low-mass galaxies. For example, the number of ($z=0$) galaxies in our
  $N^3_{\rm p}=376^3$ runs with masses $\simlt 10\times m_{\rm g}$ increases from 40 to a maximum of 2288 when
  $\epsilon$ decreases from $2800\,{\rm pc}$ to $21.8\,{\rm pc}$; there are 663 in our fiducial run (Figure~\ref{fig7}).
  Second, SF in all haloes is {\em suppressed} at later times, which decreases the total stellar mass
  of galaxies residing in the most massive DM haloes (Figure~\ref{fig2}). Combined, these effects steepen the GSMF.

  Increasing $\epsilon$ relative to our fiducial value also affects GSMFs. This occurs
  because higher values of $\epsilon$ reduce densities in halo centres, suppressing SFRs globally across all
  halo masses. But softening also affects galaxy formation models by imposing a lower-limit
  on the halo mass within which galaxy formation is able to proceed, which roughly corresponds to
  the halo mass scale below which the spline softening length $\epsilon_{\rm sp}$ is of order the 
  galaxy's physical size, or roughly ${\rm R_{max}}/2$. The main effect is to curb the GSMF below
  a softening-dependent limiting mass (see Figures~\ref{fig7} and \ref{RmaxM200}). 
  
\item At fixed particle mass, galaxy sizes are also vulnerable to changes in softening length, in part due to
  energy equipartition, i.e. the transfer of energy from massive DM particles to lower mass stars via 2-body
  scattering. The effect can be minimized by decreasing the ratio $\mu=m_{\rm DM}/m_{\rm gas}$
  \citep[see][]{Ludlow2019b}, or by increasing the softening length. Our suite of
  simulations indicate that mass segregation is present in
  our fiducial runs (which adopt $\mu\approx 5.3$), but is noticeably diminished for
  $\epsilon\simgt 2\,\epsilon_{\rm fid}$. In fact, the size-mass relations obtained from runs that used
  $\epsilon=2\,\epsilon_{\rm fid}$ agree best with observational constraints. Larger values
  tend to ``puff-up'' galaxies, regardless of their stellar mass, and runs with $\epsilon\simlt \epsilon_{\rm fid}$
  show signs of mass segregation: galaxies of a given $M_\star$ become increasingly diffuse as $\epsilon$
  is decreased, and have sizes that increase systematically with decreasing redshift, at least for
  $z\simlt 2$. The redshift dependence is erased in runs that use $\mu\approx 1$, or
  $\epsilon\simgt 2\,\epsilon_{\rm fid}$. These results are summarized in Figure~\ref{fig10}.

\item The structure of DM haloes identified in collisionless DM-only simulations is robust to changes in
  softening length at essentially all $r \simgt \epsilon$. Convergence in halo mass profiles between different
  mass resolutions is achieved at radii that exceed the well-known ``convergence radius''
  \citep[e.g.][]{Power2003,Navarro2010,Ludlow2019a}. This is {\em not} the case, in general, for cosmological
  hydrodynamical simulations, in which the innermost structure of DM haloes responds to differences in
  galaxy formation physics driven by changing $\epsilon$ (Figures~\ref{VmaxM200} and \ref{RmaxM200}).
  Smaller $\epsilon$ implies higher densities of
  SF gas in halo centres, and potentially less efficient feedback due to the corresponding increase in
  radiative energy losses. If sufficiently small, the high gas densities in halo centres draws in DM,
  which gives rise to a central DM density cusp that is steeper than the $\rho\propto r^{-1}$
  expected from DM only simulations (Figure~\ref{VcStacked}). Surprisingly, at late times this cusp is {\em not}
  baryon dominated: for example, it is a noticeable feature of many poorly-resolved haloes that are {\em today}
  baryon free, but that were not in the past (Figure~\ref{VcMAH}).
  We suspect that this result is related to the segregation of baryonic and DM particles in the inner
  regions of haloes: haloes dominated by a central stellar/gaseous component can lose it through 2-body
  scattering between DM and baryonic particles, and retain their high central concentrations of DM.  
  Nevertheless, the situation is not all dire. For a range of softening lengths,
  $1/2\simlt \epsilon_{\rm fid}\simlt 2$, circular velocity profiles are independent of $\epsilon$ at radii
  $r\simgt r_{\rm conv}$.

\end{enumerate}

The analytic constraints on softening discussed above and in Section~\ref{Expectations} are applicable to SPH simulations
\citep[and $\epsilon\gg\epsilon_{\rm eFB}$ to those employing the stochastic feedback scheme of][]{DallaVecchia2012}. Magneticum
\citep{Dolag2016}, NIHAO \citep{Wang2015,Buck2019}, CLUES \citep{Libeskind2010}, APOSTLE \citep{Sawala2016}, Hydrangea
\citep{Bahe2017}, and the galaxy clusters of the Three Hundred Project \citep{Cui2018}--recent examples of large-scale
SPH simulations--all adopt $z=0$ physical softening lengths that exceed $\epsilon^{\rm min}_{\rm 2body}$ (for halos
resolved with $\simgt 200$ particles) and $\epsilon_v$ (for
$v_\epsilon=10\,{\rm km/s}$). For results independent of softening, however, these restrictions must be
satisfied at {\em all} redshifts and the time-dependence of $\epsilon$ should be chosen with this in mind.
The resolution study \citet{Hopkins2018} carried out for the FIRE-2 project focused on a single Milky Way-mass
halo and achieved a maximum (baryonic) mass resolution of $\approx 7000\,{\rm M_\odot}$ and minimum (baryonic)
force softening of $\approx 0.4\,{\rm pc}$. According to eq.~\ref{eq:minveps}, this is sufficient to resolve
photoheating in halos with ${\rm V_{max}}\simgt 12\,{\rm km/s}$. Similar conclusions apply to the Latte
simulations presented in \citet{Wetzel2016}.

\begin{figure*}
  \includegraphics[width=0.95\textwidth]{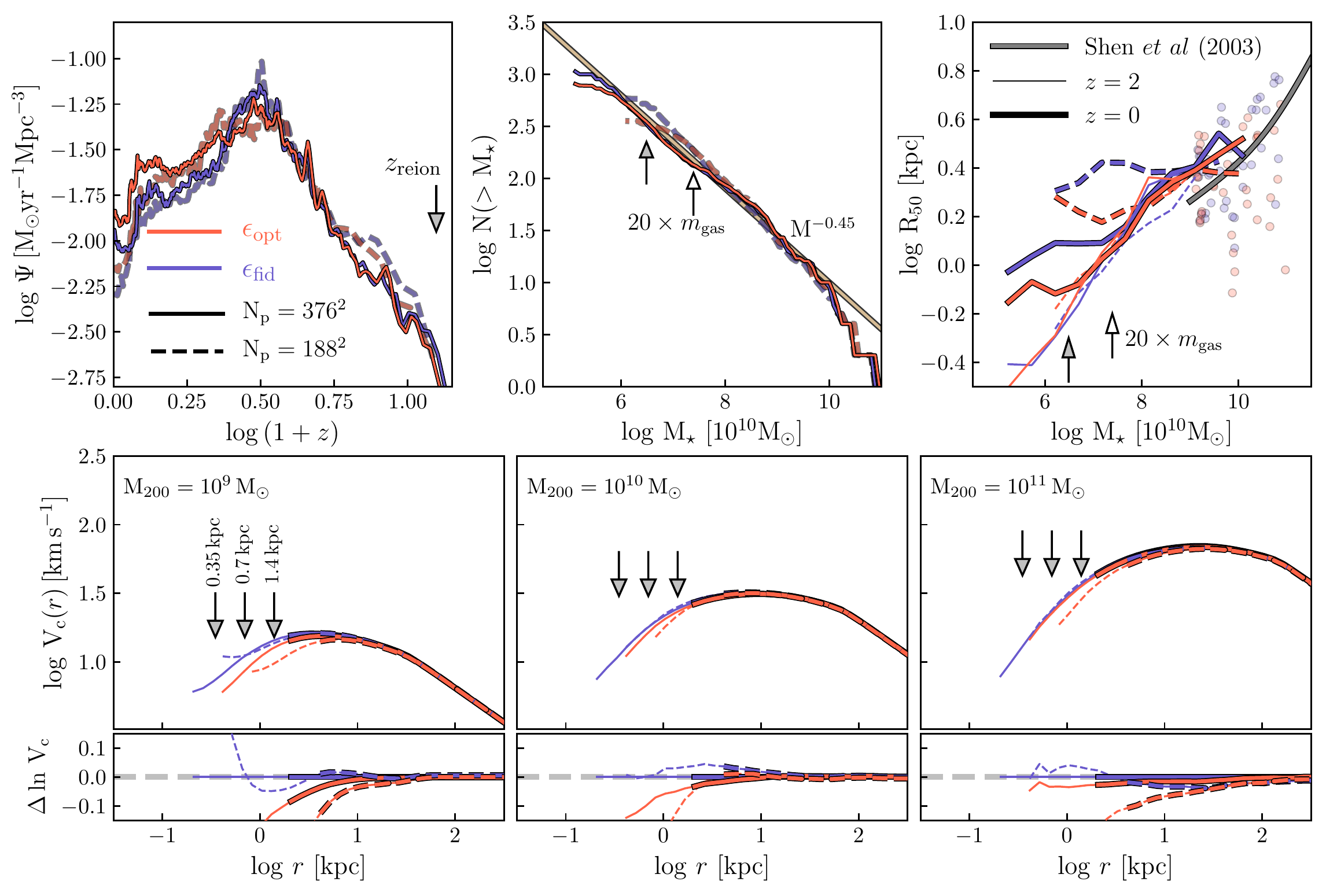}
  \caption{A summary of results from intermediate- ($N^3_{\rm p}=188^3$) and high-resolution ($N^3_{\rm p}=376^3$)
    runs that used our fiducial softening length, $\epsilon_{\rm fid}$ (blue curves), or the ``optimal'' softening
    length, $\epsilon_{\rm opt}$ (orange curves), described in Section~\ref{recc}. Top panels, from left to right,
    show the cosmic SFH, the galaxy stellar mass function, and the galaxy size-mass relations at $z=0$ (thick lines)
    and $z=2$ (thin lines). SFHs are reasonably well-converged with mass resolution at all $z$, as are GSMFs for masses that
    exceed that of $\approx 20$ primordial gas particles (vertical arrows in the middle panel). At both mass
    resolutions, the sizes of low-mass galaxies at $z=0$ are systematically {\em smaller} for $\epsilon_{\rm opt}$
    than for $\epsilon_{\rm fid}>\epsilon_{\rm opt}$, due to the suppression of 2-body scattering. Lower
    panels plot the circular velocity profiles of DM haloes in separate bins of (DM) halo mass:
    $10^9\,{\rm M_\odot}$ (left), $10^{10}\,{\rm M_\odot}$ (middle) and $10^{11}\,{\rm M_\odot}$ (right).
    Residuals are shown in the lower panels. In each case, thick lines extend down to the convergence radius
    expected for DM-only simulations \citep[see][for detail]{Ludlow2019a}; thin lines extend to $\epsilon/2$.
    The $V_c(r)$ profiles converge to better than $\approx 10$ per cent at all radii $\simgt r_{\rm conv}$.}
  \label{EpsOpt}
\end{figure*}

\subsection{Recommendation for cosmological hydrodynamical simulations}
\label{recc}

It is worth emphasizing that the values of $\epsilon$ at which the issues mentioned above first manifest
are not only of academic interest, and are close to those adopted for many existing state-of-the-art
simulations. {\textsc{eagle}}, for example, adopted a maximum physical softening length (at $z\leq 2.8$) of
$\approx 0.011\times L/N_{\rm p}=700\,{\rm pc}$ ($\epsilon=0.04\times L/N_{\rm p}$ for $z>2.8$) for the
intermediate-resolution ($L=100\,{\rm Mpc}$) simulation, only a factor of $\approx 4.5$ (1.4) larger
than the minimum value, $\epsilon_v$, required to resolve a escape velocity of $10\,{\rm km s^{-1}}$ 
at $z=0$ ($z_{\rm reion}=11.5$).
Softening lengths used in the high-resolution simulation of the \eagle{} project ($L=25\,{\rm Mpc}$ and
$N^3_{\rm p}=752^3$) were a factor of two smaller (in physical units; equivalent in units of the mean
inter-particle spacing). In that case, $\epsilon_0/\epsilon_v\approx 18$ and
$\epsilon(z_{\rm reion})/\epsilon_v\approx 5.5$. These softening lengths are just large enough to ensure
that the effects of photoionization heating associated with reionization are not artificially suppressed.

The more conservative constraint on softening, however--and the one that has affected the
outcome of our simulations the most--stems from the requirement for
efficient thermal feedback, $\epsilon\geq \epsilon_{\rm eFB}$ (eq.~\ref{eq:minnHeps}). Both the intermediate- and
high-resolution \eagle{} simulations fall short of this value, by factors of $\approx 2$ and $\approx 1.4$,
respectively (the values quoted here refer to $z=0$). This suggests that there may be some forgiveness when
overstepping this bound, as neither simulation appears to have been severely affected by inefficient feedback.
Nevertheless, there is room for improvement: we find that 28 per cent of stars formed in \eagle's
intermediate-resolution $L=100\,{\rm Mpc}$ simulation (18 per cent in the high-resolution $L=25\,{\rm Mpc}$
run) do so at densities greater than $n_{\rm H,tc}$, and therefore suffer from numerical over-cooling.
To ensure efficient thermal feedback at all resolved densities, softening lengths should be chosen
so that $\epsilon\simgt \epsilon_{\rm eFB}$, or feedback models should be {\em designed} to be efficient
at a desired $\epsilon$ and maximum resolved density. Alternatively, an equation of state that
prohibits gas particles from reaching densities $\simgt n_{\rm H,tc}$ could be imposed.

Convergence requirements for DM-only simulations suggest that $\epsilon$ should not exceed the comoving
convergence radius, $r_{\rm conv}$, of DM haloes that is dictated by 2-body scattering. In Paper I we showed that
$r_{\rm conv}\approx 0.055\times (L/N_{\rm p})$ (where $L$ is the comoving simulation box size),
regardless of halo mass or redshift. Because of this, softening lengths for hydrodynamical simulations that
ensure efficient thermal feedback but do not jeopardize the innermost structure of DM haloes may require
compromise: it will not be possible to simultaneously satisfy $\epsilon > \epsilon_{\rm eFB}$ (a condition
on the {\em physical} softening length) and $\epsilon<r_{\rm conv}$ (which restricts the {\em co-moving}
softening) at all redshifts and for arbitrary mass resolutions.

Simulations such as ours that employ baryon and DM particles of unequal mass also suffer from energy equipartition
that affects galaxy sizes \citep{Ludlow2019b}. The size-mass relations
presented in Figure~\ref{fig10} suggest that the strength of this effect increases 
as $\epsilon$ is reduced. Minimizing the spurious growth of galaxy sizes 
requires softening lengths that are sufficiently large to suppress 2-body scattering
as much as possible, but sufficiently small so that gravitational forces are unbiased on the relevant
spatial scales, in our case $R_{50}$. At present we
lack a detailed analytic framework that can guide our choices, but the results
shown in Figure~\ref{fig10} suggest that softening lengths roughly a factor of two {\em larger} than our
fiducial lengths (or $\approx 1400\,{\rm pc}$ and 700 pc for our intermediate- and high-resolution runs,
or $\approx 0.022\times L/N_{\rm p}$) provide a reasonable compromise for our fiducial mass resolution.

Our recommended ``optimal'' softening for future large-scale hydrodynamical SPH simulations is therefore to adopt a
{\em comoving} softening length of $\epsilon_{\rm opt}^{\rm CM}\approx 0.05\times (L/N_{\rm p})\simlt r_{\rm conv}$
at early times, and a {\em maximum physical} softening length of
$\epsilon_{\rm opt}^{\rm phys}\approx 0.022\times (L/N_{\rm p})$ at late times.
The transition redshift between the two regimes takes place at
$z_{\rm phys}=\epsilon_{\rm opt}^{\rm CM}/\epsilon_{\rm opt}^{\rm phys}-1\approx 1.27$. These criteria ensure that
feedback will be maximally efficient without compromising the inner-most structure of DM haloes
(since $\epsilon\simlt r_{\rm conv}$ at all times) while also minimizing the $\epsilon$-dependent gravitational
heating and force-biasing that affects galaxy sizes (Figure~\ref{fig10}). These recommendations are for
simulations that adopt equal numbers of of baryonic and DM particles, and the same softening lengths for all
particle species. Variations on this theme require additional testing.

Figure~\ref{EpsOpt} compares the results of two high- (solid lines) and intermediate-resolution (dashed lines)
simulations carried out with our fiducial softening length (blue curves) and with $\epsilon_{\rm opt}$ (orange).
Top panels show the cosmic star formation histories (SFHs; left), GSMFs (middle) and size-mass relations (right)
at $z=0$ and 2 (thick and thin lines, respectively). SFHs and GSMFs are reasonably well-converged with mass
resolution (provided galaxies are resolved with
at least 20 stellar particles for the latter; upward arrows) regardless of $\epsilon$. The sizes
of the most massive galaxies are also well-converged, but low-mass objects are smaller for the high-resolution runs
and for our optimal softening length (note that both of these changes--higher mass resolution and larger
softening--will, at fixed stellar mass, increase collisional relaxation times and therefore suppress 2-body scattering). 

The lower panels of Figure~\ref{EpsOpt} plot the median circular velocity profiles of DM haloes in three
bins of halo mass (from left to right, $10^9,\, 10^{10},\,{\rm and}\, 10^{11}{\rm M_\odot}$, respectively). Each
profile is plotted down to $r_{\rm conv}=0.055\times L/N_{\rm p}$ using thick lines, and extended to $\epsilon/2$
with thin lines. Residuals in the bottom panels confirm that, regardless of mass, the median circular velocity
profiles of DM haloes are converged to better than $\approx 10$ per cent at radii that exceed $r_{\rm conv}$.

\subsection{Outlook for future work}
\label{future}

Although our results shed light on the complex issue of ``strong convergence'' in hydrodynamical simulations,
there is clearly work to be done. 
First, our analysis focused exclusively on a limited number of galaxy properties (sizes, stellar mass, SFH)
and how they are affected by force softening and relatively modest changes to particle mass. 
But what impact would these changes have on other galaxy properties, such as their internal dynamics, their
central black hole mass, on fundamental relations such as the Tully-Fisher and Faber-Jackson relations,
or on the SFR-stellar mass relation? The impact of gravitational softening on the cosmic SFH, and galaxy masses
and sizes is not subtle, and we therefore find it unlikely that other relations will be fully absolved of
its influence.

In addition, softening and particle mass are only two of a number of important numerical parameters that
may impact the outcome of hydrodynamical simulations. Others include--but are not limited to--the number of
SPH neighbours, $N_{\rm ngb}$, used for smoothing hydrodynamic variables, the minimum smoothing length,
$l_{\rm hsml}^{\rm min}$ (fixed to $0.1\times \epsilon_{\rm sp}=0.28\times \epsilon$ in all of our runs; but see Appendix~\ref{sA3}), or
the use of adaptive instead of fixed softening lengths, or softening lengths that differ for different particle
types.

Our tests are also limited in dynamic range: all were carried out in the same $L=12.5\,{\rm Mpc}$
(comoving) cubic box and, as a result, contain only a handful of galaxies with masses exceeding
$\sim 10^{12}\,{\rm M}_\odot$, and are therefore unlikely to test convergence in the regime where AGN are expected
to strongly impact galaxy formation and evolution.

Nevertheless, some of our results are likely applicable to more massive systems, particularly
those only mildly affected by AGN. The cosmic SFHs shown in Figure~\ref{fig2}, for example, are dominated by a handful
of the most massive galaxies present in the runs, whose {\em individual} SFHs also exhibit softening-dependent
``peaks'' at early times. The peaks arise because gas particles are able to collapse to higher densities
as $\epsilon$ is decreased, resulting in enhanced SF\footnote{Enhanced SF is eventually quelled by the
  corresponding enhancement of stellar feedback, which, when softening is sufficiently small, results in a
  turnover in the SFHs and the appearance of a  ``peak''.}. At such high redshift AGN feedback is likely
unimportant in all but the rarest of collapsed structures, so the SFHs of $z=0$ galaxies much more massive than
those probed by our simulations are likely to exhibit unphysical softening-dependent peaks as well.

The DM response to galaxy formation in halo centres (Figures~\ref{VmaxM200} to \ref{VcMAH}) is another
issue of convergence that we expect extrapolates to more massive galaxies and haloes. Galaxy sizes, however, are
affected by 2-body scattering, which has diminishing importance with increasing numbers of (DM and stellar)
particles. The majority of our galaxies are affected by 2-body scattering at their half-mass radii (Figure~\ref{fig10}).
Substantially more massive systems will not be. It is difficult to anticipate the impact of numerical parameters on
galaxy size in that regime.

Resolving these issues will require a systematic study of ``zoomed''
simulations \citep[such as those recently carried out by][though without AGN]{Hopkins2018} targeting
massive haloes, or, preferably, to adopt an
approach similar to ours but using much {\rm larger} cosmological volumes. The latter, while desirable,
will be computationally costly, perhaps prohibitively.

For hydrodynamical simulations, strong convergence for arbitrary choices of numerical parameters
is unlikely to be achieved, and is perhaps undesirable. For example, subgrid models often include
physically-motivated parameters that are constrained by observation or by theory and {\em do not} take
part in calibration (for \eagle, the SF density threshold and supernova time delay are two examples).
Numerical parameters that preclude a realistic application of the subgrid model should therefore be avoided:
no amount of calibration will compensate for poorly-chosen numerical parameters.
Ideally, simulations should self-consistently model physical processes above some resolved length scale
(or below a maximum-resolved density) and emulate smaller-scale physics using subgrid models, thereby
establishing the equations that must be solved. Numerical parameters should then be chosen such that those
equations are solved accurately, but with minimal computational cost. This division of scales is
not easy to achieve but this paper makes a great stride in that direction. Changing the resolution
(for example, by modifying numerical parameters such as particle mass or force softening) introduces new
physical scales which may demand revisions to the subgrid model.

Finally, we note that our convergence tests reveal a strong junction between numerical parameters and those
governing subgrid models for star formation and stellar feedback. Co-varying these parameters may disclose other
important convergence criteria, similar to the ones presented in section~\ref{Expectations}, or may reveal ways
by which careful recalibration can compensate for changes brought about by different choices of numerical parameters.
We hope that our work inspires
future efforts to address these issues and to establish robust convergence criteria for
hydrodynamical simulations analogous to those frequently used for collisionless dynamics.

\section*{Acknowledgements}
We wish to thank Lydia Heck and the cosma support team for assistance with a variety of computing
issues. We also with to acknowledge various public {\textsc{python}} packages that have greatly
benefited our work: {\textsc{scipy}} \citep{scipy}, {\textsc{numpy}} \citep{numpy},
{\textsc{matplotlib}} \citep{matplotlib} and {\textsc{ipython}} \citep{ipython}.
ADL is supported by a Future Fellowship from the Australian Research Council (project number
FT160100250). MS is supported by the Dutch Research Council (NWO Veni 639, 041, 749).
This work was supported by the Science and Technology Facilities
Council [ST/P000541/1]. This work used the DiRAC Data Centric system at Durham University,
operated by the Institute for Computational Cosmology on behalf of the STFC DiRAC HPC
Facility (www.dirac.ac.uk. This equipment was funded by a BIS National E-infrastructure
capital grant ST/K00042X/1, STFC capital grant ST/K00087X/1, DiRAC Operations grant
ST/K003267/1 and Durham University. DiRAC is part of the National E-Infrastructure.

\appendix

\section{Sensitivity to other numerical and subgrid parameters}
\label{sec:A1}

\begin{figure*}
  \includegraphics[width=0.98\textwidth]{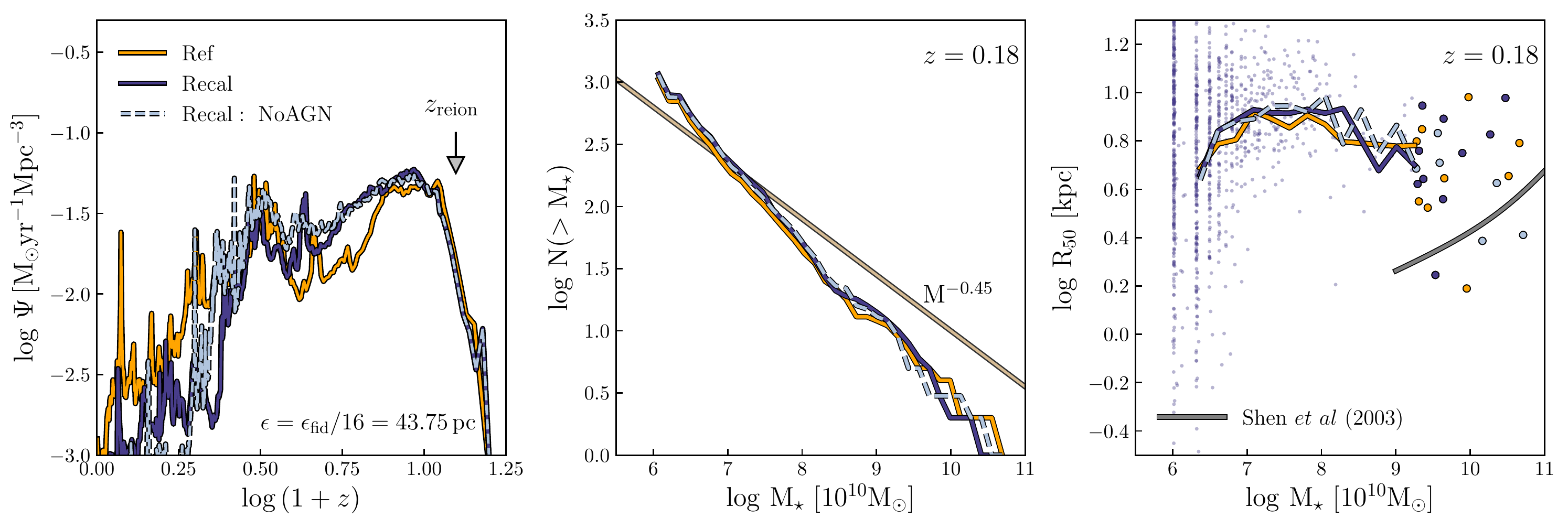}
  \caption{Cosmic star formation histories (left), cumulative galaxy stellar mass functions (middle)
    and galaxy size-mass
    relations (right) for three $N^3_{\rm p}=188^3$ runs carried out with different subgrid physics. All runs
    used the same gravitational softening length, $\epsilon=\epsilon_{\rm fid}/16=43.75\,{\rm pc}$. 
    Orange lines are used for a run carried out with Ref parameters (these parameter values were chosen
    by calibrating runs that used the same mass resolution as those shown); the dark and light (dashed)
    blue lines adopted the Recal model both with and without feedback from AGN (parameters of the Recal
    model were selected by calibrating simulations that have 8 times better mass resolution than those
    shown above, corresponding to our $N^3_{\rm p}=376^3$ runs). The results of our simulations are 
    robust to changes in subgrid parameters, and to the presence (or absence) or AGN feedback.}
  \label{A1}
\end{figure*}

A number of results of our work remain puzzling. We noted in Section~\ref{CSFH} that the early
spike in SF common to runs with $\epsilon\simlt \epsilon_v$ at $z_{\rm reion}$ is often accompanied
by a sharp decline in SF at late times. Figure~\ref{fig2} suggest that this is
true regardless of mass resolution. This result is particularly surprising given that the same runs
form the majority of their stars at densities $n_{\rm H}> n_{\rm H,tc}$ (Figure~\ref{fig5}) and should, as a result,
succumb to inefficient stellar feedback. This should {\em encourage} star formation, not disable it. 

Figures~\ref{VcStacked} and \ref{VcMAH} indicate that the central structure of dark matter haloes
in hydrodynamical simulations depend sensitively on gravitational softening, a result that is
at odds with the outcome of numerical convergence studies based on dark matter only simulations. 

What gives rise to these results? In the following sections we rule out some obvious possibilities
using the simulations in described in Table~\ref{TabSimParamApp}.

\subsection{Subgrid parameters and AGN feedback}
\label{sA1}

In Figure~\ref{A1} we plot the cosmic star formation histories in three runs carried out with
different subgrid
models: one Reference (orange lines) and two Recalibrated models that either include (solid blue lines)
or ignore AGN feedback (dashed light blue lines).
All runs used $N^3_{\rm p}=188^3$ particles (of both gas and DM) and a $z=0$ maximum physical softening
length of $\epsilon_0=43.75\,{\rm pc}$; the physical softening
lengths at $z_{\rm reion}=11.5$ ($\epsilon(z_{\rm reion})=13.3$) are therefore {\em smaller} than
$\epsilon_v\approx 156\,{\rm pc}$ by roughly a factor of $\approx 12$.
The same initial burst in SF at $z\approx z_{\rm reion}$ is clear in all runs, and is followed
by a sharp decline in SF for $z\simlt 2$, regardless of the precise details of the feedback
implementation. The same is true of the galaxy mass function and the size-mass relation. While
the results of these runs are in clear conflict with those of our fiducial model (the curves in
Figure~\ref{A1} can be compared to those in Figures~\ref{fig2}, \ref{fig7} and \ref{fig10}), they
are at least robust to small variations in subgrid parameters at {\em fixed mass resolution}, as well
as to the presence or absence of AGN feedback.

\begin{figure*}
  \includegraphics[width=0.8\textwidth]{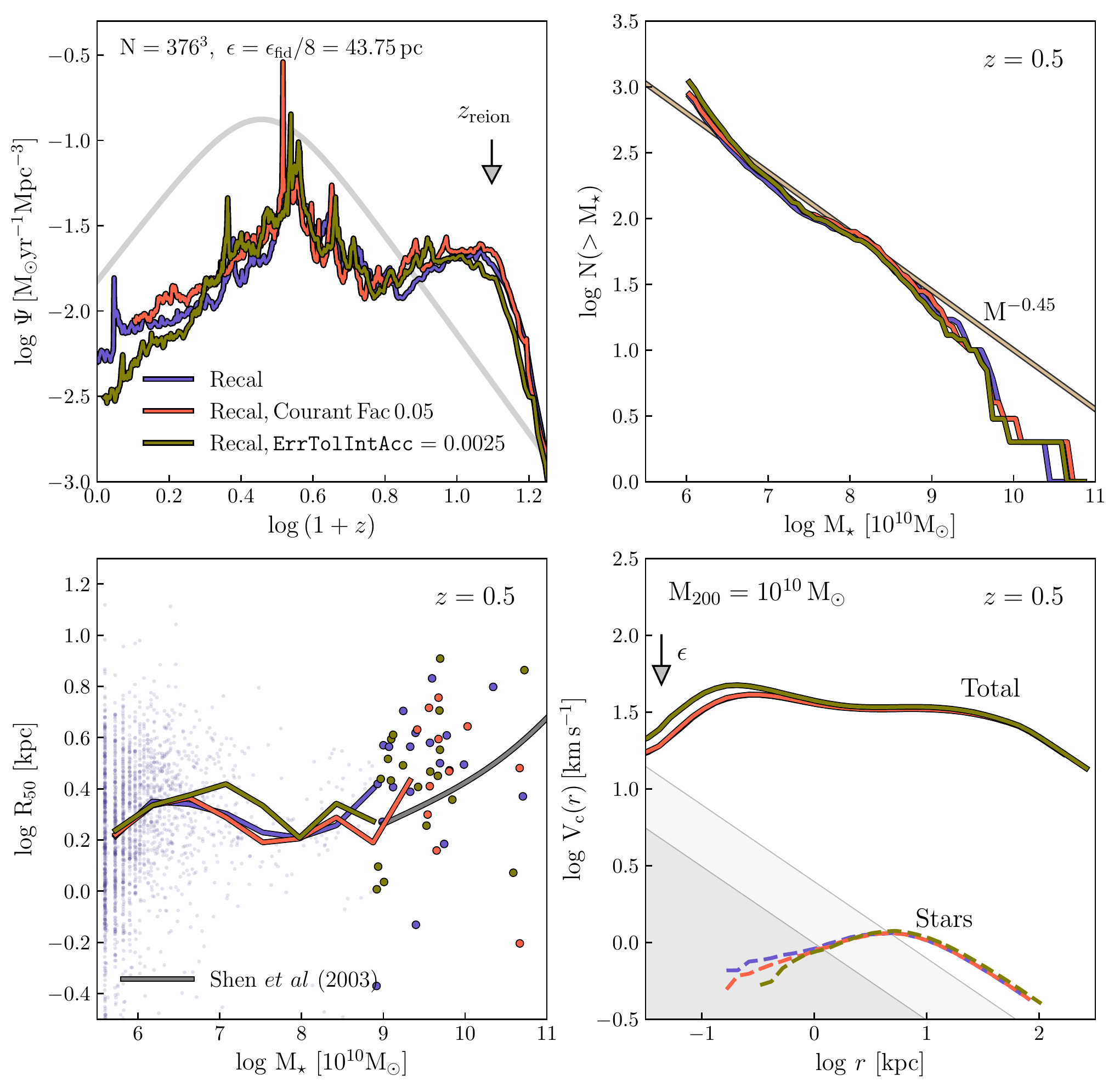}
  \caption{From top- to bottom-left, clockwise, different panels show the cosmic SFH,
    the cumulative galaxy stellar mass function, the stellar and total circular velocity profiles (for
    haloes in a narrow mass bin $-0.2\leq \log {\rm M}_{200}/[10^{10}{\rm M\odot}]\leq 0.2$), and the
    projected galaxy size-mass relation (note that the latter three panels present results at $z=0.5$).
    All runs adopted subgrid parameters of the Recal model, and used a softening
    length of $\epsilon=\epsilon_{\rm fid}/8=43.75\,{\rm pc}$ (note that this is the maximum
    physical softening length used for $z\leq 2.8$). Different colour lines correspond to
    runs that vary our default choice of \textsc{gadget}'s numerical integration parameters: the blue
    curves used a Courant factor of 0.15 and gravitational timestep parameter
    ${\tt ErrTolIntAcc}=0.025$ (our default values); the two remaining runs vary one of these parameters
    while keeping the other fixed: the orange curves used a Courant factor of 0.05; the green
    used ${\tt ErrTolIntAcc}=0.0025$.}
  \label{A2}
\end{figure*}

\subsection{Timestepping and integration accuracy}
\label{sA2}

In Paper I \citep[see also][]{Power2003,Hopkins2018} we showed that the central structure of
dark matter haloes in N-body simulations is prone to numerical artefact unless sufficient
numbers of timesteps are taken. This is particularly true when softening lengths are small,
as inter-particle accelerations scale as $a_\epsilon\propto m_{\rm DM}/\epsilon^2$ -- large
accelerations require short timesteps to ensure particle orbits are integrated accurately.
It is therefore plausible that analogous numerical effects may arise in hydrodynamical 
simulations which may impact SFHs, galaxy masses or sizes, or the structure of their DM haloes
in non-trivial ways. Indeed, Figures~\ref{VmaxM200} to \ref{VcStacked} indicate that the
central structure of DM haloes is particularly sensitivity to gravitational
softening, especially when small values are used. Figure~\ref{fig2} suggests that cosmic star formation
histories are also sensitive to $\epsilon$, when sufficiently small.

The upper-left panel of Figure~\ref{A2} verifies that cosmic SFHs are unaffected by using a smaller
integration timestep. The three curves show results for Recal models that vary the timestepping integration
parameters. All runs used $N^3_{\rm p}=376^3$ particles of gas and DM and $\epsilon_0=43.75\,{\rm pc}$
(note that we adopt a small softening length for these tests in order to to evaluate the robustness of
runs that are poorly converged with respect to our fiducial model and that are most likely
to suffer from inaccurate time integration due to large particle-on-particle accelerations).
The solid blue lines correspond to our default
choices: {\tt ErrTolIntAcc}=0.025 and a Courant Factor of 0.15; green and orange lines show,
respectively, the impact of reducing {\tt ErrTolIntAcc} by a factor of 10 (to 0.0025) or the Courant factor
by 3 (to 0.05) while keeping the default values for the other parameter (Figure~\ref{fig1} showed that the
baryon fractions of haloes in our adiabatic $N^3_{\rm p}=188^3$ runs are robust to equal changes in integration
accuracy). All three runs exhibit similar SFHs,
galaxy stellar mass functions (top-right) and size-mass relations (lower-left), implying that the statistical properties of
{\em galaxies} in our suite of convergence test-runs are robust to changes in integration
timestep. More importantly, the same is true of the innermost structure of their DM haloes
(lower-right panel). Although there are small differences in the central structure of haloes
between the runs with {\tt ErrTolIntAcc}=0.025 and 0.0025 (the latter being {\em more} centrally
concentrated than the former), they clearly cannot explain the softening dependence of halo
structure depicted in Figure~\ref{VcStacked}.

\begin{figure*}
  \includegraphics[width=0.8\textwidth]{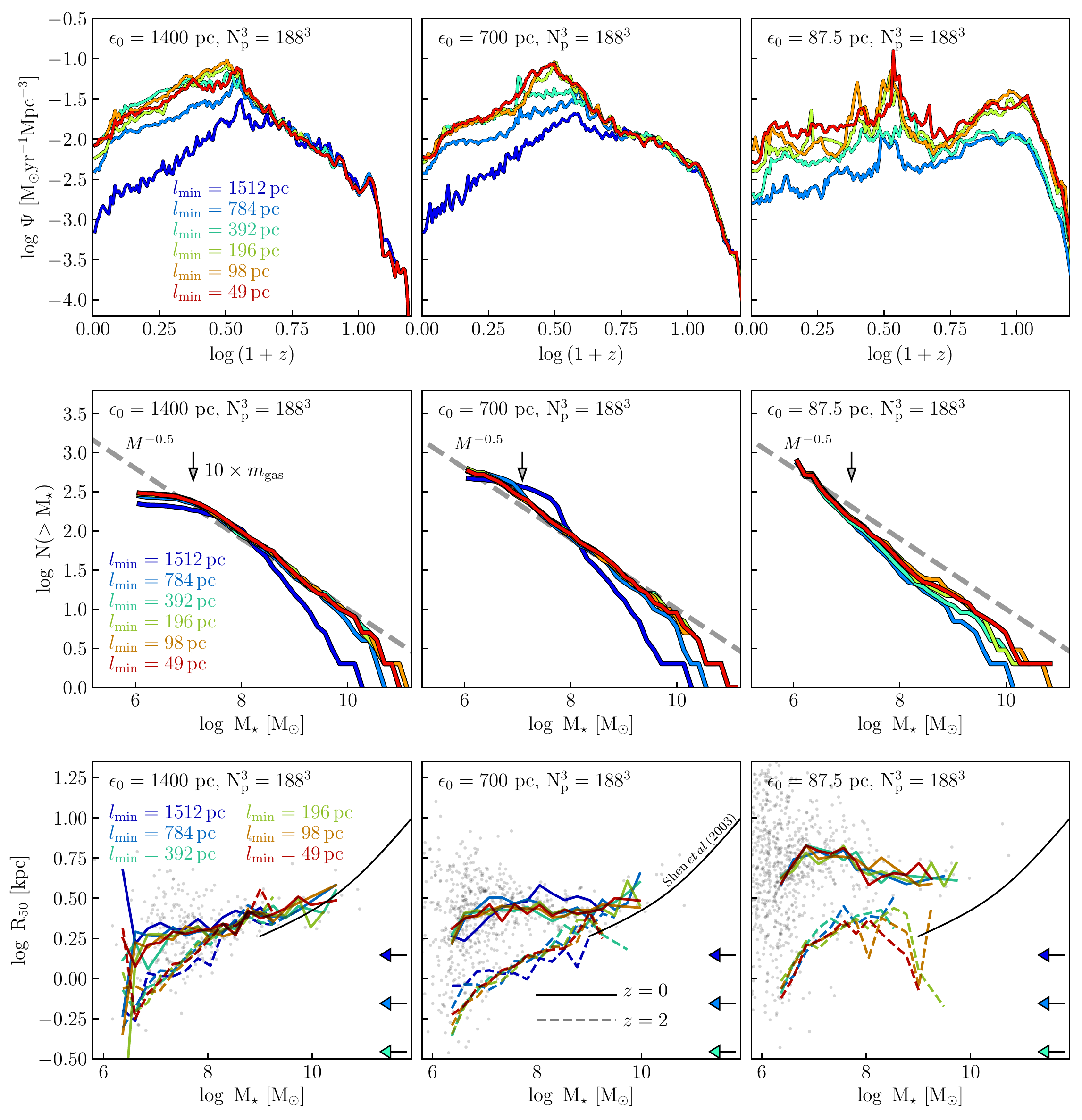}
  \caption{The impact of minimum SPH smoothing length on the cosmic star formation history
    (SFH; top row), the cumulative galaxy stellar mass function (GSMF; middle row) and the galaxy size-mass
    relation (bottom row). Different columns correspond to different gravitational softening lengths: the
    left-hand panels used $\epsilon_0=1400\,{\rm pc}$, the middle panels our fiducial softening
    length, $\epsilon_0=700\,{\rm pc}$, and the right-hand panels $\epsilon_0=87.5\,{\rm pc}$.
    All runs were carried out with $N^3_{\rm p}=188^3$ baryonic and DM particles.
    Different coloured lines are used for different minimum smoothing lengths, with values
    quoted in the legend. Note that all runs adopt softening and smoothing lengths that are
    fixed in co-moving coordinates at $z>2.8$, and fixed in physical coordinates thereafter.
    The values of $\epsilon_0$ and $l_{\rm min}$ quoted above refer to the {\em present-day}
    values. For each value of $\epsilon_0$, cosmic SFHs and GSMFs are largely independent of
    $l_{\rm min}$ provided it remains $\simlt 140\,{\rm pc}$ (with only minor deviations
    noticeable for $l_{\rm min} \simlt 280\,{\rm pc}$), which is roughly a factor of 2
    larger than our fiducial value of $l_{\rm min}=70\,{\rm pc}$ for this mass resolution.
    Galaxy sizes are independent of $l_{\rm min}$ for all values tests, but exhibit a strong
    softening-dependence, as noted in Figure~\ref{fig10}.}
  \label{A3}
\end{figure*}

\subsection{The minimum SPH smoothing length}
\label{sA3}

As discussed in Section~\ref{SS_EfficientFeedback}, all runs presented in the main body of this
paper used a minimum SPH smoothing length that co-evolves with the {\em spline} gravitational softening
length: $l_{\rm hsml}^{\rm min}(z)=0.1\times \epsilon_{\rm sp}(z)$. The most appropriate choice for $l_{\rm hsml}^{\rm min}$,
however, is not obvious: it dictates the minimum length-scale on which hydrodynamic forces can
be considered resolved, whereas $\epsilon$ governs the minimum resolved scale of gravitational
forces. The former is more closely related to the maximum resolved gas density (see eq.~\ref{eq:nmax1})
and therefore to the fraction of fluid elements that satisfy our star formation criteria, or
suffer from inefficient feedback due to numerical over-cooling. Indeed, eq.~\ref{eq:minnHeps} implies
that in order to achieve the highest possible hydrodynamic spatial resolution while simultaneously
ensuring efficient feedback, our runs should target minimum smoothing lengths of order 
$l_{\rm min}^{\rm eFB}\approx 390\,{\rm pc}$ for the intermediate-resolution, and
$\approx 140\,{\rm pc}$ for high-resolution (i.e. $0.28\times \epsilon_{\rm eFB}$, where $\epsilon_{\rm eFB}$
is the the Plummer-equivalent length of eq.~\ref{eq:minnHeps}). 

Figure~\ref{A3} investigates the impact of varying $l_{\rm hsml}^{\rm min}$ on the outcome of our simulations, focusing
on the cosmic SFH (top row), the GSMF (middle) and the median size-mass relation (bottom). Different columns
correspond to different (Plummer-equivalent) softening lengths: $\epsilon_0=1400\,{\rm pc}$ (left), 700 pc (middle) and
87.5 pc (right). Different colour lines are used for different minimum smoothing lengths. Note that
both $\epsilon$ and $l_{\rm hsml}^{\rm min}$ were fixed in physical units for $z\leq 2.8$, and in comoving units
at higher redshift; the values quoted in Figure~\ref{A3} are those at $z=0$. 

The SFHs and GSMFs are largely independent of $l_{\rm hsml}^{\rm min}$ provided is 
$\simlt 392\,{\rm pc}$, but exhibit a softening dependence already apparent in Figures~\ref{fig2}
and \ref{fig7}. Notably, the GSMF is slightly suppressed at mass scales $\simlt 10\times m_{\rm gas}$
(downward pointing arrows) when $\epsilon_0=1400\,{\rm pc}$. As discussed in Section~\ref{DMconc}, this is
a result of over-softening gravitational forces in the central regions of low-mass haloes, which
inhibits centrally-concentrated SF. When $\epsilon_0=87.5\,{\rm pc}$, the SFH develops a strong peak at
early times regardless of $l_{\rm hsml}^{\rm min}$, although it is suppressed somewhat for
$l_{\rm hsml}^{\rm min}\simgt 392\,{\rm pc}$.

Galaxy sizes are remarkably robust to changes in $l_{\rm hsml}^{\rm min}$, but exhibit a strong sensitivity
to gravitational softening, as already noted in Figure~\ref{fig10}. Galaxy projected half-mass
radii grow systematically as $\epsilon$ decreases, a result we have attributed to enhanced mass
segregation driven by the slow diffusion of energy from the more massive DM particles to the lower-mass
stellar particles \citep{Ludlow2019b}.

\begin{center}
  \begin{table*}
    \caption{As Table~\ref{TabSimParam}, but summarizing runs used in Figures~\ref{A1} to \ref{A3}. The additional
      parameters, not in Table~\ref{TabSimParam}, are the physical and comoving minimum smoothing lengths ($l_{\rm min}^{\rm phys}$
      and $l_{\rm min}^{\rm CM}$, respectively) used for Figure~\ref{A3}.}
    \begin{tabular}{c c c c c c c c c c c c c}\hline \hline
      & Model & $N_{\rm p}$ &  $m_{\rm DM}$          &  $m_{\rm g}$           &$\epsilon_{\rm phys}$  & $\epsilon_{\rm CM}$ &$l_{\rm min}^{\rm phys}$  & $l_{\rm min}^{\rm CM}$ & $z_{\rm phys}$ & \tt{ErrTolIntAcc} & Courant \\\vspace{0.15cm}
      &               &            &[$10^5\, {\rm M}_\odot$]&[$10^5\, {\rm M}_\odot$]& $[{\rm pc}]$          & $[{\rm pc}]$        & $[{\rm pc}]$          & $[{\rm pc}]$ &                &                   & Factor  \\

      \hline
      & Recal  & $376$ & 12.1 & 2.26 &   43.75  & 166.25  & 12.3 & 46.5 & 2.8 & 0.025  & 0.15  \\
      & Recal  & $376$ & 12.1 & 2.26 &   43.75  & 166.25  & 12.3 & 46.5 & 2.8 & 0.025  & 0.05  \\\vspace{0.25cm}
      & Recal  & $376$ & 12.1 & 2.26 &   43.75  & 166.25  & 12.3 & 46.5 & 2.8 & 0.0025 & 0.15 \\
      
      & Ref    & $188$ & 97.0 & 18.1 &   43.75  & 166.25  & 12.3 & 46.5 & 2.8 & 0.025  & 0.15  \\
      & Recal  & $188$ & 97.0 & 18.1 &   43.75  & 166.25  & 12.3 & 46.5 & 2.8 & 0.025  & 0.15  \\\vspace{0.25cm}
      & No AGN & $188$ & 97.0 & 18.1 &   43.75  & 166.25  & 12.3 & 46.5 & 2.8 & 0.025  & 0.15  \\

      & Ref    & $188$ & 97.0 & 18.1 &  1400.0  & 5320.0  & 1512.0 & 3433.0 & 1.3 & 0.025  & 0.15  \\
      & Ref    & $188$ & 97.0 & 18.1 &  1400.0  & 5320.0  & 784.0  & 1781.0 & 1.3 & 0.025  & 0.15  \\
      & Ref    & $188$ & 97.0 & 18.1 &  1400.0  & 5320.0  & 392.0  & 890.0  & 1.3 & 0.025  & 0.15  \\
      & Ref    & $188$ & 97.0 & 18.1 &  1400.0  & 5320.0  & 196.0  & 445.0  & 1.3 & 0.025  & 0.15  \\
      & Ref    & $188$ & 97.0 & 18.1 &  1400.0  & 5320.0  & 98.0   & 223.0  & 1.3 & 0.025  & 0.15  \\\vspace{0.1cm}
      & Ref    & $188$ & 97.0 & 18.1 &  1400.0  & 5320.0  & 49.0   & 111.0  & 1.3 & 0.025  & 0.15  \\

      & Ref    & $188$ & 97.0 & 18.1 &   700.0  & 2660.0  & 1512.0 & 3433.0  & 2.8 & 0.025  & 0.15  \\
      & Ref    & $188$ & 97.0 & 18.1 &   700.0  & 2660.0  & 784.0  & 1781.0  & 2.8 & 0.025  & 0.15  \\
      & Ref    & $188$ & 97.0 & 18.1 &   700.0  & 2660.0  & 392.0  & 890.0   & 2.8 & 0.025  & 0.15  \\
      & Ref    & $188$ & 97.0 & 18.1 &   700.0  & 2660.0  & 196.0  & 445.0   & 2.8 & 0.025  & 0.15  \\
      & Ref    & $188$ & 97.0 & 18.1 &   700.0  & 2660.0  & 98.0   & 223.0   & 2.8 & 0.025  & 0.15  \\\vspace{0.1cm}
      & Ref    & $188$ & 97.0 & 18.1 &   700.0  & 2660.0  & 49.0   & 111.0   & 2.8 & 0.025  & 0.15  \\

      & Ref    & $188$ & 97.0 & 18.1 &   87.5   & 332.5   & 784.0 & 1781.0  & 2.8 & 0.025  & 0.15  \\
      & Ref    & $188$ & 97.0 & 18.1 &   87.5   & 332.5   & 392.0 & 890.0   & 2.8 & 0.025  & 0.15  \\
      & Ref    & $188$ & 97.0 & 18.1 &   87.5   & 332.5   & 196.0 & 445.0   & 2.8 & 0.025  & 0.15  \\
      & Ref    & $188$ & 97.0 & 18.1 &   87.5   & 332.5   & 98.0  & 223.0   & 2.8 & 0.025  & 0.15  \\\vspace{0.1cm}
      & Ref    & $188$ & 97.0 & 18.1 &   87.5   & 332.5   & 49.0  & 111.0   & 2.8 & 0.025  & 0.15  \\\hline

\end{tabular}
    \label{TabSimParamApp}
  \end{table*}
\end{center}

\bibliographystyle{mn2e}
\bibliography{paper}

\begin{thebibliography}{}

\bibitem[\protect\citeauthoryear{{Agertz}, {Kravtsov}, {Leitner} \&
  {Gnedin}}{{Agertz} et~al.}{2013}]{Agertz2013}
{Agertz} O.,  {Kravtsov} A.~V.,  {Leitner} S.~N.,    {Gnedin} N.~Y.,  2013,
  \apj, 770, 25

\bibitem[\protect\citeauthoryear{{Agertz, et al}}{{Agertz, et
  al}}{2007}]{Agertz2007}
{Agertz, et al} 2007, \mnras, 380, 963

\bibitem[\protect\citeauthoryear{{Bah{\'e}}, {Barnes}, {Dalla Vecchia}, {Kay},
  {White}, {McCarthy}, {Schaye}, {Bower}, {Crain}, {Theuns}, {Jenkins},
  {McGee}, {Schaller}, {Thomas} \& {Trayford}}{{Bah{\'e}}
  et~al.}{2017}]{Bahe2017}
{Bah{\'e}} Y.~M.,  {Barnes} D.~J.,  {Dalla Vecchia} C.,  {Kay} S.~T.,  {White}
  S. D.~M.,  {McCarthy} I.~G.,  {Schaye} J.,  {Bower} R.~G.,  {Crain} R.~A.,
  {Theuns} T.,  {Jenkins} A.,  {McGee} S.~L.,  {Schaller} M.,  {Thomas} P.~A.,
    {Trayford} J.~W.,  2017, \mnras, 470, 4186

\bibitem[\protect\citeauthoryear{{Ben{\'\i}tez-Llambay}, {Frenk}, {Ludlow} \&
  {Navarro}}{{Ben{\'\i}tez-Llambay} et~al.}{2019}]{Benitez-Llambay2018}
{Ben{\'\i}tez-Llambay} A.,  {Frenk} C.~S.,  {Ludlow} A.~D.,    {Navarro} J.~F.,
   2019, \mnras, 488, 2387

\bibitem[\protect\citeauthoryear{{Binney} \& {Knebe}}{{Binney} \&
  {Knebe}}{2002}]{BinneyKnebe2002}
{Binney} J.,  {Knebe} A.,  2002, \mnras, 333, 378

\bibitem[\protect\citeauthoryear{{Binney} \& {Tremaine}}{{Binney} \&
  {Tremaine}}{1987}]{BinneyTremaine87}
{Binney} J.,  {Tremaine} S.,  1987, {Galactic dynamics}.
Princeton, NJ, Princeton University Press, 1987, 747 p.

\bibitem[\protect\citeauthoryear{{Blumenthal}, {Faber}, {Flores} \&
  {Primack}}{{Blumenthal} et~al.}{1986}]{Blumenthal1986}
{Blumenthal} G.~R.,  {Faber} S.~M.,  {Flores} R.,    {Primack} J.~R.,  1986,
  \apj, 301, 27

\bibitem[\protect\citeauthoryear{{Booth} \& {Schaye}}{{Booth} \&
  {Schaye}}{2009}]{BoothSchaye2009}
{Booth} C.~M.,  {Schaye} J.,  2009, \mnras, 398, 53

\bibitem[\protect\citeauthoryear{{Brook}, {Stinson}, {Gibson}, {Ro{\v s}kar},
  {Wadsley} \& {Quinn}}{{Brook} et~al.}{2012}]{Brook2012b}
{Brook} C.~B.,  {Stinson} G.,  {Gibson} B.~K.,  {Ro{\v s}kar} R.,  {Wadsley}
  J.,    {Quinn} T.,  2012, \mnras, 419, 771

\bibitem[\protect\citeauthoryear{{Brook}, {Stinson}, {Gibson}, {Wadsley} \&
  {Quinn}}{{Brook} et~al.}{2012}]{Brook2012a}
{Brook} C.~B.,  {Stinson} G.,  {Gibson} B.~K.,  {Wadsley} J.,    {Quinn} T.,
  2012, \mnras, 424, 1275

\bibitem[\protect\citeauthoryear{{Bryan}, {Kay}, {Duffy}, {Schaye}, {Dalla
  Vecchia} \& {Booth}}{{Bryan} et~al.}{2013}]{Bryan2013}
{Bryan} S.~E.,  {Kay} S.~T.,  {Duffy} A.~R.,  {Schaye} J.,  {Dalla Vecchia} C.,
     {Booth} C.~M.,  2013, \mnras, 429, 3316

\bibitem[\protect\citeauthoryear{{Bryan, G. et al}}{{Bryan, G. et
  al}}{2014}]{ENZO}
{Bryan, G. et al} 2014, \apjs, 211, 19

\bibitem[\protect\citeauthoryear{{Buck}, {Macci{\`o}}, {Dutton}, {Obreja} \&
  {Frings}}{{Buck} et~al.}{2019}]{Buck2019}
{Buck} T.,  {Macci{\`o}} A.~V.,  {Dutton} A.~A.,  {Obreja} A.,    {Frings} J.,
  2019, \mnras, 483, 1314

\bibitem[\protect\citeauthoryear{{Chabrier}}{{Chabrier}}{2003}]{Chabrier2003b}
{Chabrier} G.,  2003, \pasp, 115, 763

\bibitem[\protect\citeauthoryear{{Crain}, {Schaye}, {Bower}, {Furlong},
  {Schaller}, {Theuns}, {Dalla Vecchia}, {Frenk}, {McCarthy}, {Helly},
  {Jenkins}, {Rosas-Guevara}, {White} \& {Trayford}}{{Crain}
  et~al.}{2015}]{Crain2015}
{Crain} R.~A.,  {Schaye} J.,  {Bower} R.~G.,  {Furlong} M.,  {Schaller} M.,
  {Theuns} T.,  {Dalla Vecchia} C.,  {Frenk} C.~S.,  {McCarthy} I.~G.,  {Helly}
  J.~C.,  {Jenkins} A.,  {Rosas-Guevara} Y.~M.,  {White} S.~D.~M.,
  {Trayford} J.~W.,  2015, \mnras, 450, 1937

\bibitem[\protect\citeauthoryear{{Cui et al}}{{Cui et al}}{2018}]{Cui2018}
{Cui et al} 2018, \mnras, 480, 2898

\bibitem[\protect\citeauthoryear{{Dalla Vecchia} \& {Schaye}}{{Dalla Vecchia}
  \& {Schaye}}{2012}]{DallaVecchia2012}
{Dalla Vecchia} C.,  {Schaye} J.,  2012, \mnras, 426, 140

\bibitem[\protect\citeauthoryear{{Dav{\'e}}, {Angl{\'e}s-Alc{\'a}zar},
  {Narayanan}, {Li}, {Rafieferantsoa} \& {Appleby}}{{Dav{\'e}}
  et~al.}{2019}]{Dave2019}
{Dav{\'e}} R.,  {Angl{\'e}s-Alc{\'a}zar} D.,  {Narayanan} D.,  {Li} Q.,
  {Rafieferantsoa} M.~H.,    {Appleby} S.,  2019, \mnras, 486, 2827

\bibitem[\protect\citeauthoryear{{Dolag}, {Borgani}, {Murante} \&
  {Springel}}{{Dolag} et~al.}{2009}]{Dolag2009}
{Dolag} K.,  {Borgani} S.,  {Murante} G.,    {Springel} V.,  2009, \mnras, 399,
  497

\bibitem[\protect\citeauthoryear{{Dolag}, {Komatsu} \& {Sunyaev}}{{Dolag}
  et~al.}{2016}]{Dolag2016}
{Dolag} K.,  {Komatsu} E.,    {Sunyaev} R.,  2016, \mnras, 463, 1797

\bibitem[\protect\citeauthoryear{{Duffy}, {Schaye}, {Kay}, {Dalla Vecchia},
  {Battye} \& {Booth}}{{Duffy} et~al.}{2010}]{Duffy2010}
{Duffy} A.~R.,  {Schaye} J.,  {Kay} S.~T.,  {Dalla Vecchia} C.,  {Battye}
  R.~A.,    {Booth} C.~M.,  2010, \mnras, 405, 2161

\bibitem[\protect\citeauthoryear{{Dutton}, {Macci{\`o}}, {Buck}, {Dixon},
  {Blank} \& {Obreja}}{{Dutton} et~al.}{2019}]{Dutton2019}
{Dutton} A.~A.,  {Macci{\`o}} A.~V.,  {Buck} T.,  {Dixon} K.~L.,  {Blank} M.,
   {Obreja} A.,  2019, \mnras, 486, 655

\bibitem[\protect\citeauthoryear{{Efstathiou}, {Frenk}, {White} \&
  {Davis}}{{Efstathiou} et~al.}{1988}]{Efstathiou1988}
{Efstathiou} G.,  {Frenk} C.~S.,  {White} S.~D.~M.,    {Davis} M.,  1988,
  \mnras, 235, 715

\bibitem[\protect\citeauthoryear{{El-Zant}, {Everitt} \& {Kassem}}{{El-Zant}
  et~al.}{2019}]{ElZant2018}
{El-Zant} A.~A.,  {Everitt} M.~J.,    {Kassem} S.~M.,  2019, \mnras, 484, 1456

\bibitem[\protect\citeauthoryear{{Ferland}, {Korista}, {Verner}, {Ferguson},
  {Kingdon} \& {Verner}}{{Ferland} et~al.}{1998}]{Ferland1998}
{Ferland} G.~J.,  {Korista} K.~T.,  {Verner} D.~A.,  {Ferguson} J.~W.,
  {Kingdon} J.~B.,    {Verner} E.~M.,  1998, \pasp, 110, 761

\bibitem[\protect\citeauthoryear{{Gaburov} \& {Nitadori}}{{Gaburov} \&
  {Nitadori}}{2011}]{Gaburov2011}
{Gaburov} E.,  {Nitadori} K.,  2011, \mnras, 414, 129

\bibitem[\protect\citeauthoryear{{Genel}, {Bryan}, {Springel}, {Hernquist},
  {Nelson}, {Pillepich}, {Weinberger}, {Pakmor}, {Marinacci} \&
  {Vogelsberger}}{{Genel} et~al.}{2019}]{Genel2018}
{Genel} S.,  {Bryan} G.~L.,  {Springel} V.,  {Hernquist} L.,  {Nelson} D.,
  {Pillepich} A.,  {Weinberger} R.,  {Pakmor} R.,  {Marinacci} F.,
  {Vogelsberger} M.,  2019, \apj, 871, 21

\bibitem[\protect\citeauthoryear{{Gingold} \& {Monaghan}}{{Gingold} \&
  {Monaghan}}{1977}]{Gingold1977}
{Gingold} R.~A.,  {Monaghan} J.~J.,  1977, \mnras, 181, 375

\bibitem[\protect\citeauthoryear{{Governato}, {Brook}, {Mayer}, {Brooks},
  {Rhee}, {Wadsley}, {Jonsson}, {Willman}, {Stinson}, {Quinn} \&
  {Madau}}{{Governato} et~al.}{2010}]{Governato2010}
{Governato} F.,  {Brook} C.,  {Mayer} L.,  {Brooks} A.,  {Rhee} G.,  {Wadsley}
  J.,  {Jonsson} P.,  {Willman} B.,  {Stinson} G.,  {Quinn} T.,    {Madau} P.,
  2010, \nat, 463, 203

\bibitem[\protect\citeauthoryear{{Haardt} \& {Madau}}{{Haardt} \&
  {Madau}}{2001}]{Haardt_and_Madau_01}
{Haardt} F.,  {Madau} P.,  2001, in {Neumann} D.~M.,  {Tran} J.~T.~V.,  eds,
  Clusters of Galaxies and the High Redshift Universe Observed in X-rays
  {Modelling the UV/X-ray cosmic background with CUBA}.
p.~64

\bibitem[\protect\citeauthoryear{{Haas}, {Schaye}, {Booth}, {Dalla Vecchia},
  {Springel}, {Theuns} \& {Wiersma}}{{Haas} et~al.}{2013a}]{Haas2013a}
{Haas} M.~R.,  {Schaye} J.,  {Booth} C.~M.,  {Dalla Vecchia} C.,  {Springel}
  V.,  {Theuns} T.,    {Wiersma} R. P.~C.,  2013a, \mnras, 435, 2931

\bibitem[\protect\citeauthoryear{{Haas}, {Schaye}, {Booth}, {Dalla Vecchia},
  {Springel}, {Theuns} \& {Wiersma}}{{Haas} et~al.}{2013b}]{Haas2013b}
{Haas} M.~R.,  {Schaye} J.,  {Booth} C.~M.,  {Dalla Vecchia} C.,  {Springel}
  V.,  {Theuns} T.,    {Wiersma} R. P.~C.,  2013b, \mnras, 435, 2955

\bibitem[\protect\citeauthoryear{{Hopkins}}{{Hopkins}}{2015}]{Hopkins2015}
{Hopkins} P.~F.,  2015, \mnras, 450, 53

\bibitem[\protect\citeauthoryear{{Hopkins et al}}{{Hopkins et
  al}}{2018}]{Hopkins2018}
{Hopkins et al} 2018, \mnras, 480, 800

\bibitem[\protect\citeauthoryear{{Hubber}, {Falle} \& {Goodwin}}{{Hubber}
  et~al.}{2013}]{Hubber2013}
{Hubber} D.~A.,  {Falle} S.~A.~E.~G.,    {Goodwin} S.~P.,  2013, \mnras, 432,
  711

\bibitem[\protect\citeauthoryear{Hunter}{Hunter}{2007}]{matplotlib}
Hunter J.~D.,  2007, Computing In Science \& Engineering, 9, 90

\bibitem[\protect\citeauthoryear{{Jenkins}, {Frenk}, {White}, {Colberg},
  {Cole}, {Evrard}, {Couchman} \& {Yoshida}}{{Jenkins}
  et~al.}{2001}]{Jenkins2001}
{Jenkins} A.,  {Frenk} C.~S.,  {White} S.~D.~M.,  {Colberg} J.~M.,  {Cole} S.,
  {Evrard} A.~E.,  {Couchman} H.~M.~P.,    {Yoshida} N.,  2001, \mnras, 321,
  372

\bibitem[\protect\citeauthoryear{{Jiang}, {Helly}, {Cole} \& {Frenk}}{{Jiang}
  et~al.}{2014}]{Jiang2014}
{Jiang} L.,  {Helly} J.~C.,  {Cole} S.,    {Frenk} C.~S.,  2014, \mnras, 440,
  2115

\bibitem[\protect\citeauthoryear{Jones, Oliphant, Peterson et~al.,}{Jones
  et~al.}{2001}]{scipy}
Jones E.,  Oliphant T.,  Peterson P.,    et~al.,, 2001, {SciPy}: Open source
  scientific tools for {Python}

\bibitem[\protect\citeauthoryear{{Katz}, {Weinberg} \& {Hernquist}}{{Katz}
  et~al.}{1996}]{Katz1996}
{Katz} N.,  {Weinberg} D.~H.,    {Hernquist} L.,  1996, \apjs, 105, 19

\bibitem[\protect\citeauthoryear{{Keller}, {Wadsley}, {Wang} \&
  {Kruijssen}}{{Keller} et~al.}{2019}]{Keller2018}
{Keller} B.~W.,  {Wadsley} J.~W.,  {Wang} L.,    {Kruijssen} J.~M.~D.,  2019,
  \mnras, 482, 2244

\bibitem[\protect\citeauthoryear{{Kennicutt}
  Jr.}{{Kennicutt}}{1989}]{Kennicutt1989}
{Kennicutt} Jr. R.~C.,  1989, \apj, 344, 685

\bibitem[\protect\citeauthoryear{{Knebe}, {Green} \& {Binney}}{{Knebe}
  et~al.}{2001}]{Knebe2001}
{Knebe} A.,  {Green} A.,    {Binney} J.,  2001, \mnras, 325, 845

\bibitem[\protect\citeauthoryear{{Kravtsov}}{{Kravtsov}}{1999}]{Kravtsov1999}
{Kravtsov} A.~V.,  1999, PhD thesis, NEW MEXICO STATE UNIVERSITY

\bibitem[\protect\citeauthoryear{{Libeskind}, {Yepes}, {Knebe},
  {Gottl{\"o}ber}, {Hoffman} \& {Knollmann}}{{Libeskind}
  et~al.}{2010}]{Libeskind2010}
{Libeskind} N.~I.,  {Yepes} G.,  {Knebe} A.,  {Gottl{\"o}ber} S.,  {Hoffman}
  Y.,    {Knollmann} S.~R.,  2010, \mnras, 401, 1889

\bibitem[\protect\citeauthoryear{{Ludlow}, {Bose}, {Angulo}, {Wang},
  {Hellwing}, {Navarro}, {Cole} \& {Frenk}}{{Ludlow} et~al.}{2016}]{Ludlow2016}
{Ludlow} A.~D.,  {Bose} S.,  {Angulo} R.~E.,  {Wang} L.,  {Hellwing} W.~A.,
  {Navarro} J.~F.,  {Cole} S.,    {Frenk} C.~S.,  2016, \mnras, 460, 1214

\bibitem[\protect\citeauthoryear{{Ludlow}, {Navarro}, {Angulo},
  {Boylan-Kolchin}, {Springel}, {Frenk} \& {White}}{{Ludlow}
  et~al.}{2014}]{Ludlow2014}
{Ludlow} A.~D.,  {Navarro} J.~F.,  {Angulo} R.~E.,  {Boylan-Kolchin} M.,
  {Springel} V.,  {Frenk} C.,    {White} S.~D.~M.,  2014, \mnras, 441, 378

\bibitem[\protect\citeauthoryear{{Ludlow}, {Schaye} \& {Bower}}{{Ludlow}
  et~al.}{2019}]{Ludlow2019a}
{Ludlow} A.~D.,  {Schaye} J.,    {Bower} R.,  2019, \mnras, 488, 3663

\bibitem[\protect\citeauthoryear{{Ludlow}, {Schaye}, {Schaller} \&
  {Richings}}{{Ludlow} et~al.}{2019}]{Ludlow2019b}
{Ludlow} A.~D.,  {Schaye} J.,  {Schaller} M.,    {Richings} J.,  2019, \mnras,
  488, L123

\bibitem[\protect\citeauthoryear{{Macci{\`o}}, {Stinson}, {Brook}, {Wadsley},
  {Couchman}, {Shen}, {Gibson} \& {Quinn}}{{Macci{\`o}}
  et~al.}{2012}]{Maccio2012}
{Macci{\`o}} A.~V.,  {Stinson} G.,  {Brook} C.~B.,  {Wadsley} J.,  {Couchman}
  H.~M.~P.,  {Shen} S.,  {Gibson} B.~K.,    {Quinn} T.,  2012, \apjl, 744, L9

\bibitem[\protect\citeauthoryear{{Madau} \& {Dickinson}}{{Madau} \&
  {Dickinson}}{2014}]{MadauDickinsons2014}
{Madau} P.,  {Dickinson} M.,  2014, \araa, 52, 415

\bibitem[\protect\citeauthoryear{{Monaghan}}{{Monaghan}}{1992}]{Monaghan1992}
{Monaghan} J.~J.,  1992, \araa, 30, 543

\bibitem[\protect\citeauthoryear{{Navarro}, {Eke} \& {Frenk}}{{Navarro}
  et~al.}{1996}]{NEF1996}
{Navarro} J.~F.,  {Eke} V.~R.,    {Frenk} C.~S.,  1996, \mnras, 283, L72

\bibitem[\protect\citeauthoryear{{Navarro}, {Ludlow}, {Springel}, {Wang},
  {Vogelsberger}, {White}, {Jenkins}, {Frenk} \& {Helmi}}{{Navarro}
  et~al.}{2010}]{Navarro2010}
{Navarro} J.~F.,  {Ludlow} A.,  {Springel} V.,  {Wang} J.,  {Vogelsberger} M.,
  {White} S.~D.~M.,  {Jenkins} A.,  {Frenk} C.~S.,    {Helmi} A.,  2010,
  \mnras, 402, 21

\bibitem[\protect\citeauthoryear{{Ogiya} \& {Mori}}{{Ogiya} \&
  {Mori}}{2014}]{Ogiya2014}
{Ogiya} G.,  {Mori} M.,  2014, \apj, 793, 46

\bibitem[\protect\citeauthoryear{{O'Shea}, {Nagamine}, {Springel}, {Hernquist}
  \& {Norman}}{{O'Shea} et~al.}{2005}]{OShea2005}
{O'Shea} B.~W.,  {Nagamine} K.,  {Springel} V.,  {Hernquist} L.,    {Norman}
  M.~L.,  2005, \apjs, 160, 1

\bibitem[\protect\citeauthoryear{P\'erez \& Granger}{P\'erez \&
  Granger}{2007}]{ipython}
P\'erez F.,  Granger B.~E.,  2007, Computing in Science and Engineering, 9, 21

\bibitem[\protect\citeauthoryear{{Pillepich}, {Nelson}, {Hernquist},
  {Springel}, {Pakmor}, {Torrey}, {Weinberger}, {Genel}, {Naiman}, {Marinacci}
  \& {Vogelsberger}}{{Pillepich} et~al.}{2018}]{Pillepich2018}
{Pillepich} A.,  {Nelson} D.,  {Hernquist} L.,  {Springel} V.,  {Pakmor} R.,
  {Torrey} P.,  {Weinberger} R.,  {Genel} S.,  {Naiman} J.~P.,  {Marinacci} F.,
     {Vogelsberger} M.,  2018, \mnras, 475, 648

\bibitem[\protect\citeauthoryear{{Planck Collaboration}, {Ade}, {Aghanim},
  {Alves}, {Armitage-Caplan}, {Arnaud}, {Ashdown}, {Atrio-Barandela}, {Aumont},
  {Aussel} \& et al.}{{Planck Collaboration} et~al.}{2014}]{Planck2014}
{Planck Collaboration} {Ade} P.~A.~R.,  {Aghanim} N.,  {Alves} M.~I.~R.,
  {Armitage-Caplan} C.,  {Arnaud} M.,  {Ashdown} M.,  {Atrio-Barandela} F.,
  {Aumont} J.,  {Aussel} H.,    et al. 2014, \aap, 571, A1

\bibitem[\protect\citeauthoryear{{Pontzen} \& {Governato}}{{Pontzen} \&
  {Governato}}{2012}]{Pontzen2012}
{Pontzen} A.,  {Governato} F.,  2012, \mnras, 421, 3464

\bibitem[\protect\citeauthoryear{{Power}, {Navarro}, {Jenkins}, {Frenk},
  {White}, {Springel}, {Stadel} \& {Quinn}}{{Power} et~al.}{2003}]{Power2003}
{Power} C.,  {Navarro} J.~F.,  {Jenkins} A.,  {Frenk} C.~S.,  {White} S.~D.~M.,
   {Springel} V.,  {Stadel} J.,    {Quinn} T.,  2003, \mnras, 338, 14

\bibitem[\protect\citeauthoryear{{Price} \& {Monaghan}}{{Price} \&
  {Monaghan}}{2007}]{Price2007}
{Price} D.~J.,  {Monaghan} J.~J.,  2007, \mnras, 374, 1347

\bibitem[\protect\citeauthoryear{{Rosas-Guevara}, {Bower}, {Schaye}, {Furlong},
  {Frenk}, {Booth}, {Crain}, {Dalla Vecchia}, {Schaller} \&
  {Theuns}}{{Rosas-Guevara} et~al.}{2015}]{Rosas-Guevara2015}
{Rosas-Guevara} Y.~M.,  {Bower} R.~G.,  {Schaye} J.,  {Furlong} M.,  {Frenk}
  C.~S.,  {Booth} C.~M.,  {Crain} R.~A.,  {Dalla Vecchia} C.,  {Schaller} M.,
   {Theuns} T.,  2015, \mnras, 454, 1038

\bibitem[\protect\citeauthoryear{{Sawala}, {Frenk}, {Fattahi}, {Navarro},
  {Bower}, {Crain}, {Dalla Vecchia}, {Furlong}, {Helly}, {Jenkins}, {Oman},
  {Schaller}, {Schaye}, {Theuns}, {Trayford} \& {White}}{{Sawala}
  et~al.}{2016}]{Sawala2016}
{Sawala} T.,  {Frenk} C.~S.,  {Fattahi} A.,  {Navarro} J.~F.,  {Bower} R.~G.,
  {Crain} R.~A.,  {Dalla Vecchia} C.,  {Furlong} M.,  {Helly} J.~C.,  {Jenkins}
  A.,  {Oman} K.~A.,  {Schaller} M.,  {Schaye} J.,  {Theuns} T.,  {Trayford}
  J.,    {White} S. D.~M.,  2016, \mnras, 457, 1931

\bibitem[\protect\citeauthoryear{{Scannapieco}, {Tissera}, {White} \&
  {Springel}}{{Scannapieco} et~al.}{2008}]{Scannapieco2008a}
{Scannapieco} C.,  {Tissera} P.~B.,  {White} S.~D.~M.,    {Springel} V.,  2008,
  \mnras, 389, 1137

\bibitem[\protect\citeauthoryear{{Scannapieco et al}}{{Scannapieco et
  al}}{2012}]{Scannapieco2012}
{Scannapieco et al} 2012, \mnras, 423, 1726

\bibitem[\protect\citeauthoryear{{Schaller}, {Dalla Vecchia}, {Schaye},
  {Bower}, {Theuns}, {Crain}, {Furlong} \& {McCarthy}}{{Schaller}
  et~al.}{2015}]{Schaller2015b}
{Schaller} M.,  {Dalla Vecchia} C.,  {Schaye} J.,  {Bower} R.~G.,  {Theuns} T.,
   {Crain} R.~A.,  {Furlong} M.,    {McCarthy} I.~G.,  2015, \mnras, 454, 2277

\bibitem[\protect\citeauthoryear{{Schaller}, {Frenk}, {Bower}, {Theuns},
  {Jenkins}, {Schaye}, {Crain}, {Furlong}, {Dalla Vecchia} \&
  {McCarthy}}{{Schaller} et~al.}{2015}]{Schaller2015a}
{Schaller} M.,  {Frenk} C.~S.,  {Bower} R.~G.,  {Theuns} T.,  {Jenkins} A.,
  {Schaye} J.,  {Crain} R.~A.,  {Furlong} M.,  {Dalla Vecchia} C.,
  {McCarthy} I.~G.,  2015, \mnras, 451, 1247

\bibitem[\protect\citeauthoryear{{Schaye}}{{Schaye}}{2004}]{Schaye2004}
{Schaye} J.,  2004, \apj, 609, 667

\bibitem[\protect\citeauthoryear{{Schaye} \& {Dalla Vecchia}}{{Schaye} \&
  {Dalla Vecchia}}{2008}]{Schaye2008}
{Schaye} J.,  {Dalla Vecchia} C.,  2008, \mnras, 383, 1210

\bibitem[\protect\citeauthoryear{{Schaye et al}}{{Schaye et
  al}}{2015}]{Schaye2015}
{Schaye et al} 2015, \mnras, 446, 521

\bibitem[\protect\citeauthoryear{{Sellwood} \& {Debattista}}{{Sellwood} \&
  {Debattista}}{2009}]{Sellwood2009}
{Sellwood} J.~A.,  {Debattista} V.~P.,  2009, \mnras, 398, 1279

\bibitem[\protect\citeauthoryear{{Sembolini et al}}{{Sembolini et
  al}}{2016}]{Sembolini2016}
{Sembolini et al} 2016, \mnras, 459, 2973

\bibitem[\protect\citeauthoryear{{Shen}, {Mo}, {White}, {Blanton}, {Kauffmann},
  {Voges}, {Brinkmann} \& {Csabai}}{{Shen} et~al.}{2003}]{Shen2003}
{Shen} S.,  {Mo} H.~J.,  {White} S.~D.~M.,  {Blanton} M.~R.,  {Kauffmann} G.,
  {Voges} W.,  {Brinkmann} J.,    {Csabai} I.,  2003, \mnras, 343, 978

\bibitem[\protect\citeauthoryear{{Somerville} \& {Dav{\'e}}}{{Somerville} \&
  {Dav{\'e}}}{2015}]{SomervilleDave2015}
{Somerville} R.~S.,  {Dav{\'e}} R.,  2015, \araa, 53, 51

\bibitem[\protect\citeauthoryear{{Springel}}{{Springel}}{2000}]{Springel2000}
{Springel} V.,  2000, \mnras, 312, 859

\bibitem[\protect\citeauthoryear{{Springel}}{{Springel}}{2005}]{Springel2005b}
{Springel} V.,  2005, \mnras, 364, 1105

\bibitem[\protect\citeauthoryear{{Springel}}{{Springel}}{2010}]{Springel2010}
{Springel} V.,  2010, \mnras, 401, 791

\bibitem[\protect\citeauthoryear{{Springel}, {Di Matteo} \&
  {Hernquist}}{{Springel} et~al.}{2005}]{Springel2005c}
{Springel} V.,  {Di Matteo} T.,    {Hernquist} L.,  2005, \mnras, 361, 776

\bibitem[\protect\citeauthoryear{{Springel} \& {Hernquist}}{{Springel} \&
  {Hernquist}}{2003}]{Springel2003}
{Springel} V.,  {Hernquist} L.,  2003, \mnras, 339, 289

\bibitem[\protect\citeauthoryear{{Springel}, {Wang}, {Vogelsberger}, {Ludlow},
  {Jenkins}, {Helmi}, {Navarro}, {Frenk} \& {White}}{{Springel}
  et~al.}{2008}]{Springel2008b}
{Springel} V.,  {Wang} J.,  {Vogelsberger} M.,  {Ludlow} A.,  {Jenkins} A.,
  {Helmi} A.,  {Navarro} J.~F.,  {Frenk} C.~S.,    {White} S.~D.~M.,  2008,
  \mnras, 391, 1685

\bibitem[\protect\citeauthoryear{{Springel}, {White}, {Tormen} \&
  {Kauffmann}}{{Springel} et~al.}{2001}]{Springel2001b}
{Springel} V.,  {White} S.~D.~M.,  {Tormen} G.,    {Kauffmann} G.,  2001,
  \mnras, 328, 726

\bibitem[\protect\citeauthoryear{{Stadel}, {Potter}, {Moore}, {Diemand},
  {Madau}, {Zemp}, {Kuhlen} \& {Quilis}}{{Stadel} et~al.}{2009}]{Stadel2009}
{Stadel} J.,  {Potter} D.,  {Moore} B.,  {Diemand} J.,  {Madau} P.,  {Zemp} M.,
   {Kuhlen} M.,    {Quilis} V.,  2009, \mnras, 398, L21

\bibitem[\protect\citeauthoryear{{Stevens}, {Martig}, {Croton} \&
  {Feng}}{{Stevens} et~al.}{2014}]{Stevens2014}
{Stevens} A.~R.~H.,  {Martig} M.,  {Croton} D.~J.,    {Feng} Y.,  2014, \mnras,
  445, 239

\bibitem[\protect\citeauthoryear{{Stinson}, {Seth}, {Katz}, {Wadsley},
  {Governato} \& {Quinn}}{{Stinson} et~al.}{2006}]{Stinson2006}
{Stinson} G.,  {Seth} A.,  {Katz} N.,  {Wadsley} J.,  {Governato} F.,
  {Quinn} T.,  2006, \mnras, 373, 1074

\bibitem[\protect\citeauthoryear{{Tasker}, {Brunino}, {Mitchell}, {Michielsen},
  {Hopton}, {Pearce}, {Bryan} \& {Theuns}}{{Tasker} et~al.}{2008}]{Tasker2008}
{Tasker} E.~J.,  {Brunino} R.,  {Mitchell} N.~L.,  {Michielsen} D.,  {Hopton}
  S.,  {Pearce} F.~R.,  {Bryan} G.~L.,    {Theuns} T.,  2008, \mnras, 390, 1267

\bibitem[\protect\citeauthoryear{{Teyssier}}{{Teyssier}}{2002}]{Teyssier2002}
{Teyssier} R.,  2002, \aap, 385, 337

\bibitem[\protect\citeauthoryear{{Teyssier}, {Pontzen}, {Dubois} \&
  {Read}}{{Teyssier} et~al.}{2013}]{Teyssier2013}
{Teyssier} R.,  {Pontzen} A.,  {Dubois} Y.,    {Read} J.~I.,  2013, \mnras,
  429, 3068

\bibitem[\protect\citeauthoryear{{Thacker} \& {Couchman}}{{Thacker} \&
  {Couchman}}{2001}]{Thacker2001}
{Thacker} R.~J.,  {Couchman} H.~M.~P.,  2001, \apjl, 555, L17

\bibitem[\protect\citeauthoryear{{Thi{\'e}baut}, {Pichon}, {Sousbie}, {Prunet}
  \& {Pogosyan}}{{Thi{\'e}baut} et~al.}{2008}]{Thiebaut2008}
{Thi{\'e}baut} J.,  {Pichon} C.,  {Sousbie} T.,  {Prunet} S.,    {Pogosyan} D.,
   2008, \mnras, 387, 397

\bibitem[\protect\citeauthoryear{{Tinker}, {Kravtsov}, {Klypin} \&
  {Abazajian}}{{Tinker} et~al.}{2008}]{Tinker2008}
{Tinker} J.,  {Kravtsov} A.~V.,  {Klypin} A.,    {Abazajian} K.~a.,  2008,
  \apj, 688, 709

\bibitem[\protect\citeauthoryear{{van den Bosch} \& {Ogiya}}{{van den Bosch} \&
  {Ogiya}}{2018}]{vandenBosch2018a}
{van den Bosch} F.~C.,  {Ogiya} G.,  2018, \mnras, 475, 4066

\bibitem[\protect\citeauthoryear{van~der Walt, Colbert \& Varoquaux}{van~der
  Walt et~al.}{2011}]{numpy}
van~der Walt S.,  Colbert S.~C.,    Varoquaux G.,  2011, CoRR, abs/1102.1523

\bibitem[\protect\citeauthoryear{{Vandenbroucke} \& {De
  Rijcke}}{{Vandenbroucke} \& {De Rijcke}}{2016}]{Vandenbroucke2016}
{Vandenbroucke} B.,  {De Rijcke} S.,  2016, Astronomy and Computing, 16, 109

\bibitem[\protect\citeauthoryear{{Vera-Ciro}, {Sales}, {Helmi}, {Frenk},
  {Navarro}, {Springel}, {Vogelsberger} \& {White}}{{Vera-Ciro}
  et~al.}{2011}]{VeraCiro2011}
{Vera-Ciro} C.~A.,  {Sales} L.~V.,  {Helmi} A.,  {Frenk} C.~S.,  {Navarro}
  J.~F.,  {Springel} V.,  {Vogelsberger} M.,    {White} S.~D.~M.,  2011,
  \mnras, 416, 1377

\bibitem[\protect\citeauthoryear{{Wang}, {Dutton}, {Stinson}, {Macci{\`o}},
  {Penzo}, {Kang}, {Keller} \& {Wadsley}}{{Wang} et~al.}{2015}]{Wang2015}
{Wang} L.,  {Dutton} A.~A.,  {Stinson} G.~S.,  {Macci{\`o}} A.~V.,  {Penzo} C.,
   {Kang} X.,  {Keller} B.~W.,    {Wadsley} J.,  2015, \mnras, 454, 83

\bibitem[\protect\citeauthoryear{{Wetzel}, {Hopkins}, {Kim},
  {Faucher-Gigu{\`e}re}, {Kere{\v{s}}} \& {Quataert}}{{Wetzel}
  et~al.}{2016}]{Wetzel2016}
{Wetzel} A.~R.,  {Hopkins} P.~F.,  {Kim} J.-h.,  {Faucher-Gigu{\`e}re} C.-A.,
  {Kere{\v{s}}} D.,    {Quataert} E.,  2016, \apjl, 827, L23

\bibitem[\protect\citeauthoryear{{Wiersma}, {Schaye} \& {Smith}}{{Wiersma}
  et~al.}{2009}]{Wiersma2009a}
{Wiersma} R.~P.~C.,  {Schaye} J.,    {Smith} B.~D.,  2009, \mnras, 393, 99

\bibitem[\protect\citeauthoryear{{Wiersma}, {Schaye}, {Theuns}, {Dalla Vecchia}
  \& {Tornatore}}{{Wiersma} et~al.}{2009}]{Wiersma2009b}
{Wiersma} R. P.~C.,  {Schaye} J.,  {Theuns} T.,  {Dalla Vecchia} C.,
  {Tornatore} L.,  2009, \mnras, 399, 574

\bibitem[\protect\citeauthoryear{{Yepes}, {Kates}, {Khokhlov} \&
  {Klypin}}{{Yepes} et~al.}{1997}]{Yepes1997}
{Yepes} G.,  {Kates} R.,  {Khokhlov} A.,    {Klypin} A.,  1997, \mnras, 284,
  235

\end{thebibliography}

\end{document}